\documentclass[10pt, oneside]{article}
\usepackage[top=2cm,left=3cm,right=3cm]{geometry}

\makeatletter
\@addtoreset{equation}{section}
\makeatother

\newcommand{\deftlen}{\setlength\arraycolsep{3.0pt}}
\newcommand{\setL}[1]{\setlength\arraycolsep{#1pt}}

\newcommand{\gapa}{&}

\renewcommand{\t}[1]{\widetilde{#1}}
\newcommand{\h}[1]{\widehat{#1}}
\newcommand{\ch}[1]{\check{#1}}

\newcommand{\bZ}{\mathbb{Z}}

\newcommand{\bC}{\mathbb{C}}
\newcommand{\bCp}{\mathbb{C}^{\prime}}

\newcommand{\bT}{\mathbb{T}}

\renewcommand{\bf}{\mathbf{f}}
\newcommand{\bw}{\mathbf{w}}
\newcommand{\bV}{\mathbf{V}}
\newcommand{\bR}{\mathbf{R}}

\newcommand{\A}{\mathcal{A}}
\newcommand{\B}{\mathcal{B}}
\newcommand{\cC}{\mathcal{C}}
\newcommand{\F}{\mathcal{F}}
\newcommand{\cI}{\mathcal{I}}
\renewcommand{\d}{\textrm{d}}

\newcommand{\cP}{\mathcal{P}}
\newcommand{\cPh}{\h{\cP}}
\newcommand{\cT}{\mathcal{T}}
\newcommand{\cU}{\mathcal{U}}
\newcommand{\sT}{\mathsf{T}}
\newcommand{\sU}{\mathsf{U}}
\newcommand{\sg}{\mathsf{g}}
\newcommand{\sh}{\mathsf{h}}
\newcommand{\cX}{\mathcal{X}}

\newcommand{\cG}{\mathcal{G}}

\newcommand{\G}{\mathfrak{G}}
\newcommand{\fA}{\mathfrak{A}}
\newcommand{\fB}{\mathfrak{B}}
\newcommand{\fC}{\mathfrak{C}}
\newcommand{\fD}{\mathfrak{D}}
\newcommand{\fd}{\mathfrak{d}}
\newcommand{\fDp}{\mathfrak{D}^{\prime}}
\newcommand{\fM}{\mathfrak{M}}
\newcommand{\fF}{\mathfrak{F}}
\newcommand{\fJ}{\mathfrak{J}}
\newcommand{\fP}{\mathfrak{P}}
\newcommand{\fPh}{\h{\fP}}

\newcommand{\fT}{\mathfrak{T}}
\newcommand{\fU}{\mathfrak{U}}

\newcommand{\M}{\mathcal{M}}
\newcommand{\W}{\mathcal{W}}

\newcommand{\sF}{\mathsf{F}}
\newcommand{\sFh}{\h{\sF}}
\newcommand{\sA}{\mathsf{A}}
\newcommand{\sB}{\mathsf{B}}

\newcommand{\dD}{\mathcal{D}}

\newcommand{\D}{\dD}
\newcommand{\Dp}{\dD'}

\newcommand{\sD}{\mathsf{D}}
\newcommand{\sDp}{\sD'}

\newcommand{\Fh}{\h{\F}}

\newcommand{\fFh}{\h{\fF}}

\newcommand{\ZZ}{\mathbb{Z}_{2} \times \mathbb{Z}_{2}}
\newcommand{\Mz}{\M_{\bZ^{2}_{2}}}
\newcommand{\SL}{\textrm{SL}(2,\bZ)}

\newcommand{\DevenA}{M}
\newcommand{\DoddA}{N}
\newcommand{\DpevenA}{M'}
\newcommand{\DpoddA}{N'}

\newcommand{\DevenB}{\mathsf{M}}
\newcommand{\DoddB}{\mathsf{N}}
\newcommand{\DpevenB}{\mathsf{M}'}
\newcommand{\DpoddB}{\mathsf{N}'}

\newcommand{\gapA}{ \quad&\quad }

\newcommand{\Jn}[1]{\mathcal{J}^{(#1)}}
\newcommand{\Jtn}[1]{\t{\mathcal{J}}^{(#1)}}
\newcommand{\J}{\mathcal{J}}
\newcommand{\fJn}[1]{\mathfrak{J}^{(#1)}}
\newcommand{\Hn}[1]{H^{(#1)}}
\newcommand{\Hnt}[1]{\t{H}^{(#1)}}
\newcommand{\cHn}[1]{\mathcal{H}^{(#1)}}

\newcommand{\Ha}{H^{(\ast)}}
\newcommand{\La}{\Lambda^{(\ast)}}
\newcommand{\Lp}{\Lambda^{(+)}}
\newcommand{\Lm}{\Lambda^{(-)}}
\newcommand{\Ln}[1]{\Lambda^{(#1)}}

\newcommand{\dual}{_{\ast}}

\newcommand{\Hbasis}{\mathfrak{a}}
\newcommand{\HtbasisA}{\omega}
\newcommand{\HtbasisB}{\nu}
\newcommand{\Basis}{\mathbf{e}}

\newcommand{\BasisA}{\Basis_{(\Hbasis)}}
\newcommand{\BasisB}{\Basis_{(\HtbasisA)}}
\newcommand{\BasisC}{\Basis_{(\HtbasisB)}}
\newcommand{\BasisX}{\mathbf{f}}
\newcommand{\BasisY}{\BasisX_{(\Hbasis)}}

\newcommand{\BasisZ}{\BasisX_{(\HtbasisB)}}

\newcommand{\w}[1]{\omega_{#1}}
\newcommand{\wt}[1]{\t{\omega}^{#1}}
\renewcommand{\v}[1]{\nu_{#1}}
\newcommand{\vt}[1]{\t{\nu}^{#1}}
\newcommand{\vol}{\mu_{\mathrm{vol}}}
\renewcommand{\a}[1]{\alpha_{#1}}
\renewcommand{\b}[1]{\beta^{#1}}
\newcommand{\syma}[1]{\mathfrak{a}_{#1}}
\newcommand{\symb}[1]{\mathfrak{b}^{#1}}
\renewcommand{\u}[1]{\underline{#1}}
\renewcommand{\i}{\iota}

\renewcommand{\P}[2]{^{#1}P_{#2}}

\newcommand{\nn}{\nonumber}
\newcommand{\beq}{\begin{eqnarray}}
\newcommand{\eeq}{\end{eqnarray}\deftlen}
\newcommand{\ba}{\begin{array} }
\newcommand{\ea}{\end{array}}
\newcommand{\ad}{^{\dag}}
\newcommand{\la}{\langle}
\newcommand{\ra}{\rangle}
\newcommand{\IP}[1]{\big\la \hspace{-0.8mm} \big\la \, #1 \, \big\ra \hspace{-0.8mm} \big\ra}
\newcommand{\hdot}{\diamond}
\newcommand{\fm}{\phantom{-}}
\newcommand{\merge}{\fM}

\newcommand{\bpm}{\begin{pmatrix}}
\newcommand{\epm}{\end{pmatrix}}

\newcommand{\Tdual}{\mathbf{T}}
\newcommand{\ModA}{\zeta_{1}}
\newcommand{\ModB}{\zeta_{2}}

\deftlen

\usepackage{ulem}
\usepackage[pdftex]{graphicx}
\usepackage{amsfonts}
\usepackage{amsmath}

\begin{document}

\title{The Generalised Geometry of Type II Non-Geometric Fluxes Under T and S Dualities}
\date{\ }
\author{George James Weatherill\footnote{email : gjw@soton.ac.uk}\\School of Physics and Astronomy,\\University of Southampton,\\Highfield, Southampton SO17 1BJ, UK}
\maketitle

\begin{abstract}
We examine the flux structures defined by NS-NS superpotentials of Type IIA and Type IIB string theories compactified on a particular class of internal spaces which include non-geometric flux contributions due to T duality or mirror symmetry. This is then extended to the Type IIB R-R sector through the use of S duality and then finally to its mirror dual Type IIA R-R sector, with note of how this sector breaks S duality invariance in Type IIA. The nilpotency and tadpole constraints associated with the fluxes induced by both dualities are derived, explicitly demonstrated to be mirror invariant and classified in terms of S duality multiplets. These results are then used to motivate the postulation of an additional symmetry for internal spaces which are their own mirror duals and an analysis is done of the resultant constraints for such a construction. 
\end{abstract}

\tableofcontents

\section{Introduction}

Calabi-Yau manifolds have been an area of considerable interest in string theory due to their partial breaking of supersymmetry in theories with internal spaces because of their $\textrm{SU}(3)$ holonomy. However, due to their complexity few explicit cases are known in sufficient detail to allow for detailed analysis of the dynamics of the moduli of the manifold. Furthermore, a larger class of spaces with the same partial supersymmetry breaking properties is obtained from the generalisation of $\textrm{SU}(3)$ holonomy to $\textrm{SU}(3)$ structure, which arise when some of the simplifying assumptions of Calabi-Yau spaces are relaxed, with such particular cases as `half flat' spaces being a well investigated area \cite{Gurrieri:2004dt,deCarlos:2005kh}. This relaxation can be viewed in terms of torsion classes which are in turn expressible in terms of fluxes such as the inclusion of fluxes like $H_{3} = \d B_{2}$. It was realised in the second string theory revolution that these fluxes can be used to stablise the moduli of the internal space, who gain vacuum expectation values due to the non-trivial potential induced by the fluxes \cite{Grana:2005jc}.Considerable work \cite{fluxes} has been done using orbifolds and orientifolds, whose construction via Kaluza-Klein reduction onto a six dimensional torus followed by quotienting by discrete groups allows for the behaviour of fluxes and their effect on moduli to be examined explicitly. 
\\

The flux structure differs between Type IIA and Type IIB theories, but in both cases there is an NS-NS $3$-form flux $H_{3}$ and R-R fluxes which live on the branes appropriate to each theory, which in the Type IIB case is only $F_{3}$ because $F_{1}=0=F_{5}$ due to $\Hn{1}$ and $\Hn{5}$ being empty. This is related to the cohomology condition $\B \wedge \Omega = 0$ which is closely tied to the non-total breaking of the supersymmetry and the compatibility of $\textrm{SU}(3)$ spin structures for the space \cite{Grana:2006hr}. In order for there to be invariance in the moduli effective theory these fluxes must be augmented by additional terms, such as the NS-NS geometric flux $\omega$ which in the case of Type IIA is sufficient to provide stable vacua \cite{Grimm,Derendinger,vz1,DeWolfe,cfi,af} but not in Type IIB as the moduli potential is independent of the K\"{a}hler moduli and thus those moduli are not stablised. This too can be addressed by the inclusion of further NS-NS fluxes \cite{Shelton:2005cf,Shelton:2006fd,Wecht:2007wu} which are such that the geometric interpretation of the space is no longer straight forward, making the new fluxes `non-geometric' in nature \cite{Hull:2005hk}, at which point the space can no longer be categorised by torsion classes.
\\

The properties of possible orbifolds, such as the number of moduli and their complex structure, have been fully classified \cite{Lust:2005dy,Lust:2006zg}, with the $\ZZ$ case being one of the most examined cases in terms of non-geometric fluxes \cite{Shelton:2005cf,Shelton:2006fd,Wecht:2007wu}, T duality invariance \cite{Aldazabal:2008aa,Font:2008vd} and S duality invariance \cite{Aldazabal:2006up,Guarino:2008ik}. One of the reasons for this is that the orbifold discrete group actions can be such that the six dimensional torus is factorised into three two dimensional tori and this can be further restricted by isotropy to make the three tori equal, reducing the number of moduli of each type to one, reducing the number of possible independent flux terms and thus making it one of the simplest non-trivial orbifolds. 
\\

In this paper we shall consider the superpotential from which the moduli potential can be constructed in its most general form when all possible T and S duality induced fluxes are included. The polynomial form of the superpotential is reexpressed in terms of fluxes, K\"{a}hler forms and holomorphic forms and then converted into expressions involving covariant derivatives. The formulation of the superpotential in terms of covariant derivatives has been seen in the case of twisted derivatives, where the inclusion of the $H_{3}$ flux causes the exterior derivative to gain a torsion-like term, $\d \to \d_{H} = \d + H\wedge$, with the associated internal space being a twisted Calabi-Yau. With the inclusion of geometric fluxes $p$-forms which were originally closed under $\d$ are no longer as $\d\lambda = \omega \cdot \lambda$. This can be further extended to also include the non-geometric fluxes, as considered explicitly in the $\bZ_{4}$ orientifold \cite{Ihl:2007ah}, with such spaces being generalised twisted Calabi-Yaus \cite{Grana:2006hr}. In the standard Calabi-Yau case the properties of $\d$ allow for the definition of the cohomologies $\Hn{p}$ but with the inclusion of geometric fluxes this is no longer the case. However, due to $p$-form truncations the bases for the untwisted $\Hn{p}$ can still be used as bases for the generalised twisted space \cite{Grana:2006hr}. For twisted Calabi-Yaus we have $\d H_{3}=0$ but in generalised twisted Calabi-Yaus the altered action of $\d$ is such that $\d H_{3} = \omega \cdot H_{3}$ but the $3$-form expansion of $H_{3}$ remains unchanged provided the new Bianchi constraint $\d H_{3} = \omega \cdot H_{3} = 0$ is satisfied. The inclusion of additional fluxes makes the Bianchi constraints non-trivial but we may still use the fluxless and closed $\Hn{p}$ as our bases. \cite{Grana:2006hr}
\\

It is this approach we will consider, first reexpressing the polynomial form of a Type IIB superpotential in terms of a pair of covariant derivatives, one for each flux sector, and the two holomorphic forms associated to the complex structure and K\"{a}hler moduli. The covariant derivatives, $\D$ and $\Dp$, are then expressed in two different ways; the first of which uses the tensor structure of the fluxes such as $H_{3} \sim H_{abc}$ and $\omega \sim \omega^{a}_{bc}$ and the second of which defines the fluxes by their actions on the twisted generalised $\Hn{p}$ basis elements. The Bianchi constraints are then obtained by requiring the nilpotency of the derivatives, which in the former case requires us to apply $\D^{2}$ to the $\Hn{p}$ basis elements and in the latter case the matrix representation must be nilpotency, $\u{\u{\D}}^{2}=0$ and likewise for $\Dp$. We then consider the Type IIA superpotential, expressing it in terms of two covariant derivatives, $\sD$ and $\sDp$, and the holomorphic forms but noting how it treats the NS-NS and R-R flux sectors differently and how the internal space's degrees of freedom are labelled in a different manner to the Type IIB case, before then considering the relationship between the two different flux formulations of Type IIA and Type IIB and demonstrating the equivalence of the Bianchi constraints. Using results obtained in these two sections we then consider S duality in Type IIB using the matrix representation of the covariant derivatives, obtaining sets of $\SL_{S}$ triplets and singlets which are either Bianchi constraints induced by T and S duality or tadpoles which can couple to the D-branes or O-planes wrapping particular cycles of the internal space. Finally we discuss the general structure of the superpotential and note the existence of two T duality inequivalent formulations of the superpotential. Based on this we postulate an additional symmetry between the two moduli spaces for internal spaces which are their own mirror dual. Throughout the paper the $\ZZ$ orientifold is used as an explicit example to illustrate results in terms of flux entries, however because it is often more illuminating to maintain the index structure only in the Appendix is the notation commonly found in the literature \cite{Wecht:2007wu,Font:2008vd,Guarino:2008ik} used. 

\subsection*{Motivation : The $\ZZ$ orientifold}
\label{sec:Motivation}

The $\ZZ$ orientifold, which we will refer to as $\Mz$, is extensively examined in the literature due to it possessing several important features.
\begin{itemize}
\item With $h^{2,1}(\Mz)=3$ the space possesses dynamical complex structure moduli, as well as K\"{a}hler and dilaton moduli, unlike such cases as the $\bZ_{3}$ orbifold whose complex structure is determined entirely by the orbifold group. \cite{Lust:2005dy,Lust:2006zg}

\item The orbifold group action can be set such that the six dimensional torus from which $\Mz$ is built factorises into three two dimensional tori. Each torus possesses a complex structure modulus and a K\"{a}hler modulus and these two moduli possess equivalent qualitative dynamics. \cite{Shelton:2005cf,Aldazabal:2006up}

\item The degrees of freedom can be reduced in a straight forward manner, by requiring `isotropy', while maintaining the existence of a modulus of each type. This is acheived by restricting the three tori to being equal to one another. \cite{Shelton:2005cf,Font:2008vd,Guarino:2008ik}

\item It is its own mirror dual, which is inherited from the self mirror dual nature of two dimensional tori. \cite{Aldazabal:2006up}
\end{itemize}
In their discussion of non-geometric fluxes induced by T duality \cite{Shelton:2005cf,Shelton:2006fd,Wecht:2007wu} use the isotropic case as their explict example for constructing non-geometric fluxes and examining the modifications they make to the space and this has been extended to include the S duality induced constraints \cite{Aldazabal:2006up}. These formed the basis of the explcit examples used in work done on solving these T \cite{Font:2008vd} and S \cite{Guarino:2008ik} duality constraints, which are the precursors to this work. As described in \cite{Aldazabal:2006up} an IIB/O3 superpotential on the $\ZZ$ orientifold which is seperately invariant under S and T duality transformations can be constructed from four fluxes, two of which are non-geometric in nature.
\beq
\ba {ccccc}\hline\hline
\gapA 	\gapA O(T^{0}) \gapA O(T^{1}) \gapA \\\hline\hline
\gapA O(S^{0})\gapA F_{3} = F_{abc} \gapA Q = Q^{ab}_{c} \gapA \\
\gapA O(S^{1})\gapA H_{3} = H_{abc} \gapA P = P^{ab}_{c} \gapA \\ \hline\hline
\ea \nn
\eeq
The Type IIB/O$3$ non-geometric fluxes $Q$ and $P$ must be contracted with a $4$-form but it is linear in the K\"{a}hler moduli $T_{a}$ \cite{Aldazabal:2006up} and so this is taken to be $\B_{c} = T_{a}\wt{a}$, where $\wt{a}$ are the $\Hn{4}$ versions of the $\w{a}$ and their specific relationship discussed later. From this we can construct a superpotential linear in the K\"{a}hler moduli and cubic in the complex structure moduli.
\beq
\label{eqn:ZZW}
W &=& \int_{\M}\Omega \wedge \Big(\; \Jn{0}\cdot( F_{3} - S H_{3})  + \mathcal{J}_{c}\cdot(Q-SP) \;\Big) 
\eeq
The lack of higher order K\"{a}hler moduli terms in the superpotential suggests not all fluxes have been included, as the Hodge numbers\footnote{Under the orientifold action used in the literature \cite{Aldazabal:2008aa,Aldazabal:2006up,Guarino:2008ik} $h^{1,1}=h^{1,1}_{+}$, there are no $2n$-forms odd under the projection.} of the orientifold are $h^{1,1}=h^{2,1}=3$. Type IIB possesses self S duality symmetry and it is this which motivates the existence of the R-R non-geometric flux $P$ in \cite{Aldazabal:2006up,Guarino:2008ik} so that by construction $\SL_{S}$ transformation to leave (\ref{eqn:ZZW}) invariant,  but Type IIA does not possess this symmetry, its S dual theory is eleven dimensional supergravity. The superpotential which is exactly mirrror dual to (\ref{eqn:ZZW}) will have $\SL_{S}$ invariance but this is not generically true, it requires certain Type IIA fluxes to be turned on which in turn induce further fluxes in the Type IIB mirror. To that end we wish to consider an additional set of Type IIB fluxes so that all possible moduli contributions to the superpotential are, in principle, included.
\beq
\ba {ccccc}\hline\hline
\gapA 	\gapA O(T^{2}) \gapA O(T^{3}) \gapA \\\hline\hline
\gapA O(S^{0})\gapA \bw = \bw^{a}_{bc} \gapA \bV = \bV^{abc} \gapA \\
\gapA O(S^{1})\gapA \bf = \bf^{a}_{bc} \gapA \bR = \bR^{abc} \gapA \\ \hline\hline
\ea \nn
\eeq
The existence of $\bw$ and $\bV$ is motivated \cite{Aldazabal:2008aa} by the effect S duality transformations have on Type IIA theories and their tensor structure  on $\Mz$ can be obtained by modular transformations, they are the complement or duals of the $3$-form fluxes and the non-geometric fluxes. However, to account for the K\"{a}hler moduli contributions to the superpotential they are not contracted with terms proportional to $\J$ and $\J^{3}$, as this would produce terms of $O(T)$ and $O(T^{3})$. This, as well as the issue with the K\"{a}hler dependence of the non-geometric contribution, can be resolved by defining a new set of K\"{a}hler forms $\Jtn{n}$, obtained by modular transformations\footnote{The definitions of $\G_{U}$ and $\G_{S}$ follow in the same manner as $\G_{T}$.} on the K\"{a}hler moduli.
\beq
\Jtn{n} &\equiv& \G_{T}(\Jn{n}) \quad,\quad \G_{T} : T_{a} \to -\frac{1}{T_{a}} \quad \forall a=1,\ldots,h^{1,1} \nn
\eeq
The order of K\"{a}hler moduli dependency in K\"{a}hler forms is $\Jn{n} \sim O(T^{n})$, but the $\G_{T}$ transformation changes this to $\Jtn{n} \sim O(T^{3-n})$ and the contracted tensor $Q\cdot\Jtn{2}$ is therefore linear in the K\"{a}hler moduli, as is known to be required. However, if we are to write the superpotential integrand so that it's K\"{a}hler dependence is only constructed from $\Jtn{n}$ K\"{a}hler forms then the object which gives the $O(T^{0})$ contribution must contract with the $6$-form $\Jtn{3}$ and so $F_{3}$ is not viewed in terms of components $F_{abc}$ but rather $\t{F}^{abc}$, the Hodge dual $\ast_{6}$ of $F_{abc}$. Similarly the tensor structure of the term contributing the cubic K\"{a}hler moduli dependency is not viewed as $V^{abc}$ but rather it's Hodge dual, $\t{\bV}_{abc}$. The space also possesses structure which allows modular invariance in both the K\"{a}hler and complex structure moduli \cite{Aldazabal:2006up} and through this analysis precisely this kind of alternative tensor structure for the fluxes arises. Using tildes to represent the dual tensor the superpotential then can be written in a way which makes its K\"{a}hler moduli dependence much clearer.
\beq
\label{eqn:GenW}
W &=& \int_{\M}\Omega \wedge \Big(\; \t{F}\cdot\Jtn{3} + Q\cdot\Jtn{2} + \mathbf{w}\cdot\Jtn{1} + \t{\mathbf{V}}\cdot\Jtn{0} \;\Big) + \nn \\
&& \qquad\qquad + (-S)\int_{\M}\Omega \wedge \Big(\; \t{H}\cdot\Jtn{3} + P\cdot\Jtn{2} + \mathbf{f}\cdot\Jtn{1} + \t{\mathbf{R}}\cdot\Jtn{0} \;\Big) 
\eeq
Because of Bianchi constrants and possible orientifold projections not all of these fluxes can be non-zero at the same time but in principle this is the most general superpotential which can arise from consider T and S duality induced fluxes. From this point onwards we will not consider such apriori effects as the orientifold projection, rather we will assume all of these fluxes can in principle be turned on at once and construct various flux structures from them, with the assumption that once these structures are found they can be simplified down by such things as which fluxes or forms are even or odd under the projection. As will be observed, the symmetry between the two moduli spaces is more easily seen by the inclusion of all of the terms in (\ref{eqn:GenW}) and the alteration of the index structure of such fluxes as $Q$ or $H$ by the Hodge star does not alter the constraints on the fluxes.

\section{Notation}
\label{Section : Notation}

Before beginning our analysis of the superpotential in terms of fluxes we need to define how we represent the fluxes, moduli, $p$-forms, $\Hn{p}$ bases in general. Much of the notation is choosen to match \cite{ModuliSpace}. The specific case of the $\ZZ$ orientifold is done in Section \ref{sec:ZZ}.

\subsection{Cohomologies and $p$-forms}

Working on a connected orientated six dimensional manifold $\M$ we wish to define a set of bases for the $p$-form spaces $\Ln{p}(\M)$, whose direct sum form the ring of $p$-forms $\La(\M) = \oplus_{p=0}^{6}\Ln{p}(\M)$. Assuming that all fluxes are set to zero we can also straightforwardly define the cohomologies $\Hn{p}$ of $\M$ and their sum $\Ha(\M) = \oplus_{p=0}^{6}\Hn{p}(\M)$. By orientated connectedness $\M$ has a unique $6$-form $\vol$ which satisfies $\int_{\M} \vol = 1$ and so given the set of tangent $1$-forms $\eta^{\sigma}$ which span $\Ln{1}(\M)$ we also have their Hodge duals $\ch{\eta}^{\tau}$ which form the basis of $\Ln{5}(\M)$, defined by $\eta^{\sigma} \wedge \ast\eta^{\tau} = \delta^{\sigma\tau}\vol = \eta^{\sigma} \wedge \ch{\eta}^{\tau}$.
\beq 
\ba {cccc}\hline\hline
\quad p \gapA \textrm{basis} \gapA \textrm{dim} \gapA \textrm{Elements} \\\hline\hline
1 & \eta^{\sigma} 		& 6 		& \eta^{1},\cdots,\eta^{6} \\
5 & \ch{\eta}^{\sigma} 		& 6 		& \ch{\eta}^{1},\cdots,\ch{\eta}^{6} \\\hline\hline
\ea  
\eeq
Though we construct $(p>1)$-forms from wedge products of the $\eta^{\sigma}$, we are making the stipulation that $\Hn{1}(\M)$, and thus by duality $\Hn{5}(\M)$, is empty. We will also make use of interior forms $\i_{\sigma}$, which satisfy $\iota_{\sigma}(\eta^{\tau}) = \delta^{\tau}_{\sigma}$ and so belong to the space dual to $\Ln{1}$ which we denote as $\Ln{1}\dual$. Elements of the other $\Ln{p}\dual$ are defined as exterior products of these forms such as $\i_{abc} \equiv \i_{a}\i_{b}\i_{c}$, with the restriction to $\Hn{p}\dual$ being straightforward though we will later consider them in more detail.

\subsection{Complex structure moduli space}

A generic element of $\Ln{3}(\M)$ can be constructed from wedge products of three of the tangent $1$-form, $\eta^{abc} \in \Ln{3}(\M)$ and fluxes can be expanded in this form.
\beq
F = \frac{1}{3!}F_{abc}\eta^{abc}
\eeq
The properties of $\M$ determine the structure of the coefficients $F_{abc}$ but it is more convenient to work with the basis of $\Hn{3}(\M)$ for fluxes. On $\M$ we can define $h^{2,1}+1$ pairs of $3$-cycles $A^{I}$, $B_{J}$, $I,J=0,\cdots,h^{2,1}$ with a symplectic structure in their intersection numbers.
\beq
A^{I}\cap B_{J} = -B_{J}\cap A^{I} = \delta_{J}^{I} \qquad A^{I}\cap A^{J} = B_{I}\cap B_{J} = 0 \nn
\eeq
The ($\a{I}$,$\b{J}$) basis of $\Hn{3}$ is defined as the Poincare dual of the $(A^{I},B_{J})$ homology $3$-cycles.
\beq
\label{eqn:H3basis}
\int_{A^{J}} \a{I} = \int_{\M} \a{I} \wedge \b{J} = \delta_{J}^{I} \qquad \int_{B_{J}} \b{I} = \int_{\M} \b{I} \wedge \a{J} = -\delta_{J}^{I} 
\eeq
The set of $h^{2,1}+1$ coordinates $\cU_{I}$ are defined by the $3$-cycles and the holomorphic $3$-form $\Omega$, which in turn define $h^{2,1}$ projective coordinates, the complex structure moduli $U_{i}$.
\beq
\cU_{I} = \int_{A^{I}} \Omega \quad \Rightarrow \quad U_{i} = \frac{\cU_{i}}{\cU_{0}} 
\eeq
Therefore the holomorphic $3$-form can then be expanded in terms of these coordinates, with the factor of $-1$ arising due to the symplectic definitions of the basis $3$-forms.
\beq
\Omega = \frac{1}{3!}\Omega_{abc}\eta^{abc} = \cU_{I}\a{I} - \cU^{J}\b{J} \nn
\eeq
The $\cU^{I}$ are defined as the $\cU_{I}$ derivatives of the complex structure prepotential $\F$ and all functions and tensors dependent upon $\cU_{0}$ are then evaluated at $\cU_{0}=1$.
\beq
\label{eqn:Uintersection}
\cU^{I} = \int_{B_{J}} \Omega \equiv \frac{\partial \F}{\partial \cU_{I}} \qquad \F = \frac{1}{3!}\frac{1}{\cU_{0}}\F_{ijk}\;\cU_{i}\cU_{j}\cU_{k}
\eeq
The K\"{a}hler potential for the complex structure moduli is also constructable from the holomorphic $3$-form.
\beq
K^{2,1} = -\ln \left( i \int_{\M}\Omega \wedge \overline{\Omega} \right) \equiv -\ln i\left( \overline{\cU}_{I}\cU^{I} - \cU_{J}\overline{\cU}^{J} \right) = -\ln i\left( \overline{\cU}_{I}\frac{\partial \F}{\partial \cU_{I}} - \cU_{J}\frac{\partial \overline{\F}}{\partial \overline{\cU}_{J}} \right) \nn
\eeq
We will find it more convenient to work in a slightly ammended sympletic basis $(\syma{I},\symb{J})$ which is obtained from $(\a{I},\b{J})$ by a sympletic transformation and which forces us to define a set of moduli $\fU$ such that $\Omega$ is left schematically invariant.
\beq
\ba{ccccccccccccccccccccccc}
(&\syma{0}&,&\syma{i}&,&\symb{0}&,&\symb{j}&) &\equiv& (&\a{0}&,&\b{i}&,&\b{0}&,&-\a{j}&) \\
(&\fU_{0}&,&\fU_{i}&,&\fU^{0}&,&\fU^{j}&) &\equiv& (&\cU_{0}&,&-\cU^{i}&,&\cU^{0}&,&\cU_{j}&)
\ea \quad \Rightarrow \quad \Omega = \Big\{ \ba{ccc}\cU_{I}\a{I}&-&\cU^{J}\b{J}\\ \fU_{I}\syma{I}&-&\fU^{J}\symb{J} \ea 
\eeq
Though the definition of this second set of sympletic $3$-forms and related moduli has no apriori motivation the reason for this exchange of $\a{i}$ and $\b{i}$ will be given in our discussion of Type IIB fluxes, as well as the action of T duality on R-R fluxes. Also for later convenience we define two bilinear forms associated to the intersection numbers of the cohomology bases, the first of which is $g$, defined on $\phi \in \Hn{p}(\M)$ and $\varphi \in \Hn{6-p}(\M)$ mapping them to $\bZ$.
\beq
g(\phi,\varphi) \equiv \int_{\M} \phi \wedge \varphi  
\eeq
If $\phi$ and $\varphi$ are vectors of forms then $g(\phi,\varphi)$ is a matrix with entries defined by $g(\phi_{m},\varphi_{n})$. The basis vector for the complex structure moduli space we take to be $\BasisA = \bpm \syma{I} , \symb{J} \epm$ so that $g$ takes on the form of the canonical symplectic form.
\beq
\ba {cccccccccc}
g_{\Hbasis} &\equiv& g(\BasisA,\BasisA) &=& \bpm g(\syma{I}, \syma{J}) & g(\syma{I}, \symb{J}) \\ g(\symb{I}, \syma{J}) & g(\symb{I}, \symb{J}) \epm &=& \bpm g_{IJ} & g_{I}^{\phantom{I}J} \\ g^{I}_{\phantom{I}J} & g^{IJ} \epm &=& \bpm 0 & 1 \\ -1 & 0 \epm \otimes \mathbb{I}_{h^{2,1}+1} \ea \nn
\eeq
By construction this choice of basis leads to a formulation which is invariant under transformations $\BasisA \to L \cdot \BasisA$ where $L \in \textrm{Sp}(h^{2,1}+1)$ as $L^{\top} \cdot g_{\Hbasis} \cdot L = g_{\Hbasis}$. The second bilinear form is $h(\phi,\varphi) \equiv g(\phi,\sigma \cdot \varphi)$ where $\sigma = \bpm 0 & 1 \\ 1 & 0 \epm \otimes \mathbb{I}_{n}$ and $2n$ is the dimension of the $\varphi$.
\beq
\ba {cccccccccc}
h_{\Hbasis} &\equiv& g(\BasisA,\sigma \cdot \BasisA) &=& \bpm g_{I}^{\phantom{I}J} & g_{IJ} \\ g^{IJ} & g^{I}_{\phantom{I}J} \epm &=& \bpm 1 & 0 \\ 0 & -1 \epm \otimes \mathbb{I}_{h^{2,1}+1} \ea \nn
\eeq
Henceforth unless the dimensions are ambigious we shall drop the subscript on $\mathbb{I}$. With these two matrices defined by $\BasisA$ we can express superpotential-like integrals in terms of matrices and vectors, such as two elements $C$ and $D$ of $\Hn{3}$ with the expansion in terms of the symplectic basis.
\beq
\ba {cccccccc}
C &=& C_{I}\syma{I} - C^{J}\symb{J} &=& \bpm C_{I} & C^{J} \epm \bpm \mathbb{I} & 0 \\ 0 & -\mathbb{I} \epm\bpm \syma{I} \\ \symb{J} \epm &\equiv& \u{C}^{\top} \cdot h_{\Hbasis} \cdot \BasisA  \ea  \nn
\eeq
The expansion of $D$ in the symplectic basis follows in the same manner and using these vector expressions we can represent the inner product between these two $3$-forms in a particularly straight forward manner. 
\beq
g(C,D) = \int_{\M} C \wedge D = D_{I}C^{I} - D^{J}C_{J} = \u{C}^{\top} \cdot h_{\Hbasis} \cdot g_{\Hbasis} \cdot h_{\Hbasis}^{\top} \cdot \u{D} = \u{D}^{\top} \cdot g_{\Hbasis} \cdot \u{C} \nn
\eeq
Expanding $\u{D}^{\top} \cdot g_{\Hbasis} \cdot \u{C}$ in terms of the components we note that the structure of the expression is preserved, there is a factor of $-1$ on half of the terms and manifestly symplectic transformations on the $\Hn{3} $ basis leave $g(C,D)$ invariant. The vector associated to the holomorphic form is the moduli vector, $\Omega = \u{\fU}^{\top} \cdot h_{\Hbasis} \cdot \BasisA$ and so we can obtain an expression for a generic Type IIB superpotential contribution due to some $3$-form flux $G$.
\beq
\int_{\M} \Omega \wedge G &=& \int_{\M}(\fU_{I}\syma{I} - \fU^{J}\symb{J}) \wedge (G_{I}\syma{I} - G^{J}\symb{J}) =  \fU^{J}G_{J}-\fU_{I}G^{I} = g\Big( \Omega , G \Big)\nn 
\eeq
The $\d$ defined deRham cohomology $\Hn{3}$ is decomposable in terms of Dolbeault operator defined cohomologies $\Hn{3-n,n}$, which are spanned by forms related to the holomorphic $3$-form, its complex conjugate and their K\"{a}hler derivatives $D_{U_{i}} \equiv \partial_{U_{i}} + \partial_{U_{i}}K^{(2,1)}$, due to the manner in which $\Omega$ is defined as a holomorphic $3$-form.
\beq
\ba {cccccccccc}
\Hn{3} &\equiv &\Hn{3,0} &\oplus& \Hn{2,1}& \oplus & \Hn{1,2} &\oplus & \Hn{0,3} \\
&\equiv & \la \Omega \ra &\oplus& \la D_{U_{i}}\Omega \ra & \oplus & \la \overline{D_{U_{i}}\Omega} \ra &\oplus & \la \overline{\Omega} \ra 
\ea  \nn
\eeq
Rather then work these spaces we instead work with the spaces spanned by the ($\a{I}$,$\b{J}$) sympletic basis elements, which we denote by $\cHn{3-n,n}$.
\beq
\ba {cccccccccc}
\Hn{3} & = & \la \a{0} \ra &\oplus& \la \a{i} \ra& \oplus & \la \b{j} \ra &\oplus & \la \b{0} \ra \\
 &\equiv & \cHn{3,0} &\oplus& \cHn{2,1}& \oplus & \cHn{1,2} &\oplus & \cHn{0,3} & \\
 & = & \la \syma{0} \ra &\oplus& \la \symb{i} \ra& \oplus & \la \syma{j} \ra &\oplus & \la \symb{0} \ra 
\ea  \nn
\eeq

\subsection{K\"{a}hler moduli space}

We now consider the K\"{a}hler modui space, constructing it from even dimensional cycles as opposed to the odd dimensional cycles of the complex structure moduli space, but with the guiding principle that the two moduli spaces must be as close in structure as possible. The set of complex K\"{a}hler moduli are defined as the coefficients of the expansion of the complex K\"{a}hler $2$-form $\J \equiv B+iJ$, in terms of the $(1,1)$ forms $\w{a}$, $\J = T_{a}\w{a}$, although when we consider the K\"{a}hler moduli holomorphic form we will find it more convenient to work with a set of projective coordinates $\cT_{A}$ defined in the same manner as the complex structure moduli space, with $T_{a} \equiv \frac{\cT_{a}}{\cT_{0}}$. The $(1,1)$-forms have a set of $h^{1,1}$ $2$-cycles associated to them, $\A^{a}$ with $a,b=1,\cdots,h^{1,1}$, which in turn have a set of $4$-cycles $\B_{b}$ partners which define the set of $h^{1,1}$ $(2,2)$-forms, $\wt{b}$. These two sets of forms are partnered in the same manner in which $\syma{I}$ and $\symb{J}$ are partnered but have non-trivial intersection numbers $g_{a}^{\phantom{a}b}$ and $g_{\phantom{a}b}^{a}$ defined in the same manner as the sympletic interaction numbers previously.
\beq
\label{eqn:Ht3basis}
\int_{\A^{b}} \w{a} = \int_{\M} \w{a} \wedge \wt{b} = g_{a}^{\phantom{a}b} \qquad \int_{\B_{b}} \wt{a} = \int_{\M} \wt{a} \wedge \w{b} = g_{\phantom{a}b}^{a}  \nn
\eeq
We can expand this basis to include the $(0,0)$ and $(3,3)$ forms, $\w{0} \equiv 1$ and $\wt{0} \equiv \vol$. The $6$-form $\wt{0}$ is associated with $\B_{0} \equiv \M$ itself, which is the only $6$-cycle if $\M$ is connected. In the case of $\w{0}$ we have to associate it with a $0$-cycle point $\A^{0}$, which can be any point other than orbifold singularities.
\beq
\int_{\A^{0}} \w{0} = \int_{\M} \w{0} \wedge \wt{0} = 1 \qquad \int_{\B_{0}} \wt{0} = \int_{\M} \wt{0} \wedge \w{0} = 1  \nn
\eeq
For the purposes of clarity we shall sometimes regard expressions such as $\w{a} \wedge \wt{a}$ (no sum) not as $\wt{0}$ but $\vol$, so that the pair of forms $\w{0}$ and $\wt{0}$ satisfy the same notation. As in the complex structure case we can define a vector of these even forms $\BasisB = \bpm \w{0} & \w{a} & \wt{0} & \wt{b} \epm$, from which the general intersection matrix $g$ of the K\"{a}hler moduli space is built and the associated second bilinear form $h$, where we once again have written the elements such that $\wt{0}$ comes before the $\wt{a}$, allowing us to extend our index labelling to $A,B=0,1,\ldots,h^{1,1}$ in the same manner as the sympletic form indices.
\beq
\ba {cccccccc}
g_{\HtbasisA} &\equiv& g(\BasisB,\BasisB) &=& \bpm g(\w{A}, \w{B}) & g(\w{A}, \wt{B}) \\ g(\wt{A}, \w{B}) & g(\wt{A}, \wt{B}) \epm &=& \bpm 0 & g_{A}^{\phantom{A}B} \\ g^{A}_{\phantom{A}B} & 0 \epm \\\\
h_{\HtbasisA} &\equiv& h(\BasisB,\BasisB) &=& \bpm g(\w{A}, \wt{B}) & g(\w{A}, \w{B}) \\ g(\wt{A}, \wt{B}) & g(\wt{A}, \w{B}) \epm &=& \bpm g_{A}^{\phantom{A}B} & 0 \\ 0 & g^{A}_{\phantom{A}B} \epm \ea \nn
\eeq
Unlike the complex structure case the intersection matrix $g_{\HtbasisA}$ is symmetric due to the commutative properties of the even forms but we have not assumed any specific form of $g_{A}^{\phantom{A}B} = \left( g^{B}_{\phantom{B}A} \right)^{\top}$ and a general change of basis on either the $(1,1)$ or $(2,2)$-forms would alter the intersection numbers and thus $g_{\HtbasisA}$. We can explicitly construct a convenient set of intersection numbers by considering the K\"{a}hler moduli version of the holomorphic $3$-form, obtained by exponentiating the complexified K\"{a}hler form.
\beq
e^{\J} \equiv \mho &=& \sum_{n=0}^{\infty}\frac{1}{n!}\J^{n} = \sum_{n=0}^{\infty}\Jn{n} = \Jn{0} + \Jn{1} + \Jn{2} + \Jn{3} \nn \\
&=& \cT_{0}\,\w{0} + \cT_{a}\,\w{a} + \frac{1}{2!}\cT_{a}\cT_{b}\,\w{a}\wedge\w{b} + \frac{1}{3!}\cT_{a}\cT_{b}\cT_{c}\,\w{a}\wedge\w{b}\wedge\w{c} 
\eeq
The first and second terms in $\mho$ have a simple expansion in terms of the $\Hn{0,0}$ and $\Hn{1,1}$ bases and bear a close resemblence to the $\syma{I}$ terms in $\Omega$. However, the latter two terms are not expressed in terms of the $\Hn{2,2}$ and $\Hn{3,3}$ forms we have already seen, but combinations of the $(1,1)$-forms, where the latter term defines a set of intersection numbers by $\w{a}\wedge\w{b}\wedge\w{c} = \kappa_{abc}\,\vol$. The $(2,2)$-forms can be written in terms of the $(1,1)$-forms by $\wt{a} = f^{abc}\w{b}\wedge\w{c}$, which relate to the intersection numbers $\kappa_{abc}$ by $g^{a}_{\phantom{a}b} = f^{acd}\kappa_{bcd}$. In order to make the expansion of $\mho$ as close to the structure of the expansion of $\Omega$ we make an ansatz for $\mho$ so that the latter two terms are written in the more natural basis of the $\wt{A}$.
\beq
\Omega = \fU_{0}\syma{0} + \fU_{i}\syma{i} - \fU^{j}\symb{j} - \fU^{0}\symb{0} \quad , \quad \mho = \cT_{0}\,\w{0} + \cT_{a}\,\w{a} + \cT^{b}\wt{b} + \cT^{0}\wt{0} \nn 
\eeq
The lack of minus signs is due to the $h_{\HtbasisA}$ being positive definite, compared to $h_{\Hbasis}$, and because of this there is no pre-potential associated to $\mho$, which requires a sympletic structure to the forms. Such a structure can be given to these forms by the use of Grassmannian variables \cite{ModuliSpace} but this would introduce the notion of Grassmannian fluxes, an unwanted complication. If we treat the $A,B=0$ cases as seperate from the $A,B = 1,\ldots,h^{1,1}$ cases then we can define a reduced pre-potential, in that $\cT^{0}$ is not the $\cT_{0}$ derivative of a pre-potential but infact plays the role of the reduced pre-potential. To obtain this we compare the coefficients of each of the seperate $2n$-form terms and express $\cT^{a}$ and $\cT^{0}$ in terms of intersection numbers and $\cT_{A}$. Comparing coefficients we can express $\cT^{a}$ in terms of the derivatives of $\cT^{0}$, in an analogous way to the complex structure moduli space.
\beq
\cT^{0} = \frac{1}{3!}\kappa_{abc}\cT_{a}\cT_{b}\cT_{c} \quad \Rightarrow \quad \cT^{b}g_{\phantom{b}a}^{b} = \frac{1}{2!}\kappa_{abc}\cT_{b}\cT_{c} = \frac{\partial \cT^{0}}{\partial \cT_{a}}
\eeq
Therefore, if we set $\cT^{a} = \frac{\partial \cT^{0}}{\partial \cT_{a}}$, inline with the complex structure case of (\ref{eqn:Uintersection}), then the intersection matrix $g_{\phantom{b}a}^{b}$ reduces to $\delta_{a}^{b}$ and the associated $h_{\HtbasisA}$ to the identity.
\beq
\mho = \cT_{A}\w{A} + \cT^{B}\wt{B} = \w{0} + T_{a}\w{a} + \frac{\partial \cG}{\partial T_{b}} \wt{b} + \cG \wt{0} \qquad \cG = \frac{1}{3!}\kappa_{abc}\;\cT_{a}\cT_{b}\cT_{c} \nn 
\eeq
With $g_{\HtbasisA} = \sigma \otimes \mathbb{I}_{h^{1,1}+1}$ the symmetry group associated to $\BasisB$ is not the sympletic group but is isomorphic to $\textrm{O}(h^{1,1}+1,h^{1,1}+1)$ instead because of the invariant under transformations $\BasisB \to L_{0} \cdot L \cdot \BasisA$ where $L \in \textrm{O}(h^{1,1}+1,h^{1,1}+1)$. This follows from the fact $L_{0}^{\top} \cdot g_{\HtbasisA} \cdot L_{0} = \eta$ where $\eta$ is the Lorentzian metric with $h^{1,1}+1$ positive and negative entries. As a result of this the K\"{a}hler potential for the moduli can be expressed generically in terms of the moduli only.
\beq
K^{1,1} = -\ln \left( \int_{\M}\mho \wedge \overline{\mho} \right) = -\ln \left( \overline{\cT}_{A}\cT^{A} + \cT_{B}\overline{\cT}^{B} \right)\nn
\eeq
The factor of $i$ has been dropped because $\overline{\cT}_{A}\cT^{A} + \cT_{B}\overline{\cT}^{B}$ is real, unlike $\overline{\fU}_{I}\fU^{I} - \fU_{J}\overline{\fU}^{J}$, which is purely imaginary. Given $X \in \Hnt{3}$ we define its components by the expansion $X = X_{A}\w{A} + X^{B}\wt{B}$, and likewise for $Y \in \Hnt{3}$, choosen so that integration over $\M$ preserves the structure of the expansion.
\beq
g(X,Y) = g_{\HtbasisA}(X,Y) = \int_{\M} X \wedge Y = X_{A}Y^{A} + X^{B}Y_{B} \nn
\eeq
As in the complex structure case we can associated a vector to any element of $\Hnt{3}$ by its coefficients in the $\BasisB$ basis.
\beq
\ba {cccccccc}
X &=& X_{A}\w{A} + X^{B}\wt{B} &=& \bpm X_{A} & X^{B} \epm \bpm \mathbb{I} & 0 \\ 0 & \mathbb{I} \epm\bpm \w{A} \\ \wt{B} \epm &\equiv& \u{X}^{\top} \cdot h_{\HtbasisA} \cdot \BasisB \ea  \nn
\eeq
We again express the inner product between two elements of a cohomology in terms of vectors.
\beq
g(X,Y) = \int_{\M} X \wedge Y = X_{A}Y^{A} + X^{B}Y_{B} = \u{X}^{\top} \cdot h_{\HtbasisA} \cdot g_{\HtbasisA} \cdot h_{\HtbasisA}^{\top} \cdot \u{Y} = \u{Y}^{\top} \cdot g_{\HtbasisA} \cdot \u{X} \nn
\eeq
In the definition of the vector associated to $\Omega$ the explicit $h_{\Hbasis}$ factorisation was required for the entries of the vector to be the complex structure moduli and while we have defined the $\Hnt{3}$ basis to make $h_{\HtbasisA}$ the identity we include $h_{\HtbasisA}$ in the definition of the $\Hnt{3}$ vectors on grounds of symmetry. As a result in both moduli spaces the holomorphic form has a vector associated to it whose entries are the moduli, $\mho = \u{\cT}^{\top} \cdot h_{\HtbasisA} \cdot \BasisB$. As with $\Hn{3}$ the cohomology $\Hnt{3}$ decomposes into subspaces of $\Ha$ which are spanned by forms constructed from $\mho$ in the same manner as in the $\Hn{3}$ case by the use of a K\"{a}hler derivative of the K\"{a}hler moduli, $D_{T_{a}} = \partial_{T_{a}} + \partial_{T_{a}}K^{1,1}$.
\beq
\ba {cccccccccc}
\Hnt{3} &\equiv & \la \mho \ra &\oplus& \la D_{T_{a}}\mho \ra & \oplus & \la \overline{D_{T_{a}}\mho} \ra &\oplus & \la \overline{\mho} \ra 
\ea  \nn
\eeq
As in the case of $\Hn{3}$ it is more convenient to work with the Dolbeault operator defined cohomologies but unlike the complex structure case they are synonymous with deRham cohomologies due to the properties of spaces with SU$(3)$ structure such as $h^{2,0}=0$, resulting in such identities as $b_{2n} = h^{n,n}$ where $b_{p}$ is the $p$'th Betti number.
\beq
\ba {cccccccccc}
\Hnt{3} & = & \la\w{0}\ra &\oplus& \la\w{a}\ra& \oplus & \la\wt{b}\ra &\oplus &  \la\wt{0}\ra \\
 & = & \Hn{0} &\oplus& \Hn{2}& \oplus & \Hn{4} &\oplus & \Hn{6} & \\
 &\equiv & \cHn{0,0} &\oplus& \cHn{1,1}& \oplus & \cHn{2,2} &\oplus & \cHn{3,3} &
\ea  \nn
\eeq

\subsection{Additional definitions} 

The two moduli spaces split $\Ha$ into $\Hn{3}$ and $\Hnt{3}$ and so we define the $\La$ version of this splitting, but due to the non-emptiness of $\Ln{1}$ and $\Ln{5}$ we denote them as $\Lm$ and $\Lp$ respectively.
\beq
\ba {cccccccccc}
\Lm & = & \Ln{1} &\oplus& \Ln{3}& \oplus & \Ln{5} \\
 & = & \la \eta^{\tau} \ra &\oplus& \la \eta^{\tau\sigma\rho} \ra& \oplus & \la \eta^{\tau\sigma\rho\lambda\upsilon} \ra \\
\Lp & = & \Ln{0} &\oplus& \Ln{2}& \oplus & \Ln{4} &\oplus & \Ln{6} & \\
 & = & \la 1 \ra &\oplus& \la \eta^{\tau\sigma} \ra& \oplus & \la \eta^{\tau\sigma\rho\lambda} \ra &\oplus &  \la \eta^{\tau\sigma\rho\lambda\upsilon\chi} \ra 
\ea  \nn
\eeq
The compatibility condition $\J \wedge \Omega = 0$ is automatically satisfied if $\Hn{5}$ is empty and can be reexpressed in terms of the $\Hn{3}$ and $\Hnt{3}$ bases.
\beq
\label{eqn:compat}
\ba{ccc} \d \w{a} &=& 0 \\
\d \syma{I} &=& 0 \\
\d \symb{J} &=& 0 \ea \quad \Leftrightarrow \quad \ba{ccc} \d (\w{a}\wedge \syma{I}) &=& 0 \\
\d(\w{a} \wedge \symb{J}) &=& 0 \ea \quad \Leftrightarrow \quad \ba{ccc} \w{a}\wedge \syma{I} \\ \w{a}\wedge \symb{J}  \ea \Big\} \in \Hn{5} \quad \Rightarrow \quad \ba{ccc} \w{a}\wedge \syma{I} &=& 0 \\ \w{a}\wedge \symb{J} &=& 0 \ea
\eeq
The general expansion $\chi = \u{\chi}\cdot h \cdot \Basis$ is used to define the vector associated to any element of $\Ha$, irrespective of the cohomologies $\chi$ has support in. The general elements of $\Ha$ can be written as $C+X$ and $D+Y$ and since $g \equiv g(\Basis,\Basis)$ is the direct sum of $g_{\Hbasis}$ and $g_{\HtbasisA}$ we have that $g(C+X,D+Y) = g_{\Hbasis}(C,D)+g_{\HtbasisA}(X,Y)$. The natural, positive definite, inner product on $\La$ follows from $g$ by use of the Hodge star $\ast \equiv \ast_{6}$, whose action on the cohomologies can be written as a matrix $\u{\u{\ast}}$ by defining $\ast(\chi) = \u{\chi}^{\top} \cdot h \cdot \u{\u{\ast}} \cdot \Basis$. 
\beq
\IP{\phi,\chi} \equiv g\Big(\phi,\ast(\chi)\Big) &=& \int_{\M} \Big( \u{\phi}^{\top} \cdot h \cdot \Basis \Big) \wedge \Big( \u{\chi}^{\top} \cdot h \cdot \u{\u{\ast}} \cdot \Basis \Big) \nn \\
&=& \int_{\M} \Big( \u{\phi}^{\top} \cdot h \cdot \Basis \Big) \wedge \Big( \u{\chi}^{\top} \cdot \textrm{Ad}(\u{\u{\ast}}) \cdot h \cdot \Basis \Big) \nn \\
&=& \u{\chi}^{\top} \cdot \textrm{Ad}(\u{\u{\ast}}) \cdot g \cdot \u{\phi} = \u{\chi}^{\top} \cdot \u{\phi} \nn 
\eeq
With $\textrm{Ad}(\u{\u{\ast}}) \equiv h \cdot \u{\u{\ast}} \cdot h^{\top} = g^{\top}$ we can make use of the definitions of of $g_{\omega}$ and $g_{\alpha}$ to obtain the result $\u{\u{\ast}} = g$.
\\

In Type IIB theories the superpotential integrand is written with the holomorphic $3$-form $\Omega$, while in the NS-NS sector of Type IIA theories the complexified holomorphic $3$-form $\Omega_{c}$ is used, a point we will return to shortly. $\Omega_{c}$ is dilaton independent and $\Omega$ not, with the specific dependency being easily expressible in terms of the symplectic basis of $\Hn{3}$.
\beq
\Omega = \fU_{0}\syma{0} + \fU_{i}\syma{i} - \fU^{j}\symb{j} - \fU^{0}\symb{0} \qquad 
\Omega_{c} = -S\fU_{0}\syma{0} + \fU_{i}\syma{i} - \fU^{j}\symb{j} + S\fU^{0}\symb{0}\nn
\eeq
Using this as a guide we define four complexified holomorphic forms, two for each moduli type.
\beq
\ba {ccccccccccccccccccccccc}
\Omega_{c} & \;=\; & -S\,\fU_{0}\syma{0} &+& \fU_{j}\syma{j} &-& \fU^{j}\symb{j} &+& S\,\fU^{0}\symb{0} & \quad & \Omega^{\prime}_{c} & \;=\; & \fU_{0}\syma{0} &-& S\,\fU_{j}\syma{j} &+& S\,\fU^{j}\symb{j} &-& \fU^{0}\symb{0} \\
\mho_{c} & \;=\; & -S\,\cT_{0}\w{0} &+& \cT_{a}\w{a} &+& \cT^{b}\wt{b} &-& S\,\cT^{0}\wt{0} & \quad & \mho^{\prime}_{c} & \;=\; & \cT_{0}\w{0} &-& S\,\cT_{a}\w{a} &-& S\,\cT^{b}\wt{b} &+& \cT^{0}\wt{0} 
\ea \nn
\eeq

\section{Neveu-Schwarz-Neveu-Schwarz Superpotentials}

\subsection{Type IIA Fluxes}

With the explicit example of the $\ZZ$ orientifold superpotential (\ref{eqn:GenW}) for motivation we wish to consider the most general superpotential of this form on a general internal space $\M$. However, it is preferable to first begin with the Type IIA NS-NS sector where the subtlies of taking duals of the fluxes are not required but the schematic form of the superpotential is of the same form. In order to systematize our analysis of the fluxes we drop the notation commonly used to denote the various fluxes \cite{Shelton:2005cf,Shelton:2006fd,Wecht:2007wu} and instead label the fluxes which couple to the dilaton as $\Fh_{n}$ and those which do not as $\F_{m}$.
\beq
W  &=& \int_{\M} \Omega_{c} \wedge \Big(\F_{0}\cdot\Jn{0} + \F_{1}\cdot\Jn{1} + \F_{2}\cdot\Jn{2} + \F_{3}\cdot\Jn{3}\Big) \equiv \int_{\M} \Omega_{c} \wedge G \nn
\eeq
Since $G$ contributes to the superpotential via exterior multiplication with the complexified holomorphic $3$-form only $3$-form terms in $G$ will contribute, all others will be integrated out. By allowing this additional freedom in how we view $G$ we are able to factorise it into a pair of terms, one of whom involves the moduli in $\Jn{n}$ and the other of which is dependent entirely upon the fluxes only. This factorisation is not unique as there is a degeneracy in terms of the signs of the fluxes and the terms they contract with but we set the signs such that the factorisation is written in terms of $\mho$.
\beq
G = \Big(\F_{0} + \F_{1} + \F_{2} + \F_{3}\Big).\left(\Jn{0} + \Jn{1} + \Jn{2} + \Jn{3}\right) \equiv \mathbf{G}\cdot \mho\nn  
\eeq
Without fluxes the action of $\mathbf{G}$ is trivialised to zero and this is synonymous with the fact that all basis elements of the $\Hn{p}$ are closed and though fluxes break the closure we shall still refer to such spaces as cohomologies. Therefore we can view the action of $\mathbf{G}$ as equivalent to that of the exterior derivative, $\d$. With the inclusion of the $3$-form field strength $H_{3}$ $\mathbf{G}$ is no longer trivial and the exterior derivative is twisted by the inclusion of a torsion term defined by $H_{3}$, $\d \to \d_{H} = \d + H \wedge$, further suggesting a link between the flux actions and exterior derivatives. This can be extended \cite{Grana:2006hr,Ihl:2007ah} to include the T duality images of the NS-NS field strength, making the twisted Calabi-Yau into a generalised non-geometric one with covariant exterior derivative $\D$. By considering $\mathbf{G} \sim \D$ the action of $\d$ on $\Ln{2}$, $\Ln{4}$ and $\Ln{6}$ is $\F_{1}$, $\F_{2}$ and $\F_{3}$ respectively.
\beq
\D \equiv \F_{0} + \F_{1} + \F_{2} + \F_{3} \quad \Rightarrow \quad G = \D(\mho) \nn
\eeq
This is only a schematic factorisation, the precise forms of $\D$ as a differential operator can be found by considering its individual flux terms. As a result of the definition of the superpotential's integrand in terms of the $3$-form $\Omega_{c}$ it acts on the subspaces of $\Hnt{3}$ as $\F_{n} : \cHn{n,n} \to \Hn{3}$ . 
\beq
\label{eqn:Wint}
W = \int_{\M} \; \Omega_{c} \wedge G = \int_{\M} \; \Omega_{c} \wedge \D(\mho) \nn
\eeq

\subsubsection{$p$-form defined flux components}

In this section the indices $a,b$ range from $1$ to $\textrm{dim}(\M)=6$, not from $1$ to $h^{1,1}$. We can define the components of the various fluxes using the bases of $\La$ and $\La\dual$ and from here we drop the $j$ index on $\F_{j}$ when working with such components since it is clear to which flux is being refered to by the index structure. Doing the same for the $\Jn{n}$ then allows us to express $G$ in terms of the components of the $\Jn{n}$ and the fluxes.
\beq
G = \D(\mho) = \Big(\lambda_{0}\F_{abc} + \lambda_{1}\F^{d}_{ab}\J_{dc} + \lambda_{2}\F^{de}_{a}\J_{debc}  +  \lambda_{3}\F^{def}\J_{defabc} \Big) \eta^{abc} \nn 
\eeq
We will refer to this way of expressiong the components of fluxes as the $\La$ components defined in the $(\La,\La\dual)$ basis. The $\lambda_{i}$ are combinatorical factors which are determined by the specific index structure of the fluxes and K\"{a}hler forms.  Denoting a T duality in the $\eta^{a}$ direction as $\Tdual_{a}$ we can obtain the T duality sequence of fluxes. \cite{Shelton:2005cf,Shelton:2006fd,Wecht:2007wu}
\beq
\label{eqn:Modseq}
\F_{abc} \; \longleftrightarrow \; \F_{bc}^{a} \; \longleftrightarrow \; \F^{ab}_{c} \; \longleftrightarrow \F^{abc} 
\eeq
The covariant derivative can be expressed explicitly as a differential operator by making use of the interior forms $\i_{\sigma}$ which allow us to factorise contracted indices by $\i_{a}(\eta^{b}) = \delta^{b}_{a}$. The specific coefficients of each terms in $\D$ are determined by the requirement that the Bianchi constraints of $\D$ are equivalent to the Jacobi constraints of the twelve dimensional gauge algebra whose structure constants are the fluxes.
\beq
\label{eqn:Dform1}
\D &=& \frac{1}{3!}\F_{abc}\eta^{abc} +  \frac{1}{2!}\F^{a}_{bc}\eta^{bc}\i_{a} + \frac{1}{2!}\F^{ab}_{c}\eta^{c}\i_{b}\i_{a} + \frac{1}{3!}\F^{abc}\i_{c}\i_{b}\i_{a}  
\eeq
By construction we have defined the derivative action as $\D : \Lp \to \Ln{3} \subset \Lm$ as these are the only contributions to the superpotential if the integrand involves $\Omega_{c} \in \Hn{3}$. However, it can be see that the expressions in (\ref{eqn:Dform1}) also act as $\D : \Ln{3} \to \Lp$, the two different kinds of action of the derivatives are not independent. With this in mind we consider a general element of $\Ln{3}$, $\varphi = \frac{1}{3!}\varphi_{abc}\eta^{abc}$ and apply $\D$ so as to get a more explicit understanding of this different action of the derivatives. It is clear that the action of the fluxes is $\F_{3-n} : \Ln{3} \to \Ln{2n}$ and so the action of the derivatives on $\Ln{3}$ naturally splits into four cases relating to the decomposition of $\Lp$.

\begin{itemize}

\item Case 1 : $\F_{3} : \Ln{3} \to \Ln{0}$
\beq
\frac{1}{3!}\frac{1}{3!}\F^{abc}\varphi_{ijk}\i_{c}\i_{b}\i_{a}(\eta^{ijk}) = \frac{1}{3!}\frac{1}{3!}\F^{abc}\varphi_{ijk}\big(\P{3}{3}\delta^{a}_{i}\delta^{b}_{j}\delta^{c}_{k}\big) = \frac{1}{3!}\F^{abc}\varphi_{abc} \nn
\eeq

\item Case 2 : $\F_{2} : \Ln{3} \to \Ln{2}$
\beq
\frac{1}{3!}\frac{1}{2!}\F^{ab}_{c}\varphi_{ijk}\eta^{c}\i_{b}\i_{a}(\eta^{ijk}) = \frac{1}{3!}\frac{1}{3!}\F^{ab}_{c}\varphi_{ijk}\eta^{c} \big(\P{3}{2} \delta^{a}_{i}\delta^{b}_{j}\eta^{k}\big) = \frac{1}{2!}\F^{ab}_{i}\varphi_{abj}\eta^{ij}\nn
\eeq

\item Case 3 : $\F_{1} : \Ln{3} \to \Ln{4}$
\beq
 \frac{1}{3!}\frac{1}{2!}\F^{a}_{bc}\varphi_{ijk}\eta^{bc}\i_{a}(\eta^{ijk}) = \frac{1}{3!}\frac{1}{2!}\F^{a}_{bc}\varphi_{ijk}\eta^{bc}\big(\P{3}{1} \delta^{a}_{i}\eta^{jk}\big) = \frac{1}{2!}\frac{1}{2!}\F^{a}_{ij}\varphi_{akl}\eta^{ijkl}\nn
\eeq

\item Case 4 : $\F_{0} : \Ln{3} \to \Ln{6}$
\beq
 \frac{1}{3!}\frac{1}{3!}\F_{abc}\varphi_{ijk}\eta^{abc}(\eta^{ijk}) = \frac{1}{3!}\frac{1}{3!}\F_{abc}\varphi_{ijk}\eta^{abc}(\eta^{ijk})\big(\P{3}{0}\big) = \frac{1}{3!}\frac{1}{3!}\F_{abc}\varphi_{ijk}\eta^{abcijk} \nn
\eeq

\end{itemize}

This can be summarised by noting the defining action of $\F_{n}$, $\F_{n} : \Ln{2n} \to \Ln{3} \to \Ln{6-2n}$ can be rewritten as $\F_{n} : \Ln{2n} \to \Ln{3} \to \ast\Ln{2n}$, a result which will be of importance later on. This is not the complete action of $\D$ as the image of some flux terms acting on elements of $\Lp$ reside in $\Ln{1}$ or $\Ln{5}$ but as pointed out in our factorisation of the superpotential they do not contribute to the superpotential because such terms are projected out from the integrand by $\Omega_{c} \in \Hn{3} \subset \Ln{3}$. 
\beq
\label{eqn:SpuriousF}
\ba{cccccccccccccccccc}
\F_{0} \,:\, \ba{ccc}\Ln{2} &\to& \Ln{5}\ea &,& \F_{1} \,:\, \ba{ccc}\Ln{4} &\to& \Ln{5}\\\Ln{0} &\to& \Ln{1}\ea &,& \F_{1} \,:\, \ba{ccc}\Ln{6} &\to& \Ln{5}\\\Ln{2} &\to& \Ln{1}\ea &,& \F_{3} \,:\, \ba{ccc}\Ln{4} &\to& \Ln{1}\ea 
\ea
\eeq
These actions are not relevant to constructing the superpotential but are to the Bianchi constraints of $\D$. In the fluxless case the exterior derivative is trivially nilpotent, $\d^{2}=0$, and defines an exact sequence on the even and odd $p$-form bundles. The same is true for the covariant derivatives in the case of non-zero fluxes but in a less straightforward way. This is because the action $\d : \Ln{p} \to \Ln{p+1}$ is no longer true for $\D$, giving rise to non-trivial Bianchi constraints, as done in full generality in the appendix of \cite{Ihl:2007ah}. 
\beq
\cdots \; \xrightarrow{\; \D \;} \;\Lp\; \xrightarrow{\; \D \;} \;\Lm\; \xrightarrow{\; \D \;} \;\Lp\; \xrightarrow{\; \D \;} \; \cdots \nn
\eeq
However, expanding the fluxes, K\"{a}hler forms and their contractions in terms of elements of $\Ln{p}$ does not lend itself to a straightforward analysis of the flux obtained from the superpotential. An example of this is how $\F_{1}$ can be defined in terms of the geometry of the space by $\d \eta^{\tau} \propto \F^{\tau}_{\sigma\rho}\eta^{\sigma\rho}$ and the properties of $\M$ determine the number and location of non-zero fluxes in $\F^{\tau}_{\sigma\rho}$. This formulation, while general to any internal space, is not manifest in its independent fluxes or how they contribute to the superpotential. In fluxless cases the moduli are massless so moduli dependent forms are expanded in the $\Hn{p}$ basis and those $p$-forms not in the cohomologies can be ignored in the effective theory because they are associated to `heavy modes'. Though the fluxes break this massless/heavy splitting of the degrees of freedom of $\M$ through the non-closure of elements originally in $\Hn{p}$ giving the moduli masses, it is possible to use the original $\Hn{p}$ bases to expand flux contributions to the superpotential integrand \cite{Grana:2006hr}. The SU(3) structure is that which reduces to massless field modes in the fluxless case, while the `heavy modes' are those which have mass even in the absence of fluxes. We therefore have two approaches we might wish to consider; analysing the `light' modes using the SU(3) structure or analysing all modes using all $p$-forms as our basis. The latter case has the advantage of not excluding anything from our considerations, removing the issue of just how light is light, how heavy is heavy or which $p$-forms or expressions we neglect. However, as we will show, the cohomologies approach is considerably more `natural' when it comes to describing the superpotential. This will allow us to construct general superpotentials and see the effect T and S duality have on them, rather than restricting ourselves to a single particular $\M$. We cannot completely ignore the fact an analysis in terms of SU(3) structure will not be a complete description of the fluxes of $\M$ but assuming the moduli are sufficiently lighter than other fields we will mostly restrict our discussion to the cohomology construction only.

\subsubsection{Cohomology defined flux components}

With our attention restricted to the $\Hn{p}$ the $p$-forms ($p$ odd) of interest are only those in $\Hn{3}$ and the generic exact sequence for $\D$ previously considered simplifies down to being in terms of $\Hn{3}$ and $\Hnt{3}$.
\beq
\cdots \; \xrightarrow{\; \D \;} \;\Hn{3}\; \xrightarrow{\;\D \;} \;\Hnt{3}\; \xrightarrow{\;\D\;} \;\Hn{3}\; \xrightarrow{\;\D\;} \; \cdots \nn
\eeq
To facilitate an analysis of the flux actions on the cohomologies we reformulate the fluxes and derivatives such that they can be viewed as matrices (or components of matrices) acting upon vectors of basis elements of cohomologies. To that end we define a $2h^{2,1}+2h^{1,1}+4$ dimensional vector of $p$-forms $\Basis$ by combining $\BasisA$ and $\BasisB$.
\beq
\ba {ccccc}
\Basis & \equiv & \bpm \BasisB & \BasisA \epm &\equiv& \bpm \w{0} & \w{a} & \wt{0} & \wt{b} & \syma{0} & \syma{i} & \symb{0} & \symb{j} \epm\ea  \nn
\eeq
With the entries of $\Basis$ forming the basis for any differential form in $\M$ we can express the $\D$ image of any given form as a linear combination of other cohomology basis elements, thus giving a matrix representation to $\D$. We choose to write this action in such a manner that the $h$ matrices are explicit factors, for later convenience.
\beq
\D\bpm  \BasisB \\ \BasisA \epm = \bpm 0 & \DevenA \\ \DoddA & 0 \epm \bpm h_{\HtbasisA} & 0 \\ 0 & h_{\Hbasis}\epm \bpm  \BasisB \\ \BasisA \epm \qquad \u{\u{\D}} \equiv \bpm 0 & \DevenA \\ \DoddA & 0 \epm \nn
\eeq
As a result of the action of the derivatives exchanging the two cohomologies, $\Hn{3} \leftrightarrow \Hnt{3}$, we have that $\u{\u{\D}}$ must be block-off-diagonal, hence only the $\DevenA$ and $\DoddA$ sub-matrices are non-zero and and we can express the exact sequence in terms of these flux matrices.
\beq
\cdots \; \xrightarrow{\; \DevenA \cdot h_{\HtbasisA} \;} \;\Hn{3}\; \xrightarrow{\;\DoddA \cdot h_{\Hbasis}\;} \;\Hnt{3}\; \xrightarrow{\;\DevenA \cdot h_{\HtbasisA}\;} \;\Hn{3}\; \xrightarrow{\;\DoddA \cdot h_{\Hbasis}\;} \; \cdots \nn
\eeq
We also observed in the previous section that the action of the fluxes on $\Hnt{3}$ defines the action of the fluxes on $\Hn{3}$ and vice versa, implying that the entries of $\DevenA$ define those of $\DoddA$ and it is this dependence which we aim to derive. In the NS-NS superpotential the flux multiplets $\F_{n}$ are defined by the action of $\D$ on the basis elements of $\Hnt{3}$ by $\F_{n} : \cHn{n,n} \to \Hn{3}$. As a result the action of $\D$ is used to define the flux entries of the multiplets by $\D(\BasisB) = \DevenA \cdot h_{\Hbasis} \cdot \BasisA$. As in the previous subsection we consider the $\F_{n}$ in order of increasing $n$, beginning with $\F_{0} : \cHn{0,0} \to \Hn{3}$, whose action is equivalent to multiplying by a $3$-form, expanded in the sympletic basis.
\beq
\D(\w{0}) = \F_{0} = (\F_{0})_{I}\syma{I} - (\F_{0})^{J}\symb{J}   = (\DevenA \cdot h_{\Hbasis} \cdot \BasisA)_{1} \nn
\eeq
Therefore the first row of $\DevenA$ is written in terms of the coefficients of $\F_{0}$, providing us with a vector of flux entries.
\beq
(\DevenA)_{1\,\cdot} = \bpm(\F_{0})_{0} & (\F_{0})_{i} & (\F_{0})^{0} & (\F_{0})^{j} \epm\nn
\eeq
The next case, $\D(\w{a})$, defines $\F_{1}$ as it is the part of $\D$ whose action is of the form $\cHn{1,1} \to \Hn{3}$ and we define the flux entries of $\F_{1}$ such that its expansion in terms of the $\Hn{3}$ basis is of the same format as the $\F_{0}$ case.
\beq
\F_{1}(\w{a}) = (\F_{1})_{(a)I}\syma{I} - (\F_{1})_{(a)}^{\phantom{(a)}J}\symb{J} \equiv (\DevenA\cdot h_{\Hbasis} \cdot\BasisA)_{a+1}  \nn 
\eeq
Each of the $\w{a}$ has a corresponding vector of flux entries and so overall $\F_{1}$ can be associated with a $h^{1,1} \times 2(h^{2,1}+1)$ submatrix of $\DevenA$, specifically the $2$nd through to the $(h^{1,1}+1)$'th rows of $\DevenA$. 
\beq
(\DevenA)_{a+1\,\cdot} &=& \bpm (\F_{1})_{(a)0} & (\F_{1})_{(a)i}  & (\F_{1})_{(a)}^{\phantom{(a)}0} &  (\F_{1})_{(a)}^{\phantom{(a)}j} \epm \nn
\eeq
By considering coefficients of the moduli in the superpotential it is clear there are no other independent flux components in $\F_{1}$, a result which will be demonstrated explicitly for the $\ZZ$ orientifold. Continuing with this method for $\F_{2}$ and $\F_{3}$ we can define all of the entries for $\DevenA$. Due to our choice of ordering in the entries of $\BasisB$ the components of $\F_{3}$ are before those of $\F_{2}$. 
\beq
\label{eqn:MatrixM}
\DevenA = \bpm 
(\F_{0})^{\phantom{(a)0}}_{0} & (\F_{0})^{\phantom{(a)}}_{i}  & (\F_{0})^{0\phantom{(a)}} & (\F_{0})^{j\phantom{(a)}} \\ 
(\F_{1})_{(a)0} & (\F_{1})_{(a)i}  & (\F_{1})^{\phantom{(a)}0}_{(a)} & (\F_{1})^{\phantom{(a)}j}_{(a)} \\ 
(\F_{3})^{\phantom{(a)0}}_{0} & (\F_{3})^{\phantom{(a)}}_{i} & (\F_{3})^{0\phantom{(a)}} & (\F_{3})^{j\phantom{(a)}} \\ 
(\F_{2})^{(b)}_{\phantom{(b)}0} & (\F_{2})^{(b)}_{\phantom{(b)}i} & (\F_{2})^{(b)0} & (\F_{2})^{(b)j} 
\epm = \bpm 
(\F_{0})^{\phantom{(a)I}}_{I} & (\F_{0})^{J\phantom{(a)}} \\ 
(\F_{1})_{(a)I} & (\F_{1})^{\phantom{(a)}J}_{(a)}  \\ 
(\F_{3})^{\phantom{(a)I}}_{I} & (\F_{3})^{J\phantom{(a)}} \\ 
(\F_{2})^{(b)}_{\phantom{(b)}I} & (\F_{2})^{(b)J}  
\epm
\eeq
Using this definition of the components of each of the fluxes and this matrix we can construct a more natural differential operator representation for $\D$ than that given in (\ref{eqn:Dform1}). This can be done by using a set of interior forms defined in terms of the basis elements of $\Hn{3}$ and $\Hnt{3}$ and their intersection numbers, which in turn we take to define the basis elements of $\Hn{p}\dual$ such as $\Hn{2}\dual = \la \i_{\w{a}} \ra$.
\beq
\i_{\w{a}}(\w{b}) = g_{b}^{a} = \delta_{b}^{a} = \i_{\wt{b}}(\wt{a}) \quad , \quad \i_{\syma{i}}(\syma{j}) = g_{j}^{i} = \delta_{j}^{i} = \i_{\symb{j}}(\symb{i}) \nn 
\eeq
The volume form is factorisable into pairs of Hodge dual forms, such as $\vol = \syma{0} \wedge \symb{0}$, which we denote generally as $\vol = \xi \wedge \zeta$. Therefore an interior form applied to $\vol$ is non-zero. 
\beq
\vol = \xi \wedge \zeta \quad \Rightarrow \quad \i_{\xi}(\vol) = \zeta \nn
\eeq
All other combinations do not need to be specified because they lead to forms of degree higher than six when combined with the other parts of the superpotential integrand and so do not contribute. It is also worth noting that $\J \wedge \Omega = 0$ excludes any $\Hn{5}$ contributions of the form $\w{a} \wedge \syma{I}$ and $\w{a} \wedge \symb{J}$. This is in contrast to the action of $\F_{0}$ on $\Ln{2}$ in (\ref{eqn:SpuriousF}), where a non-zero $5$-form might exist but simply not contribute to the superpotential.
\beq
\ba{cccccccc}
\F_{0} &\,=\,& \frac{1}{3!}\F_{\sigma\tau\rho}\,\eta^{\sigma\tau\rho} &\,=\,& (\F_{0})_{I}\syma{I}\i_{\w{0}} \;-\; (\F_{0})^{J}\symb{J}\i_{\w{0}}  \\
\F_{1}  &=& \frac{1}{2!}\F^{\sigma}_{\tau\rho}\,\eta^{\tau\rho}\,\i_{\sigma}  &=& (\F_{1})_{(a)I}\syma{I}\i_{\w{a}} \;-\; (\F_{1})_{(a)}^{\phantom{(a)}J}\symb{J}\i_{\w{a}}  \\
\F_{2}  &=& \frac{1}{2!}\F^{\sigma\tau}_{\rho}\,\eta^{\rho}\,\i_{\tau\sigma}  &=& (\F_{2})_{\phantom{(a)}I}^{(b)}\syma{I}\i_{\wt{b}} \;-\; (\F_{2})^{(a)J}\symb{J}\i_{\wt{b}}  \\
\F_{3} &=& \frac{1}{3!}\F^{\sigma\tau\rho}\,\i_{\rho\tau\sigma}  &=& (\F_{3})_{I}\syma{I}\i_{\wt{0}} \;-\; (\F_{3})^{J}\symb{J}\i_{\wt{0}}
\ea \label{eqn:AltF1F2}
\eeq
This reformulation of $\D$, with its action on the basis elements of the $\Hnt{3}$ cohomologies manifest, can be written in a more compact form once we define $\i_{\Basis}$ to be a vector of interior forms, with the $n$'th entry being the dual of the $n$'th entry of $\Basis$ and likewise for $\BasisA$ and $\BasisB$.
\beq
\label{eqn:Dmatrix1}
\D = \F_{0} + \F_{1} + \F_{2} + \F_{3} \equiv \BasisA^{\top} \cdot h_{\Hbasis}^{\top} \cdot \DevenA^{\top} \cdot \i_{\BasisB}
\eeq
This manner of expressing $\D$ has a clear action on $\Hnt{3}$ but the $\D$ of (\ref{eqn:Dform1}) can also be applied to $\Hn{3}$ elements, with the action being related in some way to $\DoddA$. On the grounds of symmetry between the two cohomologies and the definition of $\D$ on $\Basis$ we can postulate the alternative form of $\D$ with manifest action on $\Hn{3}$.
\beq
\label{eqn:Dmatrix1a}
\D = \F_{0} + \F_{1} + \F_{2} + \F_{3} &\equiv& \BasisB^{\top} \cdot h_{\HtbasisA}^{\top} \cdot \DoddA^{\top} \cdot \i_{\BasisA} 
\eeq
In order to find the entries of $\DoddA$ in terms of the entries of $\DevenA$ we can reformulate each of the fluxes in (\ref{eqn:AltF1F2}) individually. We first consider $\F_{0}$ and $\F_{3}$ which have the simplest actions, reducing to either removing or adding $3$-forms.
\beq
\ba {ccccccc}
\F_{0} & \quad : \quad & \cHn{0,0} & \quad \rightarrow \quad & \Hn{3} & \quad \rightarrow \quad & \cHn{3,3} \\
\F_{3} & \quad : \quad & \cHn{0,0} & \quad \leftarrow \quad & \Hn{3} & \quad \leftarrow \quad & \cHn{3,3} 
\ea  \nn
\eeq
In the case of $\F_{0}$ the differential operator action is simply $\F_{0} \wedge$, whose action on both $\Hn{3}$ and $\Hnt{3}$ is easily found and we use this as a guide in reformulating the expressions. 
\beq
\ba{cccccccccccc}
\F_{0} &:& \cHn{0,0} \to \Hn{3} &\quad,\quad& \F_{0}  &=&  (\F_{0})_{I}\syma{I}\i_{\w{0}} &\;-\;& (\F_{0})^{J}\symb{J}\i_{\w{0}}  \\
\F_{0} &:& \Hn{3} \to \cHn{3,3} &\quad,\quad& \F_{0}  &=&  (\F_{0})^{J}\wt{0}\i_{\syma{J}} &\;+\;& (\F_{0})_{I}\wt{0}\i_{\symb{I}} 
\ea \nn
\eeq
Conversely the removal of $3$-forms by the action of $\F_{3}$ on $\Hnt{3}$ is easily converted into an action on $\Hn{3}$.
\beq
\ba{cccccccccccc}
\F_{3} &:& \cHn{3,3} \to \Hn{3} &\quad,\quad& \F_{3}  &=&  \phantom{-}(\F_{3})_{I}\syma{I}\i_{\wt{0}} &\;-\;& (\F_{3})^{J}\symb{J}\i_{\wt{0}} \\
\F_{3} &:& \Hn{3} \to \cHn{0,0} &\quad,\quad& \F_{3}  &=&  -(\F_{3})^{J}\w{0}\i_{\syma{J}} &\;-\;& (\F_{3})_{I}\w{0}\i_{\symb{I}} 
\ea\nn
\eeq
Comparing the two different actions of $\F_{0}$ and $\F_{3}$ we have the following expressions : 
\beq
\label{eqn:F0F3Cliff}
\ba {ccccccc}
\syma{I}\i_{\w{0}} &\simeq& \phantom{-}\wt{0}\i_{\symb{I}} &\qquad& \symb{J}\i_{\w{0}} &\simeq& -\wt{0}\i_{\syma{J}} \\
\syma{I}\i_{\wt{0}} &\simeq& -\w{0}\i_{\symb{I}} &\qquad&  \symb{J}\i_{\wt{0}} &\simeq& \phantom{-}\w{0}\i_{\syma{J}}
\ea  
\eeq
In the cases of $\F_{1}$ and $\F_{2}$ this simple addition or removal of $3$-forms does not occur so we try to reexpress the $p$-form and interior form components of $\F_{1}$, $\F_{2}$ in (\ref{eqn:AltF1F2}) into something resembling the right hand sides of (\ref{eqn:F0F3Cliff}). $\F_{1}$ arises from the non-closed nature of the basis forms and due to its geometric nature it can be expressed in terms of the exterior derivative $\d$. We have defined the entries of $\F_{1}$ by its action on $\Hnt{3}$ and using integration by parts \cite{Ihl:2007ah} the alternative action on $\Hn{3}$ can be found. 
\beq
\d(\w{a}) = \F_{1}(\w{a}) \equiv \F_{(a)I}\syma{I} - \F_{(a)}^{\phantom{(a)}J}\symb{J}
\eeq
The two sets of coefficients can be extracted from this expression by integrating over the approriate $3$-cycles and then converting these to integrals over the entire space.
\beq
\int_{A^{J}}\d(\w{a}) = \int_{A^{J}}\F_{(a)I}\syma{I} \quad \Rightarrow \quad \int_{\M}\d(\w{a})\wedge \symb{J} = \int_{\M}\F_{(a)I}\syma{I}\wedge \symb{J} \nn
\eeq
By Stokes theorem and $\Hn{5}$ being empty the left hand side integral with integrand $\d(\w{a})\wedge \symb{J}$ converts to an integral over $\M$ with integrand $-\w{a}\wedge\d(\symb{J})$ but this can be reexpressed as an intergral over a $4$-cycle, that which is associated to the $\w{a}$ $2$-form. Thus the right hand side is related to the non-closure of $\symb{J}$.
\beq
\F_{(a)J} = -\int_{\M}\w{a}\wedge\d(\symb{J}) = \int_{\M}\d(-\symb{J}) \wedge \w{a} = \int_{\B_{a}}\d(-\symb{J}) \nn
\eeq
By writing $\d\symb{J}$ in terms of the $\Hnt{3}$ basis we obtain the contribution to the non-closure of $\symb{J}$ spanned by the $(2,2)$-forms $\wt{b}$ in terms of the fluxes defining the non-closure of the $\w{a}$ and the $\Hnt{3}$ intersection numbers. However, due to our definition of the $(2,2)$-form basis these intersection numbers are to $\delta^{a}_{b}$, simplify the end result.
\beq
\F_{(a)J} = \F_{(b)J}\delta^{b}_{a} = \F_{(b)J}\int_{\M}\w{a}\wedge\wt{b} = \int_{\M}\w{a}\wedge(\F_{(b)J}\wt{b}) = \int_{\M}\w{a}\wedge\d(-\symb{J})  \nn
\eeq
Therefore the action of $\F_{1}$ on the $\symb{J}$, $\F_{1}(\symb{J})$, or equivalently the non-closure of $\symb{J}$, has a contribution in the $(2,2)$-form cohomology of $-\F_{(b)J}\wt{b}$. Repeating this method but integrating over the $B_{I}$ $3$-cycle gives the contribution of the non-closure of $\syma{I}$, $\d \syma{I} \sim \F_{1}(\syma{I})$ in the $(2,2)$-form cohomology, $\F_{(b)}^{\phantom{(a)}I}\wt{b}$ and we can summarise this in the same way as (\ref{eqn:F0F3Cliff}).
\beq
\label{eqn:F1Cliff}
\ba {ccccccc}
\syma{I}\i_{\w{a}} &\simeq& -\wt{a}\i_{\symb{I}} &\qquad& \symb{J}\i_{\w{a}} &\simeq& \phantom{-}\wt{a}\i_{\syma{J}} \\
\ea  
\eeq 
In terms of their explicit action on different cohomologies we can represent $\F_{1}$ in two different ways.
\beq
\label{eqn:F1actions}
\ba{ccccccccc}
\F_{1} &:& \cHn{1,1} \to \Hn{3} &\quad,\quad& \F_{1} &=& \phantom{-}(\F_{1})_{(a)I}\syma{I}\i_{\w{a}} &\;-\;& (\F_{1})_{(a)}^{\phantom{(a)}J}\symb{J}\i_{\w{a}} \\
\F_{1} &:& \Hn{3} \to \cHn{2,2} &\quad,\quad& \F_{1}  &=&  -(\F_{1})_{(a)}^{\phantom{(a)}J}\wt{a}\i_{\syma{J}} &\;-\;& (\F_{1})_{(a)I}\wt{a}\i_{\symb{I}} \ea 
\eeq
The remaining case of $\F_{2}$ does not immediately lend itself to the same methodology since the schematic action of the flux is $\F_{2} : \Hn{p} \to \Hn{p-1}$, the opposite behaviour of $\F_{1}$ and the exterior derivative $\d$, a reflection of its non-geometric nature. However, such action is seen in adjoint derivatives and so we can rephrase $\F_{2}$ in terms of the adjoint action of an exterior derivative. To that end we define $\fd$ and its $\IP{,}$ adjoint $\fd\ad$ by the action of $\F_{2}$ on $\cHn{2,2}$.
\beq
\fd\ad(\wt{b}) \equiv \F_{2}(\wt{b}) \equiv \F^{(a)}_{\phantom{(a)}I}\syma{I} - \F^{(a)J}\symb{J} \nn
\eeq
As in the $\F_{1}$ case we can project out particular coefficients of $\F_{2}(\wt{b})$ by taking its inner product with particular $\Hn{3}$ basis elements, allowing us to then make use of the adjoint properties of the inner product.
\beq
\F^{(a)}_{\phantom{(a)}I} = \IP{ \F_{2}(\wt{a}) , \syma{I} } = \IP{ \fd\ad\wt{a} , \syma{I} } \equiv \IP{ \wt{a} , \fd\syma{I} } \nn
\eeq
Making use of Stokes theorem again we can change which form the derivative $\fd$ acts upon, which is not possible to do when working with $\fd\ad$.
\beq
0 = \int_{\M}\fd(\syma{I} \wedge \w{a}) &=& \int_{\M}\fd\syma{I} \wedge \w{a} - \int_{\M}\syma{I} \wedge \fd\w{a} \nn \\
&=& \IP{ \fd\syma{I},\wt{a} } - \int_{\M}\syma{I} \wedge \ast(\ast^{-1}\fd \ast) \wt{b} \nn 
\eeq
Having obtained an expression for $\fd$ acting on an element of $\Hn{3}$ we need to revert back to expressing derivatives as $\fd\ad$. This is done by using the definition of adjoint derivatives in terms of Hodge stars and derivatives, taking note that the action of the derivatives on the symplectic basis elements acquires an additional factor of $-1$.
\beq
\fd\ad \quad=\quad \Big\{ \ba{rcc}\ast^{-1}\fd \ast & \quad & \fd : \Hnt{3} \to \Hn{3} \\ -\ast^{-1}\fd \ast & \quad & \fd : \Hn{3} \to \Hnt{3} \ea \nn
\eeq
Inverting this relationship, to express $\fd$ in terms of $\fd\ad$, we obtain the alternative action of $\fd\ad$ on $\Hn{3}$ by noting that $\ast = \ast^{-1}$ on $\Hnt{3}$ due to the intersection numbers of the basis elements being the Kronecker delta.
\beq
\F^{(a)}_{\phantom{(a)}I} = \IP{ \fd\ad\wt{a} , \syma{I} } = \IP{ \wt{a} , \fd\syma{I} } = \IP{ \wt{a} , -\ast\fd\ad \ast^{-1}\syma{I} } = \IP{ \w{a} , \fd\ad \symb{I} } \nn
\eeq
Repeating this method but projecting out the remaining fluxes in $\F_{2}$ gives terms related to the $\F_{2}$ image of the $\syma{I}$.
\beq
\F^{(a)J} = \IP{ \fd\ad\wt{a} , \symb{J} } = \IP{ \wt{a} , \fd\symb{J} } = \IP{ \wt{a} , -\ast\fd\ad \ast^{-1}\symb{J} } = \IP{ \w{a} , \fd\ad (-\syma{J}) } \nn
\eeq
Putting these results together we obtain the remaining set of relations, in line with (\ref{eqn:F0F3Cliff}) and (\ref{eqn:F1Cliff}).
\beq
\label{eqn:F2Cliff}
\ba {ccccccc}
\syma{I}\i_{\wt{a}} &\simeq& \w{a}\i_{\symb{I}} &\qquad&  \symb{J}\i_{\wt{a}} &\simeq& -\w{a}\i_{\syma{J}}
\ea  
\eeq
In terms of their explicit action on different cohomologies we can represent $\F_{2}$ in two different ways.
\beq
\ba{ccccccccc}
\F_{2} &:& \cHn{2,2} \to \Hn{3} &\quad,\quad& \F_{2}  &=&  (\F_{2})_{\phantom{(b)}I}^{(b)}\syma{I}\i_{\wt{b}} &\;-\;& (\F_{2})^{(b)J}\symb{J}\i_{\wt{b}}  \\
\F_{2} &:& \Hn{3} \to \cHn{1,1} &\quad,\quad& \F_{2}  &=&  (\F_{2})^{(b)J}\w{b}\i_{\syma{J}} &\;+\;& (\F_{2})_{\phantom{(b)}I}^{(b)}\w{b}\i_{\symb{I}} 
\ea \nn
\eeq
Despite the derivation of the alternative actions of each of the $\F_{n}$ using a different method, they all share the feature that if a particular term $\mathfrak{f} \in \F_{n}$ has an action $\mathfrak{f} : \la \xi \ra \to \la \zeta \ra$ then it will also have an action $\mathfrak{f} : \la \ast\zeta \ra \to \la \ast\xi \ra$, where $\la \chi \ra$ is the space spanned by the form $\chi$. To illustrate this and the results themselves we consider the explicit case of the $\ZZ$ orientifold. Using these results the expansion of $\D$ in (\ref{eqn:Dmatrix1}) is converted into the expansion postulated in (\ref{eqn:Dmatrix1a}), giving us the manifest action of $\D$ on $\Hn{3}$ and the entries of $\DoddA$ in terms of those of $\DevenA$.
\setL{5}
\beq
\label{eqn:Dmatrix2}
\D \equiv \BasisB^{\top} \cdot h_{\HtbasisA}^{\top} \cdot \DoddA^{\top} \cdot \i_{\BasisA} \quad \Rightarrow \quad \DoddA &=& \bpm -\F^{(0)J} & \F^{(b)J} & \F_{(0)}^{\phantom{(0)}J} & -\F_{(a)}^{\phantom{(a)}J} \\ -\F^{(0)}_{\phantom{(0)}I} & \F^{(b)}_{\phantom{(b)}I} & \F_{(0)I} & -\F_{(a)I} \epm 
\eeq
We are abusing notation somewhat, as $\DoddA$ is related to the transpose of $\DevenA$ and so all the flux coefficient matrices in (\ref{eqn:Dmatrix2}) should be transposed, such as $-\F^{(0)J}$ defining part of a row in $\DevenA$ but part of a column in $\DoddA$. However, due to their explicit index labels it is unambiguous how they combine with other terms once matrix expressions are written out in terms of their entries. Before moving onto how the superpotential might be expressed in terms of these flux matrices it is convenient to define a more compact notation for the flux components. In (\ref{eqn:MatrixM}) we are able to express $\DevenA$ in a more convenient manner by running the complex structure indices over the range $\{0,\ldots,h^{2,1}\}$, where the $\syma{0}$ and $\symb{0}$ contributions are easily combined with the other sympletic basis terms. In the case of the elements of $\Hnt{3}$ there was an obstruction to being able to run K\"{a}hler indices over the ranges $A,B \in \{0,1,\ldots,h^{1,1}\}$ as the expressions for $\w{0}$ in (\ref{eqn:F0F3Cliff}) do not follow the pattern given in (\ref{eqn:F1Cliff}) for the $\w{a}$ and likewise for $\wt{0}$ in (\ref{eqn:F0F3Cliff}) compared to the $\wt{b}$ of (\ref{eqn:F2Cliff}). 
\\

Any change of basis $\Basis \to \Basis^{\prime}$ which allows these expressions to be combined must preserve the associated bilinear forms $g$ and $h$ and ideally not alter how the complexified holomorphic forms are expanded in terms of the dilaton. It is clear that because the differences in (\ref{eqn:F0F3Cliff}), (\ref{eqn:F1Cliff}) and (\ref{eqn:F2Cliff}) occur in terms of the K\"{a}hler indices the change of basis must be in the $\BasisB$ section of $\Basis$. There are two ways in which to do this which preserve the schematic layout of the K\"{a}hler moduli dependent complexified holomorphic forms. In the case of the non-complexified holomorphic form $\mho = \u{\cT} \cdot h_{\HtbasisA} \cdot \BasisB$ any non-singular transformation in $\BasisB$ can be countered by a corresponding transformation in $\u{\cT}$.
\begin{itemize}
\item Exchange $\w{0}$ and $\wt{0}$ and therefore $\cT_{0}$ and $\cT^{0}$
\beq
\label{eqn:wtov1}
\ba {cccc}
& \bpm \w{0} \\ \w{a} \\ \wt{0} \\ \wt{b} \epm \to 
\bpm \v{0} \\ \v{a} \\ \vt{0} \\ \vt{b} \epm = \bpm \wt{0} \\ \w{a} \\ \w{0} \\ \wt{b} \epm &\quad,\quad& \bpm \cT_{0} \\ \cT_{a} \\ \cT^{0} \\ \cT^{b} \epm \to 
\bpm \fT_{0} \\ \fT_{a} \\ \fT^{0} \\ \fT^{b} \epm = \bpm \cT^{0} \\ \cT_{a} \\ \cT_{0} \\ \cT^{b} \epm 
\ea 
\eeq

\item Exchange $\w{a}$ and $\wt{a}$ and therefore $\cT_{a}$ and $\cT^{a}$.
\beq
\label{eqn:wtov2}
\ba {cccc}
\bpm \w{0} \\ \w{a} \\ \wt{0} \\ \wt{b} \epm \to 
\bpm \v{0} \\ \v{a} \\ \vt{0} \\ \vt{b} \epm = \bpm \w{0} \\ \wt{a} \\ \wt{0} \\ \w{b} \epm &\quad,\quad& \bpm \cT_{0} \\ \cT_{a} \\ \cT^{0} \\ \cT^{b} \epm \to 
\bpm \fT_{0} \\ \fT_{a} \\ \fT^{0} \\ \fT^{b} \epm = \bpm \cT_{0} \\ \cT^{a} \\ \cT^{0} \\ \cT_{b} \epm
\ea 
\eeq
\end{itemize}

It is noteworthy that this kind of exchanging of moduli $\cT_{0} \leftrightarrow \cT^{0}$ or $\cT_{a} \leftrightarrow \cT^{a}$ is precisely that which was used in Section \ref{sec:Motivation} to argue that the Type IIB superpotential could be written in a particular way. The difference between these two choices in how we might change the $\BasisB$ basis reduces to a difference in sign, with the change of basis in (\ref{eqn:wtov1}) giving the $+$ sign in the $\pm$ of the following expressions.
\beq
\label{eqn:CliffShort}
\ba {ccccccc}
\pm\syma{I}\i_{\v{A}} &\simeq& -\vt{A}\i_{\symb{I}} &\qquad& \pm\symb{J}\i_{\v{A}} &\simeq& \phantom{-}\vt{A}\i_{\syma{J}} \\
\pm\syma{I}\i_{\vt{B}} &\simeq& \phantom{-}\v{B}\i_{\symb{I}} &\qquad&  \pm\symb{J}\i_{\vt{B}} &\simeq& -\v{B}\i_{\syma{J}}
\ea  
\eeq 
Though this change in basis is prompted by our examination of $\DoddA$ it also transforms $\DevenA$, which is the more convenient flux matrix in which to examine the flux components. As we can now run the K\"{a}hler indices over the range $\{0,\ldots,h^{1,1}\}$ we can view the fluxes of $\F_{0}$ and $\F_{3}$ as part of the fluxes of $\F_{1}$ and $\F_{2}$, though which belongs to which depends upon which change of basis we choose. The change of basis in (\ref{eqn:wtov1}) leads us to view $\F_{3}$ as part of $\F_{2}$ and $\F_{0}$ as part of $\F_{1}$.
\beq
\label{eqn:Mshort1}
\ba {ccccccccc}
\DevenA &=& \bpm  (\F_{0})_{I\phantom{(a)}} & (\F_{0})^{J\phantom{(a)}} \\ (\F_{1})_{(a)I} & (\F_{1})^{\phantom{(a)}J}_{(a)} \\ (\F_{3})_{I\phantom{(b)}} & (\F_{3})^{J\phantom{(b)}} \\  (\F_{2})_{\phantom{(b)}I}^{(b)} & (\F_{2})^{(b)J}  \epm &\to& \bpm (\F_{3})_{I\phantom{(b)}} & (\F_{3})^{J\phantom{(b)}}  \\  (\F_{1})_{(a)I} & (\F_{1})^{\phantom{(a)}J}_{(a)} \\ (\F_{0})_{I\phantom{(a)}} & (\F_{0})^{J\phantom{(a)}}   \\ (\F_{2})_{\phantom{(b)}I}^{(b)} & (\F_{2})^{(b)J}\epm 
\ea 
\eeq
Conversely, in the case of the change of basis in (\ref{eqn:wtov2}) $\F_{0}$ is part of $\F_{2}$ and $\F_{3}$ part of $\F_{1}$ and the index structure of the components of $\DevenA$ are altered accordingly.
\beq
\label{eqn:Mshort2}
\ba {ccccccccc}
\DevenA &=& \bpm  (\F_{0})_{I\phantom{(a)}} & (\F_{0})^{J\phantom{(a)}} \\ (\F_{1})_{(a)I} & (\F_{1})^{\phantom{(a)}J}_{(a)} \\ (\F_{3})_{I\phantom{(b)}} & (\F_{3})^{J\phantom{(b)}} \\  (\F_{2})_{\phantom{(b)}I}^{(b)} & (\F_{2})^{(b)J}  \epm &\to& \bpm  (\F_{0})_{I\phantom{(a)}} & (\F_{0})^{J\phantom{(a)}}  \\  (\F_{2})_{\phantom{(b)}I}^{(b)} & (\F_{2})^{(b)J}  \\ (\F_{3})_{I\phantom{(b)}} & (\F_{3})^{J\phantom{(b)}}  \\ (\F_{1})_{(a)I} & (\F_{1})^{\phantom{(a)}J}_{(a)}\epm
\ea 
\eeq
With the compact $2 \times 2$ form of $\DevenA$ it is possible to conveniently express the action of $\D$ on $\Hnt{3}$ using a set of matrices $C_{i}$, rather than having to refer to how the flux components arrange themselves into T duality multiplets.
\beq
\label{eqn:Dnu}
\DevenA = \bpm C_{1} & C_{2} \\ C_{3} & C_{4} \epm \quad \Rightarrow \quad 
\ba {cccccc}
\D(\nu) &=& C_{1} \cdot \alpha &-& C_{2} \cdot \beta  \\
\D(\t{\nu}) &=& C_{3} \cdot \alpha &-& C_{4} \cdot \beta\ea   
\eeq
Using (\ref{eqn:CliffShort}) we convert (\ref{eqn:Dnu}) from the action $\D : \Hnt{3} \to \Hn{3}$ to $\D : \Hn{3} \to \Hnt{3}$, where the sign is as previously stated.
\beq
\ba {cccccc}
\pm\D(\alpha) &=&  C_{4}^{\top} \cdot \nu &-&  C_{2}^{\top} \cdot \t{\nu} \\
\pm\D(\beta) &=&   C_{3}^{\top} \cdot \nu &-&   C_{1}^{\top} \cdot \t{\nu} \ea  \quad \Rightarrow \quad \D \bpm \alpha \\ \beta\epm =  \pm\bpm C_{4}^{\top} & -C_{2}^{\top} \\ C_{3}^{\top} & -C_{1}^{\top} \epm \bpm \nu \\ \t{\nu} \epm\nn
\eeq
Using the definition of $\DoddA$, $\D(\BasisA) = \DoddA \cdot h_{\HtbasisB} \cdot \BasisC$ we obtain the expression for $\DoddA$ in terms of the $C_{i}$.
\beq
\label{eqn:ZetaM1}
\DoddA &=& \pm\bpm C_{4}^{\top} & -C_{2}^{\top} \\ C_{3}^{\top} & -C_{1}^{\top} \epm 
\eeq 
Using the ansatz that the expression for $\D$ in terms of $\DoddA$ will be of the same form as the expression for $\D$'s natural action on $\Hnt{3}$ in (\ref{eqn:Dmatrix1}). 
\beq
\label{eqn:Dmatrix3}
\D = \F_{0} + \F_{1} + \F_{2} + \F_{3} &=& \pm \bpm \nu & \t{\nu} \epm \cdot \bpm C_{4} & C_{3} \\ -C_{2} & - C_{1} \epm \cdot \bpm \iota_{\alpha} \\ \iota_{\beta} \epm 
\eeq 
From this point we will work with the explicit case of the second basis, that of (\ref{eqn:wtov2}), as converting to the first case is simply a matter of relabelling and altering index structures on the fluxes, and the second redefinition will be more convenient in our analysis of Type IIB.
\beq
\label{eqn : New M and Cliff}
\ba {ccccccccc}
\DevenA &=& \bpm  (\F_{0})_{I\phantom{(a)}} & (\F_{0})^{J\phantom{(a)}}  \\  (\F_{2})_{\phantom{(b)}I}^{(b)} & (\F_{2})^{(b)J}  \\ (\F_{3})_{I\phantom{(b)}} & (\F_{3})^{J\phantom{(b)}}  \\ (\F_{1})_{(a)I} & (\F_{1})^{\phantom{(a)}J}_{(a)}\epm &\equiv&
\bpm  \F_{(A)I} & \F^{\phantom{(A)}J}_{(A)} \\
\F_{\phantom{(B)}I}^{(B)} & \F^{(B)J}  \epm \quad,\quad \ba {ccccccc}
\syma{I}\i_{\v{A}} &\simeq& \phantom{-}\vt{A}\i_{\symb{I}} & \symb{J}\i_{\v{A}} &\simeq& -\vt{A}\i_{\syma{J}} \\
\syma{I}\i_{\vt{B}} &\simeq& -\v{B}\i_{\symb{I}} & \symb{J}\i_{\vt{B}} &\simeq& \phantom{-}\v{B}\i_{\syma{J}}
\ea  
\ea
\eeq
Hence we have the following two alternative actions for $\D$, in terms of the flux components.
\beq
\label{eqn:AlternateD}
\ba {cccccc}
\D(\v{A}) &=& \F_{(A)I}\syma{I} &-& \F_{(A)}^{\phantom{(A)}J}\symb{J}  \\
\D(\vt{B}) &=& \F_{\phantom{(B)}I}^{(B)}\syma{I} &-& \F^{(B)J}\symb{J} 
\ea   \quad \Leftrightarrow \quad \ba {cccccc}
\D(\syma{I}) &=& -\F^{(A)I}\v{A} &+& \F_{(B)}^{\phantom{(B)}I}\vt{B}  \\
\D(\symb{J}) &=& -\F_{\phantom{(A)}J}^{(A)}\v{A} &+& \F_{(B)J}\vt{B} 
\ea 
\eeq 
The contribution to the superpotential integrand $\D(\mho)$ splits into four parts, each associated to one of the flux multiplets $\F_{n}$.
\beq
\label{eqn:TypeIIA fluxes}
\ba{ccccccccccc}
\F_{0}\cdot \Jn{0} &=& \fT_{0} \Big( & \F_{(0)I}\syma{I} &-& \F_{(0)}^{\phantom{(0)}J}\symb{J} & \Big) &=& \frac{1}{3!}\F_{pqr}\eta^{pqr} \\
\F_{1}\cdot \Jn{1} &=& \fT^{b} \Big( & \F_{\phantom{(b)}I}^{(b)}\syma{I} &-& \F^{(b)J}\symb{J} & \Big) &=& \frac{1}{2!}\F_{pq}^{i}\J_{ir}\eta^{pqr} \\
\F_{2}\cdot \Jn{2} &=& \fT_{a} \Big( & \F_{(a)I}\syma{I} &-& \F_{(a)}^{\phantom{(a)}J}\symb{J} & \Big) &=& \frac{1}{3!}\F_{p}^{ij}\J_{ijqr}\eta^{pqr} \\
\F_{3}\cdot \Jn{3} &=& \fT^{0} \Big( & \F_{\phantom{(0)}I}^{(0)}\syma{I} &-& \F^{(0)J}\symb{J} & \Big) &=& \frac{1}{3!}\frac{1}{3!}\F^{ijk}\J_{ijkpqr}\eta^{pqr} 
\ea
\eeq
With these we can construct the general Type IIA superpotential in this formulation.
\beq 
\label{eqn:W1components}
\int_{\M} \Omega_{c} \wedge \D(\mho) &=& \left(
\ba{ccccccccccccccccc}
 & \fT_{0} \Big( & -S\,\F_{(0)0}\fU^{0} &+& \F_{(0)j}\fU^{j} &+& S\,\F_{(0)}^{\phantom{(0)}0}\fU_{0} &-& \F_{(0)}^{\phantom{(0)}i}\fU_{i} & \Big) &+&\\
+& \fT^{b} \Big( & -S\,\F^{(b)}_{\phantom{(b)}0}\fU^{0} &+& \F^{(b)}_{\phantom{(b)}J}\fU^{J} &+& S\,\F^{(b)0}\fU_{0} &-& \F^{(b)i}\fU_{i} & \Big) &+&\\
+& \fT^{0} \Big( & -S\,\F^{(0)}_{\phantom{(0)}0}\fU^{0} &+& \F^{(0)}_{\phantom{(0)}j}\fU^{j} &+& S\,\F^{(0)0}\fU_{0} &-& \F^{(0)i}\fU_{i} & \Big) &+&\\
+& \fT_{a} \Big( & -S\,\F_{(a)0}\fU^{0} &+& \F_{(a)j}\fU^{j} &+& S\,\F_{(a)}^{\phantom{(a)}0}\fU_{0} &-& \F_{(a)}^{\phantom{(a)}i}\fU_{i} & \Big) 
\ea \right) \nn
\eeq
Though we did not originally give a motive for our redefining of the sympletic basis and associated moduli $(\a{I},\b{J},\cU) \to (\syma{I},\symb{J},\fU)$ we have now derived precisely the same type of exchange in the K\"{a}hler moduli. While a symplectic transformation on the defining $\Hn{3}$ basis of (\ref{eqn:CliffShort}) does not alter the identities the K\"{a}hler moduli space transformation does and it has motivated $\BasisB \to \BasisC$. By considering mirror symmetry we can view this Type IIA $\Hn{3}$ redefinition in the context of a Type IIB $\Hnt{3}$ redefinition, where the cohomologies are defined on the mirror pair $(\M,\W)$, with Type IIA on $\M$ and we will denote the map associated to mirror symmetry by $\Upsilon$. Under mirror symmetry the degrees of freedom of the Type IIA theory on $\M$ are labelled differently to those of the Type IIB on $\W$. The complex structure of $\M$ is defined to be equivalent to the K\"{a}hler structure of $\W$ and so the $\cT$ of $\M$ relate to the $\cU$ of $\W$. Due to properties of the $\Hn{3}(\M)$ and $\Hnt{3}(\M)$ basis elements we have found it convenient to exchange $\w{a}$ and $\wt{a}$ so as to obtain (\ref{eqn : New M and Cliff}). This results in a redefinition $\cT \to \fT$ and the effect such changes have on $\W$ is to prompt its complex structure moduli and $\Hn{3}(\W)$ basis elements to be altered. Conversely, in order to obtain the results of (\ref{eqn : New M and Cliff}) on $\Hn{p}(\W)$ we would be required to exchange the $\w{a}$ and $\wt{a}$ of $\Hnt{3}(\W)$, whose $\Upsilon$ image is the redefinition of the $\Hn{3}(\M)$ elements $\a{a}$ and $\b{a}$. Thus the choice we made of how to denote the complex structure moduli, $\cU \to \fU$, was in anticipation of this redefinition of the K\"{a}hler moduli in Type IIA, allowing us to forego further redefinitions of flux components due to the exchange of such terms as $\F_{(A)i}$ and $\F_{(A)}^{\phantom{(A)}i}$. This kind of exchange in the basis of $\Hnt{3}$ will be seen to arise in Type IIB due to the manner in which the fluxes couple to the moduli. Before that we find it useful to further develop the way of expressing integrals in terms of matrices and vectors in anticipation of comparing the superpotentials of the Type II theories. We previously noted that the superpotential can be written in terms of $g$ and the vectors associated to the elements of $\Hn{3}$ which make up the integrand.

\subsection{Matrix defined superpotential}

This section, unless otherwise stated, is done in Type IIA. The superpotential can be expressed in terms of the flux matrices by the use of an inner product involving a vector of moduli.
\beq
\label{eqn:Phidef}
\ba {ccccc}
\u{\Phi}^{\top} & \equiv & \bpm \u{\fT}^{\top} & \u{\fU}^{\top} \epm &\equiv& \bpm \fT_{0} & \fT_{a} & \fT^{0} & \fT^{b} & \fU_{0} & \fU_{i} & \fU^{0} & \fU^{j} \epm\ea  
\eeq
Given any form $\chi$ we recall that the associated vector is defined via the general factorisation $\chi = \u{\chi}\cdot h \cdot \Basis$, though this is a basis dependent expression, as was seen in the case of $\mho$, whose moduli vector was redefined by the change of basis $\BasisB \to \BasisC$. Hence we defined our moduli by $\u{\fT} \equiv \u{\mho}$ and $\u{\fU} \equiv \u{\Omega}$ 
\beq
\ba {ccccccccc}
\Omega &=& \fU_{I}\syma{I} - \fU^{J}\symb{J} &=& \bpm \fU_{I} & \fU^{J} \epm \cdot h_{\Hbasis} \cdot \bpm \syma{I} \\ \symb{J} \epm &=& \u{\Omega}^{\top} \cdot h_{\Hbasis} \cdot \BasisA &=& \u{\fU}^{\top} \cdot h_{\Hbasis} \cdot \BasisA \\
\mho &=& \fT_{A}\v{A} + \fT^{B}\vt{B} &=& \bpm \fT_{A} & \fT^{B} \epm \cdot h_{\HtbasisB} \cdot \bpm \v{A} \\ \vt{B} \epm &=& \u{\mho}^{\top} \cdot h_{\HtbasisB} \cdot \BasisC &=& \u{\fT}^{\top} \cdot h_{\HtbasisB} \cdot \BasisC
\ea  \nn
\eeq
Before considering how the dilaton couples to particular fluxes, which would complicated our analysis through the use of complexified holomorphic forms, we first consider the toy model of a superpotential whose integrand is of the form $\Omega \wedge \D(\mho)$ so as to neglect dilaton dependence. We restrict our attention to the NS-NS fluxes and for comparision also consider the superpotential which would be obtained from the integrand $\mho \wedge \D(\Omega)$ as this will arise later. To this end we recall the two alternative actions of $\D$ on the basis elements of $\Hnt{3}$ and $\Hn{3}$.
\beq
\D\bpm \nu \\ \t{\nu}\epm \equiv \DevenA \cdot h_{\Hbasis}\cdot\BasisA \quad,\quad \D\bpm \alpha \\ \beta \epm \equiv \DoddA \cdot h_{\HtbasisB}\cdot\BasisC \nn
\eeq
By writing the holomorphic forms in terms of their vector factorisations and using the above expressions for the images of $\Hnt{3}$ and $\Hn{3}$ basis elements under $\D$, we can construct the vectors associated to $\D(\mho)$ and $\D(\Omega)$.
\beq
\ba {ccl}
\D(\mho) &=& \D(\u{\fT}^{\top} \cdot h_{\HtbasisB} \cdot \BasisC) \\
&=& \u{\fT}^{\top} \cdot h_{\HtbasisB} \cdot \D(\BasisC) \\
&=& \u{\fT}^{\top} \cdot h_{\HtbasisB} \cdot \DevenA \cdot h_{\Hbasis} \cdot \BasisA \\
&=& \u{\D(\mho)}^{\top} \cdot h_{\Hbasis} \cdot \BasisA 
\ea  \qquad \ba {ccl}
\D(\Omega) &=& \D(\u{\fU}^{\top} \cdot h_{\Hbasis} \cdot \BasisA) \\
&=& \u{\fU}^{\top} \cdot h_{\Hbasis} \cdot \D(\BasisA) \\
&=& \u{\fU}^{\top} \cdot h_{\Hbasis} \cdot \DoddA \cdot h_{\HtbasisB} \cdot \BasisC \\
&=& \u{\D(\Omega)}^{\top} \cdot h_{\HtbasisB} \cdot \BasisC
\ea 
\eeq
Given the vectors associated to any two wedge-paired elements in $\Ha$ their integral can be written in terms of $g$ and in the cases of integrands $\Omega \wedge \D(\mho)$ and $\mho \wedge \D(\Omega)$ the expressions reduce to being in terms of $g_{\Hbasis}$ or $g_{\HtbasisB}$ only.
\beq
\label{eqn:WofM}
\int \Omega \wedge \D(\mho) &=& g\Big(\Omega,\D(\mho)\Big) = \u{\D(\mho)}^{\top}  \cdot g_{\Hbasis} \cdot \u{\Omega} = \u{\fT}^{\top} \cdot h_{\HtbasisB} \cdot \DevenA  \cdot g_{\Hbasis} \cdot \u{\fU} \\
\label{eqn:WofN}
\int \mho \wedge \D(\Omega) &=& g\Big(\mho,\D(\Omega)\Big) = \u{\D(\Omega)}^{\top} \cdot g_{\HtbasisB} \cdot \u{\mho} = \u{\fU}^{\top} \cdot h_{\Hbasis} \cdot \DoddA \cdot g_{\HtbasisB} \cdot \u{\fT}
\eeq
It is worth noting that the sum of these expressions can be expressed in a very natural way in terms of $\u{\Phi}$ and $\u{\u{\D}}$ and the bilinear forms defined on $\Basis$, putting the two moduli spaces into a single expression, because of the identity $\mho+\Omega = \u{\Phi}^{\top} \cdot h \cdot \Basis$.
\beq
\label{eqn:symW}
\int \mho \wedge \D(\Omega) + \int \Omega \wedge \D(\mho) &=& g_{\HtbasisB}(\mho,\D(\Omega)) + g_{\Hbasis}(\Omega,\D(\mho)) \nn \\
&=& g(\mho+\Omega,\D(\Omega)+\D(\mho)) \nn \\
&=& \u{\Phi}^{\top} \cdot h \cdot \u{\u{\D}} \cdot g \cdot \u{\Phi} 
\eeq
This expression for superpotential-like terms treats the two moduli spaces in exactly the same manner but this symmetry is broken when we consider the complexified holomorphic forms, as T duality related fluxes do not all couple to the dilaton in the same manner. In spaces where the fluxes are obtained by requiring modular invariance in the complex structure and K\"{a}hler moduli it is possible to express the superpotential in such a way that all those fluxes which define a covariant derivative couple to the dilaton in the same manner, as discussed for the $\ZZ$ orientifold in \cite{Aldazabal:2008aa}. We shall return to this result later when considering a particular set of internal spaces where such a symmetric formalism is possible even for the complexified holomorphic forms. To examine precisely how the inclusion of dilaton couplings in the complexified holomorphic forms breaks this symmetry we define a set of matrices associated to the holomorphic forms in the $\BasisA$ and $\BasisC$ bases.
\beq
\ba {ccccccccccccc}
\Omega &=& \u{\Omega}^{\top} \cdot h_{\Hbasis} \cdot \BasisA &=& \u{\fU}^{\top} \cdot \uwave{\Omega} \cdot \BasisA &\quad,\quad&
\mho &=& \u{\mho}^{\top} \cdot h_{\HtbasisB} \cdot \BasisC &=& \u{\fT}^{\top} \cdot \uwave{\mho} \cdot \BasisC 
\ea  \nn
\eeq
A second set of matrices, of the same dimensions as $\u{\u{\D}}$, can be defined for the holomorphic forms by taking the basis to be $\Basis$ and the moduli vector $\u{\Phi}$. In this way both holomorphic forms can be written in the same manner, $\Omega = \u{\Phi}^{\top}\cdot\u{\u{\Omega}}\cdot\Basis$ and $\mho = \u{\Phi}^{\top}\cdot\u{\u{\mho}}\cdot\Basis$. These matrices can be expressed in terms of $\uwave{\Omega}$ and $\uwave{\mho}$ through the use of projection matrices $\cP^{\pm} = \frac{1}{2}(\mathbb{I}_{2} \pm \Gamma_{2})$. We note how $\uwave{\Omega}$ and $\uwave{\mho}$ relate to $h_{\Hbasis}$ and $h_{\HtbasisB}$.
\beq
\u{\u{\Omega}} = \cP^{-}\otimes \uwave{\Omega} = \cP^{-}\otimes  h_{\Hbasis}  &\quad,\quad& \u{\u{\mho}} = \cP^{+}\otimes \uwave{\mho} =  \cP^{+}\otimes h_{\HtbasisA} \nn
\eeq
Though much of our analysis will be done in terms of $\uwave{\Omega}$, $\uwave{\mho}$ and their complexified forms the results and methods for $\u{\u{\Omega}}$, $\u{\u{\mho}}$ and their complexified forms follow in the same way and will be used in later sections so we consider both matrix types in tandem. Although the Type IIA integrand has the complex structure moduli complexified we consider the complexification of $\mho$ instead, with the $\Omega$ cases following the same general method, so as to have a simpler sign convention in our algebra. We define its vector/matrix factorisation in the same way as $\mho$, $\mho_{c} \equiv \u{\fT}^{\top} \cdot \uwave{\mho_{c}} \cdot \BasisC$. 
\beq
\mho_{c} = \bpm \fT_{0} \\ \fT_{a} \\ \fT^{0} \\ \fT^{b} \epm^{\top} \bpm -S \mathbb{I}_{1} \\ & \mathbb{I}_{h^{1,1}}  \\ && -S\mathbb{I}_{1}  \\ &&& \mathbb{I}_{h^{1,1}} \epm \bpm \v{0} \\ \v{a} \\ \vt{0} \\ \vt{b} \epm \nn
\eeq
This can be factorised so that the complexification is due to a single\footnote{Infact there are two complexified holomorphic forms, $\mho_{c}$ and $\mho_{c}^{\prime}$ and so we distinguish their complexification matrices with a prime, $\bC$ and $\bCp$ respectively.} matrix, $\bC$, which modifies the original expressions for the holomorphic forms.
\beq
\mho = \u{\fT}^{\top} \cdot \uwave{\mho} \cdot \BasisC \quad \to \quad \mho_{c} = \u{\fT}^{\top} \cdot \uwave{\mho_{c}} \cdot \BasisC \equiv \u{\fT}^{\top} \cdot \bC \cdot \uwave{\mho} \cdot \BasisC
\eeq
We can extract the expression for $\bC$ from this factorisation but it will be useful to write it, and $\bCp$, as a linear combination of a set of projection operators for examining S duality. The projection operators are such that they seperate out the $\cHn{0,0}$ and $\cHn{3,3}$ basis elements from the $\cHn{1,1}$ and $\cHn{2,2}$ bases and are built from $SO(n,m)$ metrics with signature $(+,\cdots,-,\cdots)$, which we shall denote as $\eta_{(n,m)}$.
\beq
\A_{n} &\equiv& \mathbb{I}_{2} \otimes \frac{1}{2}\Big( \eta_{(n+1,0)} - \eta_{(1,n)} \Big) = \mathbb{I}_{2} \otimes \bpm 0 & \\ & \mathbb{I}_{n} \epm = \mathbb{I}_{2} \otimes \fA_{n} \nn \\
\B_{n} &\equiv& \mathbb{I}_{2} \otimes \frac{1}{2}\Big( \eta_{(n+1,0)} + \eta_{(1,n)} \Big) = \mathbb{I}_{2} \otimes \bpm 1 & \\ & 0_{n} \epm  = \mathbb{I}_{2} \otimes \fB_{n} 
\nn
\eeq
Of note are the following set of identities for combining $\fA_{m}$ and $\fB_{n}$ and since the dimensionalities of the matrices are unambigious we suppress the indices.
\beq
\label{eqn:Matrixidents}
\fA \cdot \fA = \fA \quad , \quad \fB \cdot \fB = \fB  \quad , \quad \fA \cdot \fB = 0 = \fB \cdot \fA
\eeq
The $\A$ and $\B$ inherit the same set of identities due to their definitions in terms of $\fA$ and $\fB$ and it is these matrices which define the two K\"{a}hler complexified holomorphic forms.
\beq
\uwave{\mho_{c}} = \A_{h^{1,1}} -S \B_{h^{1,1}}  \qquad \uwave{\mho_{c}^{\prime}} = \B_{h^{1,1}} - S \A_{h^{1,1}} \nn
\eeq
In the case of the non-complexified holomorphic form matrices they can be written as a tensor product of matrices of the form $\eta_{(n,m)}$ and a similar tensor product formulation can be done for the complexified forms but with projection operators $\mathbb{P}_{n}^{(\prime)}$. 
\beq
\uwave{\mho_{c}}  &=& \mathbb{I}_{2} \otimes \Big( \fA_{h^{1,1}} -S \fB_{h^{1,1}} \Big) \equiv\mathbb{I}_{2}  \otimes \mathbb{P}_{h^{1,1}} \nn \\
\uwave{\mho_{c}^{\prime}} &=& \mathbb{I}_{2}  \otimes \Big( \fB_{h^{1,1}} -S \fA_{h^{1,1}} \Big)  \equiv \mathbb{I}_{2}  \otimes \mathbb{P}_{h^{1,1}}^{\prime} \nn
\eeq
With this pair of projection operators we can construct the matrix representations for the full complexified holomorphic forms
\beq
\u{\u{\mho_{c}}} \equiv \cP^{-}\otimes \uwave{\mho_{c}} = \cP^{-}\otimes \mathbb{I}_{2} \otimes \mathbb{P}_{h^{1,1}} &\quad,\quad& \u{\u{\mho^{\prime}_{c}}} \equiv \cP^{+}\otimes \uwave{\mho^{\prime}_{c}} = \cP^{+}\otimes \mathbb{I}_{2} \otimes \mathbb{P}_{h^{1,1}}^{\prime} \nn 
\eeq
By our definitions of the complexified holomorphic forms, the matrix representations are factorisable into complexification matrices and the original expressions for $\uwave{\mho}$ \textit{etc}. 
\beq
\ba {ccrclcccc}
\uwave{\mho_{c}} &=& \mathbb{I}_{2} \otimes \mathbb{P}_{h^{1,1}} &=& \Big(\mathbb{I}_{2} \otimes \mathbb{P}_{h^{1,1}}\Big) \cdot \Big(\mathbb{I}_{2} \otimes \mathbb{I}_{h^{1,1}+1}\Big) &=& \bC_{h^{1,1}} \cdot \uwave{\mho} &=& \bC_{h^{1,1}} \cdot h_{\HtbasisA} \\
\uwave{\Omega_{c}} &=& \eta_{(1,1)} \otimes \mathbb{P}_{h^{1,1}} &=& \Big(\mathbb{I}_{2} \otimes \mathbb{P}_{h^{2,1}}\Big) \cdot \Big(\eta_{(1,1)} \otimes \mathbb{I}_{h^{2,1}+1}\Big) &=& \bC_{h^{2,1}} \cdot \uwave{\Omega} &=& \bC_{h^{2,1}} \cdot h_{\Hbasis} 
\ea  \nn
\eeq
As the dimensions of the matrices are unambigious due to their context we will henceforth drop the subscripts unless required. This is extended to the full $2h^{1,1}+2h^{2,1}+4$ dimensional case by defining the combination combining $\cC^{(\prime)} \equiv \bC^{(\prime)}_{h^{1,1}}\oplus\bC^{(\prime)}_{h^{2,1}}$, resulting in factorisations such as $\u{\u{\Omega_{c}^{\prime}}} = \mathcal{C}^{\prime}\cdot\u{\u{\Omega}}$ or $\u{\u{\mho_{c}}} = \mathcal{C}\cdot\u{\u{\mho}}$. With complexification having the effect of $h_{\HtbasisB} \to \bC^{(\prime)}\cdot h_{\HtbasisB}$\footnote{As $[\bC,h_{\HtbasisB}] = 0$ this could alternatively be written as $h_{\HtbasisB} \cdot \bC$ and likewise for $\Omega \to \Omega_{c}$ and other complexification matrices.} on the expansion of $\mho$ the inner product expression for the T duality induced superpotential is obtained by altering the toy model expression previously found in (\ref{eqn:WofM}), with the R-R fluxes following from the fact Type IIB treats them in the same manner as NS-NS fluxes. 
\beq
\label{eqn:fullWinnerprod}
W &=& \int_{\M} \Omega_{c} \wedge \D(\mho) = \u{\fT}^{\top} \cdot h_{\HtbasisB} \cdot \DevenA \cdot g_{\Hbasis} \cdot \bC \cdot \u{\fU} 
\eeq

\subsection{Type IIB Fluxes}
\label{sec:TypeIIA}

Type IIA and Type IIB theories are related by T duality, with the particular case of three T dualities (in different directions) being equivalent to the mirror map $\Upsilon$. In general the compact space obtained from $\M$ under this map, which we shall denote as $\W$, is not the same as $\M$ which can be most readily seen from the Hodge numbers of the two spaces.
\beq
h^{1,1}(\M) = h^{2,1}(\W) \quad , \quad h^{2,1}(\M) = h^{1,1}(\W) \nn
\eeq
This exchange of Hodge numbers is closely related to the fact the K\"{a}hler and complex structure moduli spaces of the two theories are exchanged, though for the purposes of clarity we will continue to use $h^{1,1}$ to mean the Type IIB Hodge number, rather than using additional $A$ and $B$ labels to distinguish which Type II theory we are working in. The Type IIB complex structure moduli space therefore has $h^{1,1}$ dimensions and in keeping with previous index notation the index labels $a$, $b$ etc will range over $1$ to $h^{1,1}$. Conversely, the dimension of the Type IIB K\"{a}hler moduli space is $h^{2,1}$ and the indices $i$, $j$ etc will range over $1$ to $h^{2,1}$. 
\beq
\ba {ccccccc}
\hline \hline
\quad & \textrm{Type IIA} & \quad & & \quad & \textrm{Type IIB} & \quad \\ \hline \hline 
& \fU_{I} \,,\, \fU^{J} & & I,J = 0,\cdots,h^{2,1} & &  \sT_{I} \,,\, \sT^{J} & \\
& \fT_{A} \,,\, \fT^{B} & & A,B = 0,\cdots,h^{1,1} & &  \sU_{A} \,,\, \sU^{B} & \\
\hline \hline
\ea  \nn
\eeq
This will be of particular convenience when comparing flux expansions or matrix dimensions between the two theories and we therefore have a new $p$-form vector $\BasisX$ moduli vector due to the new moduli vector $\u{\Psi}$.
\beq
\ba {ccccc}
\BasisX^{\top} & \equiv & \bpm \BasisY^{\top} & \BasisZ^{\top} \epm &\equiv& \bpm \syma{0} & \syma{a} & \symb{0} & \symb{b} & \v{0} & \v{i} & \vt{0} & \vt{j}\epm \\
\u{\Psi}^{\top} & \equiv & \bpm \u{\sU}^{\top} & \u{\sT}^{\top} \epm &\equiv& \bpm \sU_{0} & \sU_{a} & \sU^{0} & \sU^{b} & \sT_{0} & \sT_{i} & \sT^{0} & \sT^{j}\epm\ea  \nn
\eeq
The Type II theories represent the degrees of freedom of the internal space in different manners, so the degrees of freedom denoted by the Type IIA $\sU$ are represented by the $\sT$ in Type IIB. In anticipation of having to construct the two alternative actions of the covariant derivatives we have taken our $\Hnt{3}$ basis $\BasisZ$ to be analogous to the second Type IIA change of basis $\BasisB \to \BasisC$ and work with the $\sT_{J}$ K\"{a}hler moduli. With this redefinition of the form vector $\Basis \to \BasisX$ we also have a new bilinear form $g = g(\Basis) \to g(\BasisX) \equiv \sg$. Due to our choice of the $\BasisZ$ basis we can use the mirror of (\ref{eqn:CliffShort}), obtained by relabelling $\syma{I} \to \v{I}$ \textit{etc} and setting the $\pm$ sign choice to $-$ and overall this gives a set of expressions which are schemicatically unchanged.
\beq
\label{eqn:CliffShortMirror}
\ba {ccccccc}
\v{I}\i_{\syma{A}} &\simeq& \phantom{-}\symb{A}\i_{\vt{I}} &\qquad& \vt{J}\i_{\syma{A}} &\simeq& -\symb{A}\i_{\v{J}} \\
\v{I}\i_{\symb{B}} &\simeq& -\syma{B}\i_{\vt{I}} &\qquad& \vt{J}\i_{\symb{B}} &\simeq& \phantom{-}\syma{B}\i_{\v{J}}
\ea  
\eeq 
Type IIA and Type IIB have the NS-NS sector `in common' and in terms of the superpotential this amounts to the natural flux multiplets being defined as covariant derivative images of the $\Hnt{3}$ sub-cohomologies \cite{Ihl:2007ah,Grana:2006hr}. In the case of the $3$-form flux $H_{3} \to \sFh_{0}$ the superpotential contribution is proportional to $\cT_{0} = \sT_{0}$ and in the same methodology as the Type IIA theory the extension of the NS-NS sector by T duality we would expect to be of a particular form. 
\beq
\textrm{IIB}  \; : \; W_{NS} = \int_{\W} \Omega \wedge \sD(\mho_{c})  \quad,\quad \textrm{IIA}  \; : \;  W_{NS} = \int_{\M} \Omega_{c} \wedge \D(\mho) \nn
\eeq
The fact both NS-NS sectors are defined by the derivatives acting on elements of their respective $\Hnt{3}$ suggests we are able to make the same $\La$ component expansion of $\sD$ as for $\D$ in terms of four fluxes induced by T duality; $\sFh_{0}$, $\sF_{1}$, $\sF_{2}$ and $\sFh_{3}$, where the hatted fluxes couple to the dilaton in their entirity. 
\beq
\sFh_{0} = \frac{1}{3!}\sFh_{pqr}\eta^{pqr} \quad,\quad \sF_{1} = \frac{1}{2!}\sFh_{pq}^{r}\eta^{pqr}\i_{r} \quad,\quad \sF_{2} = \frac{1}{2!}\sFh_{r}^{pq}\eta^{r}i_{qp} \quad,\quad \sFh_{0} = \frac{1}{3!}\sFh^{pqr}\i^{rqp} 
\eeq
These form precisely the same kind of gauge Lie algebra as in Type IIA and so the natural choice for the Type IIB NS-NS derivative is of the same form as $\D$ in Type IIA but for the time being we shall denote the sum of the fluxes as $\mathbb{G}$, in line with the $\mathbf{G}$ of Type IIA.
\beq
\label{eqn:G fluxes}
\mathbb{G} = \frac{1}{3!}\sFh_{pqr}\eta^{pqr} + \frac{1}{2!}\sF_{pq}^{r}\eta^{pqr}\i_{r} + \frac{1}{2!}\sF_{r}^{pq}\eta^{r}\i_{qp} + \frac{1}{3!}\sFh^{pqr}\i_{rqp} = \sFh_{0} + \sF_{1} + \sF_{2} + \sFh_{3}
\eeq
However, it is known that the non-geometric flux $\sF_{2}$ (commonly denoted as $Q$ in the literature) contributes a linear K\"{a}hler moduli dependency to the superpotential, in contrast to $\F_{2}$ which gave quadratic K\"{a}hler dependence in Type IIA. Therefore the integrand of the superpotential cannot be written as $\mathbb{G}(\mho_{c})$, the fluxes of $\sF_{1}$ and $\sF_{2}$ couple to the K\"{a}hler moduli in the wrong manner. To rectify this in a manner which leaves the Bianchi constraints invariant we consider two holomorphic forms $\ch{\Omega}$ and $\ch{\mho}$, which are modifications of the standard expressions over and above simple relabellings.
\beq\ba{ccccccccccc}
\mho &=& \u{\sT}^{\top} \cdot \sh_{\HtbasisB} \cdot \BasisC &\to&  \u{\sT}^{\top}\cdot L^{\top} \cdot \sh_{\HtbasisB} \cdot \BasisC &=& \ch{\mho}\\
\Omega &=& \u{\sU}^{\top} \cdot \sh_{\Hbasis} \cdot \BasisA &\to&  \u{\sU}^{\top}\cdot K^{\top} \cdot \sh_{\Hbasis} \cdot \BasisC &=& \ch{\Omega}\ea
\eeq
A Type IIB superpotential whose $\sF_{1}$ contributes quadratic K\"{a}hler moduli and $\sF_{2}$ complex structure moduli is then easily constructed from these new holomorphic forms.
\beq
\label{eqn:Different IIB NSNS}
 W_{NS} = \int_{\W} \ch{\Omega} \wedge \mathbb{G}(\ch{\mho}_{c})  = \u{\sT}^{\top}\cdot L^{\top} \cdot \sh_{\HtbasisB} \cdot \bC \cdot \mathsf{G} \cdot \sg_{\Hbasis}\cdot K \cdot \u{\sU}  
\eeq
We make the assumption that $L$ commutes with $\bC$, which will be justified later, as allows us to refactorise the scalar product expression for the superpotential.
\beq
\label{eqn : Different IIB}
W_{NS} = \int_{\W} \ch{\Omega} \wedge \mathbb{G}(\ch{\mho}_{c})  = \u{\sT}^{\top}\cdot \sh_{\HtbasisB} \cdot \bC \cdot \Big(L^{\top} \cdot \mathsf{G} \cdot \sg_{\Hbasis}\cdot K \cdot \sg_{\Hbasis}^{\top}\Big) \cdot \sg_{\Hbasis} \cdot \u{\sU}  \equiv \int_{\W} \Omega \wedge \sD(\mho_{c})
\eeq
This provides us with the matrix representation of $\sD$ acting on $\Hnt{3}$ elements in terms of the entries of $\mathsf{G}$, where we define the matrix representation in the same manner as Type IIA.
\beq
\label{eqn:Df}
\sD (\BasisX) = \sD\bpm \BasisY \\ \BasisZ \epm = \bpm 0 & \DoddB \\ \DevenB & 0 \epm \bpm \sh_{\Hbasis} & 0 \\ 0 & \sh_{\HtbasisB}\epm\bpm \BasisY \\ \BasisZ\epm  \equiv \u{\u{\sD}} \cdot \sh \cdot \BasisX
\eeq
The components of $\mathsf{G}$ are defined by (\ref{eqn:G fluxes}), in that they have the same relationship that the fluxes of $\F$ have with the components of $\DevenA$. However, it is more convenient to define flux components in terms of $\DevenB$ than to define them from $\mathsf{G}$ and then construct how they define $\DevenB$.
\beq
\DevenB \equiv L^{\top} \cdot \mathsf{G} \cdot \sg_{\Hbasis}\cdot K \cdot \sg_{\Hbasis}^{\top} \quad \Rightarrow \quad \mathsf{G} = (L^{\top})^{-1} \cdot \DevenB \cdot \sg_{\Hbasis}\cdot K^{-1} \cdot \sg_{\Hbasis}^{\top}
\eeq
The relationship between $\DoddB$ and $\DevenB$ is the same as that between $\DoddA$ and $\DevenA$, up to an exchange of Hodge numbers. Having defined a set of bilinear forms, $\sg$ and $\sh$, in addition to the Type IIA $g$ and $h$, we are able to express $\DoddA$ and $\DoddB$ in terms of $\DevenA$ and $\DevenB$ and these bilinear forms.
\beq
\label{eqn:ZetaofM}
\DoddA = \sg_{\HtbasisB} \cdot \DevenA^{\top} \cdot \sg_{\Hbasis} \quad,\quad \DoddB = g_{\HtbasisB} \cdot \DevenB^{\top} \cdot g_{\Hbasis}
\eeq 
Using these identities we can construct the action of $\sD$ on $\Hn{3}$ from its defining action on $\Hnt{3}$.
\beq
\ba {cccccc}
\sD(\v{I}) &=& \sF_{(I)A}\syma{A} &-& \sF_{(I)}^{\phantom{(I)}B}\symb{B}  \\
\sD(\vt{J}) &=& \sF_{\phantom{(J)}A}^{(J)}\syma{A} &-& \sF^{(J)B}\symb{B} 
\ea   \quad \Leftrightarrow \quad \ba {cccccc}
\sD(\syma{A}) &=& -\sF^{(I)A}\v{I} &+& \sF_{(J)}^{\phantom{(J)}A}\vt{J}  \\
\sD(\symb{B}) &=& -\sF_{\phantom{(I)}B}^{(I)}\v{I} &+& \sF_{(J)B}\vt{J} 
\ea 
\eeq 
In order to associate these components with the $\sF_{n}$ and $\sFh_{m}$ of the NS-NS sector we need to determine the specific form of both $L$ and $K$ but schematically it can be seen that $\sF_{1}$ defines the fluxes of $\sD(\v{i})$ and $\sF_{2}$ the fluxes of $\sD(\vt{j})$. As will be done explicitly later the requirement that both $\sD$ and $\mathbb{G}$ give rise to the same Bianchi constraints determines the entries of $K$ and $L$. Having constructed scalar product expressions for both Type II NS-NS superpotentials we can obtain the relationship between the fluxes by comparing these scalar products. 

\subsection{Flux Interdependency}

We now turn our attention to finding the relationship between the entries of $\DevenA$ and $\DoddB$, which is done by comparing the Type IIA superpotential with that of the Type IIB superpotential, though presently only in the NS-NS sector. In preparation for the R-R sector we consider an additional superpotential-like term in Type IIA, whose form is motivated by the moduli space exchange symmetry $\zeta_{1} : \mho \leftrightarrow \Omega$.
\begin{itemize}
\item Type IIB NS-NS superpotential.
\beq
\label{eqn:W1}
\int_{\W} \Omega \wedge \D(\mho_{c}) = \sg\Big( \Omega,\D(\mho_{c}) \Big)  = \u{\sT}^{\top} \cdot \sh_{\HtbasisB}  \cdot \bC \cdot \DevenB \cdot \sg_{\Hbasis} \cdot \u{\sU} 
\eeq
\item Type IIA NS-NS superpotential moduli dual to the Type IIB NS-NS superpotential.
\beq
\label{eqn:W2}
\int_{\M} \mho \wedge \sD(\Omega_{c}) = g\Big( \mho,\D(\Omega_{c}) \Big) = \u{\fU}^{\top} \cdot h_{\Hbasis} \cdot \bC \cdot \DoddA \cdot g_{\HtbasisB} \cdot \u{\fT}
\eeq

\item Type IIA NS-NS superpotential mirror dual to Type IIB NS-NS superpotential.
\beq
\label{eqn:W3}
\int_{\M} \Omega_{c} \wedge \sD(\mho) = g\Big( \Omega_{c},\D(\mho) \Big)  = \u{\fT}^{\top} \cdot h_{\HtbasisB} \cdot \DevenA \cdot g_{\Hbasis} \cdot \bC \cdot \u{\fU}
\eeq

\end{itemize}

In each case the dimension of the complexification matrix is the same, a point which we will refer back to later. If the Type II superpotentials were related by a simple exchange of their moduli spaces then (\ref{eqn:W1}) is equal, up to a relabelling of the moduli, to (\ref{eqn:W2}) and by comparing the matrix expressions we can determine $\DoddA$ in terms of $\DevenB$.
\beq
\label{eqn:Mirrorrelation}
\sh_{\HtbasisB} \cdot \DevenB \cdot \sg_{\Hbasis} = h_{\Hbasis} \cdot \DoddA \cdot g_{\HtbasisB}  \quad \Rightarrow \quad  \DevenB = h_{\Hbasis} \cdot \DoddA \cdot \sh_{\Hbasis}
\eeq
However, as the Type IIA superpotential is defined as the mirror dual of the Type IIB superpotential we must equate (\ref{eqn:W1}) with (\ref{eqn:W3}) instead, once we relabel the moduli and take the transpose of the matrix expression in (\ref{eqn:W3}).
\beq
\label{eqn: NSNS IIB to IIA}
\sh_{\HtbasisB} \cdot \DevenB \cdot \sg_{\Hbasis} = \Big(h_{\HtbasisB} \cdot \DevenA \cdot g_{\Hbasis}\Big)^{\top}  \quad \Rightarrow \quad \DevenB  = g_{\Hbasis} \cdot \DevenA^{\top} \cdot \sg_{\Hbasis} 
\eeq
Recalling the general expression for $\DevenA$ in terms of the $\F$ flux we now have the explicit dependence of $\DoddB$ in terms of these fluxes.
\beq
\DevenA \equiv \bpm  \F_{(A)I} & \F^{\phantom{(A)}J}_{(A)} \\  \F_{\phantom{(B)}I}^{(B)} & \F^{(B)J} \epm \quad \Rightarrow \quad \DevenB = \bpm \sF_{(I)A} & \sF^{\phantom{(I)}B}_{(I)} \\ \sF_{\phantom{(J)}A}^{(J)} & \sF^{(J)B} \epm = \bpm -\F^{(A)I} & \fm\F^{\phantom{(B)}I}_{(B)} \\ \fm\F_{\phantom{(A)}J}^{(A)} & -\F_{(B)J} \epm \nn
\eeq
Putting these results into (\ref{eqn:AlternateD}) we have the Type IIA NS-NS derivative action given the Type IIB NS-NS derivative action. The choose of expressing Type IIA fluxes in terms of Type IIB fluxes, rather than vice versa, is convenient for our later analysis of tadpoles.
%
%\beq
%\label{eqn:AlternateD3}
%\ba {ccccccccccccccccccccc}
%\D(\v{A})  &=& \phantom{-}\F_{(A)I}\syma{I} &-& \F_{(A)}^{\phantom{(A)}J}\symb{J}    &\quad&&\quad& \D(\syma{I})  &=& -\F^{(A)I}\v{A} &+& \F_{(B)}^{\phantom{(B)}I}\vt{B} \\
%\D(\vt{B}) &=& \phantom{-}\F_{\phantom{(B)}I}^{(B)}\syma{I} &-& \F^{(B)J}\symb{J}   &\quad&&\quad& \D(\symb{J})  &=& -\F_{\phantom{(A)}J}^{(A)}\v{A} &+& \F_{(B)J}\vt{B} \\\\
%\sD(\v{I}) &=& -\F^{(A)I}\syma{A} &-& \F_{(B)}^{\phantom{(B)}I}\symb{B}  &\quad&&\quad&   \sD(\syma{A}) &=& \fm\F_{(A)I}\v{I} &+& \F_{(A)}^{\phantom{(A)}J}\vt{J} \\
%\sD(\vt{J}) &=& \phantom{-}\F_{\phantom{(A)}J}^{(A)}\syma{A} &+& \F_{(B)J}\symb{B} &\quad&&\quad& \sD(\symb{B}) &=& -\F_{\phantom{(B)}I}^{(B)}\v{I} &-& \F^{(B)J}\vt{J} 
%\ea   
%\eeq 
%
\beq
\label{eqn:AlternateD3}
\ba {ccccccccccccccccccccc}
\sD(\v{I}) &=& \sF_{(I)A}\syma{A} &-& \sF_{(I)}^{\phantom{(I)}B}\symb{B}  &\quad&&\quad& \sD(\syma{A}) &=& -\sF^{(I)A}\v{I} &+& \sF_{(J)}^{\phantom{(J)}A}\vt{J} \\
\sD(\vt{J}) &=& \sF_{\phantom{(J)}A}^{(J)}\syma{A} &-& \sF^{(J)B}\symb{B} &\quad&&\quad& \sD(\symb{B}) &=& -\sF_{\phantom{(I)}B}^{(I)}\v{I} &+& \sF_{(J)B}\vt{J}  \\\\
\D(\v{A})  &=& -\sF^{(I)A}\syma{I} &-& \sF_{(J)}^{\phantom{(J)}A}\symb{J}    &\quad&&\quad& \D(\syma{I})  &=& \sF_{(I)A}\v{A} &+& \sF_{(I)}^{\phantom{(I)}B}\vt{B} \\
\D(\vt{B}) &=& \phantom{-}\sF_{\phantom{(I)}B}^{(I)}\syma{I} &+& \sF_{(J)B}\symb{J}   &\quad&&\quad& \D(\symb{J})  &=& -\sF_{\phantom{(J)}A}^{(J)}\v{A} &-& \sF^{(J)B}\vt{B} 
\ea   
\eeq 
This interdependence between the fluxes is not what would have been obtained if the relationship between the different superpotentials was that of $\zeta_{1}$, the exchange of moduli spaces. Under such a transformation we would have expected one of the Type II theories to have their flux multiplets defined by the covariant derivatives acting a particular $\cHn{3-n,n}$ cohomology. To refer to these two different kinds of ways of reformulating the superpotential we define a pair of operators $\ModA$ and $\ModB$ by their action on integrands, where $D_{A}$ and $D_{B}$ are derivatives defined in Type IIB and Type IIA respectively.
\beq
\label{eqn:ModMirror}
\ModA &\;:\;& \int_{\W} \Omega \wedge D_{A}(\mho_{c}) \leftrightarrow \int_{\M} \Omega_{c} \wedge D_{B}(\mho) \nn \\
\ModB &\;:\;& \int_{\W} \Omega \wedge D_{A}(\mho_{c}) \leftrightarrow \int_{\M} \mho \wedge D_{B}(\Omega_{c}) \nn
\eeq
Therefore, on the NS-NS sector we have that $\Upsilon = \ModA$. We refer to the $\Upsilon$ image of a superpotential as its mirror dual and we distinguish between that and the moduli space exchanging nature of $\ModB$ by calling the $\ModB$ image of a superpotential as its moduli dual. 

\section{Ramond-Ramond Sector}

\subsection{Type IIB}

Since Type IIB is self S dual its R-R sector is constructed by from the NS-NS sector by the dilaton inversion $S \to -\frac{1}{S}$.
\beq
\int_{\W} \Omega \wedge \D(\mho_{c}) \to \int_{\W} \Omega \wedge \Dp(\mho_{c}') = \int_{\W} \ch{\Omega} \wedge \mathbb{G}'(\ch{\mho}_{c}') 
\eeq
The flux components of $\mathbb{G}'$ follow the same structure as in the $\mathbb{G}$ expansion but with fluxes coupling to the dilaton in a different manner and so the hatted and unhatted fluxes are exchanged.
\beq
\mathbb{G}' &=& \sF_{0} + \sFh_{1} + \sFh_{2} + \sF_{3} = \frac{1}{3!}\sF_{abc}\eta^{abc} +  \frac{1}{2!}\sFh^{a}_{bc}\eta^{bc}\i_{a} + \frac{1}{2!}\sFh^{ab}_{c}\eta^{c}\i_{b}\i_{a} + \frac{1}{3!}\sF^{abc}\i_{c}\i_{b}\i_{a} 
\eeq
These components are related to the entries of the pair of flux matrices which determine the action of $\sD$ on the cohomology bases in the same way $\mathsf{G}$ and $\DevenB$ were. 
\beq
\sDp\bpm \BasisY \\ \BasisZ \epm = \bpm 0 & \DpoddB \\ \DpevenB & 0 \epm \bpm \sh_{\Hbasis} & 0 \\ 0 & \sh_{\HtbasisB}\epm\bpm \BasisY \\ \BasisZ\epm  \equiv \u{\u{\sDp}} \cdot \sh \cdot \BasisX \quad,\quad \DpevenB \equiv \bpm \sFh_{(I)A} & \sFh^{\phantom{(I)}B}_{(I)} \\ \sFh_{\phantom{(J)}A}^{(J)} & \sFh^{(J)B} \epm
\eeq
With the relationship between $\DpevenB$ and $\DpoddB$ being the same as that between $\DevenB$ and $\DoddB$ the construction of the alternative action of $\Dp$ on $\Hn{3}$ follows in a straightforward manner.
\beq
\ba {cccccc}
\sDp(\v{I}) &=& \sFh_{(I)A}\syma{A} &-& \sFh_{(I)}^{\phantom{(I)}B}\symb{B}  \\
\sDp(\vt{J}) &=& \sFh_{\phantom{(J)}A}^{(J)}\syma{A} &-& \sFh^{(J)B}\symb{B} 
\ea   \quad \Leftrightarrow \quad \ba {cccccc}
\sDp(\syma{A}) &=& -\sFh^{(I)A}\v{I} &+& \sFh_{(J)}^{\phantom{(J)}A}\vt{J}  \\
\sDp(\symb{B}) &=& -\sFh_{\phantom{(I)}B}^{(I)}\v{I} &+& \sFh_{(J)B}\vt{J} 
\ea 
\eeq 

\subsection{Type IIA}

Type IIA does not possess a self $\SL_{S}$ symmetry and so we cannot construct the R-R flux sector by simply taking the $\SL_{S}$ partner of the NS-NS sector. The NS-NS superpotential of (\ref{eqn:W3}) involves $\Omega_{c}$ and if the Type IIA superpotential as a whole is to follow a schematically similar form to the Type IIB superpotential then the R-R superpotential would be expected to involve $\Omega_{c}'$ in some way. Since $F_{RR}$ is independent of the complex structure moduli we deduce that in terms of the $\cU$ it would be associated to the $\cU_{0}$ modulus, which is not a degree of freedom due to its projective definition. The $3$-form associated to this complex structure modulus in the expansion of $\Omega_{c}'$ is $\a{0} = \syma{0}$ and we have that $F_{RR} = \cU_{0}F_{RR} \equiv \fF_{0}\cdot(\cU_{0}\a{0})$, which suggests that $F_{RR}$ is the first term in a more general flux dependent expression $\sDp(\Omega_{c}')$.
\beq
\int_{\M} e^{\J} \wedge F_{RR} = \int_{\M} \mho \wedge F_{RR} \to  \int_{\M} \mho \wedge \Dp(\Omega_{c}') \nn
\eeq
This demonstrates that the R-R sector has $\Upsilon = \ModB$, the moduli spaces are exchanged in a manifest way. As in the NS-NS case, we generalise this to include fluxes relating to the action of $\Dp$ on the other $\Hn{3}$ basis elements and it is clear that this expression for the Type IIA R-R superpotential is the mirror of the Type IIB R-R superpotential
\beq
\label{eqn:TypeIIARR}
\int_{\M} \mho \wedge F_{RR} \to \int_{\M} \mho \wedge \Big( \cU_{0}\,\fF_{0}\cdot\a{0}  - S\,\cU_{i}\,\fFh_{1}\cdot\a{i} + S\,\cU^{j}\,\fFh_{2}\cdot\b{j} -  \cU^{0}\,\fF_{3}\cdot\b{0} \Big)
\eeq
We have defined the $\fF_{n}$ and $\fFh_{m}$ by the way in which they couple to the $U_{i}$ in the same way in which $\sF_{n}\cdot \Jn{n}$ contribute $O(T_{a}^{n})$ terms to the superpotential in the Type IIB case. This labelling is entirely 
a choice of notation and has no consequence in terms of the components of the flux matrices which define the action of $\Dp$. Even when considering an orientifold projection it is possible to work entirely in terms of which moduli coefficients are included or projected out, rather than determining the action of the projection on the fluxes themselves. As will be done in a more explicit manner all T and S duality transformations or induced structures can be written entirely in terms of the flux matrices, allowing us to forgo the issue of considering precisely how to regard these new R-R contributions in terms of physical constructions such as branes. This extended R-R superpotential prompts us to define a complex structure counterpart to the K\"{a}hler forms $\Jn{n}$ so that $\Omega$ has a simplified expansion similar to $\mho$, but we use the original bases of the cohomologies.
\beq
\label{eqn:OmegaJ}
\ba{ccccccccc}
\mho &=& \cT_{0}\w{0} &+& \cT_{a}\w{a} &+& \cT^{b}\wt{b} &+& \cT^{0}\wt{0} \\ 
 &=& \Jn{0} &+& \Jn{1} &+& \Jn{2} &+& \Jn{3}
\ea \quad \Leftrightarrow \quad 
\ba{ccccccccc}
\Omega &=& \cU_{0}\a{0} &+& \cU_{i}\a{i} &-& \cU^{j}\b{j} &-& \cU^{0}\b{0} \\ 
 &=& \fJn{0} &+& \fJn{1} &-& \fJn{2} &-& \fJn{3}
\ea
\eeq
We have defined the Type IIA version here but the Type IIB version follows in precisely the same manner and is something we will make use of in a later section. Using these we are able to reexpress the Type IIA R-R superpotential so as to make the equality of mirror duality with moduli duality more manifest.
\beq
\int_{\M} \mho \wedge \Dp(\Omega_{c}') = \int_{\M} \mho \wedge \Big( \fF_{0}\cdot + \fFh_{1}\cdot + \fFh_{2}\cdot + \fF_{3}\cdot \Big) \Big( \fJn{0} - S\,\fJn{1} + S\,\fJn{2} - \fJn{3} \Big)
\eeq
We define the entries of $\DpoddB$ in terms of the components of the $\fF$ by the action of $\Dp$ on the basis element of $\Hn{3}$ but it must be done in the $(\syma{I},\symb{J},\fU)$ symplectic basis in order to conform to the Type IIB results.
\beq
\label{eqn:fFdef}
\int_{\M} \mho \wedge \Dp(\Omega_{c}') = \int_{\M} \mho \wedge \Big( \fU_{0}\,\fF_{0}\cdot\syma{0} - S\,\fU_{i}\,\fFh_{2}\cdot\syma{i} + S\,\fU^{j}\,\fFh_{1}\cdot\symb{j} - \fU^{0}\,\fF_{3}\cdot\symb{0} \Big)
\eeq
Given that the $\fF$ multiplets are defined as $\Dp$ images of $\Hn{3}$ basis elements they define the rows of $\DpoddA$ and as in previous cases we absorb $\fF_{0}$ and $\fF_{3}$ into $\fF_{1}$ and $\fF_{2}$.
\beq
\DpoddA \equiv \bpm \fFh_{(I)A} & \fFh^{\phantom{(I)}B}_{(I)} \\ \fFh_{\phantom{(J)}A}^{(J)} & \fFh^{(J)B}\epm \nn
\eeq
Due to the fact the fluxes are $\Dp$ images of $\Hn{3}$ elements it is not possible to express $\Dp$ in a component expansion of the form previously considered for the NS-NS sector and the Type IIB R-R sector. This also negates being able to easily construct a gauge algebra whose structure constants are the Type IIA R-R fluxes.
\beq
\Dp =  \fF_{0}\cdot + \fFh_{1}\cdot + \fFh_{2}\cdot + \fF_{3}\cdot \neq \frac{1}{3!}\fF_{abc}\eta^{abc} +  \frac{1}{3!}\fFh^{a}_{bc}\eta^{bc}\i_{a} + \frac{1}{3!}\fFh^{ab}_{c}\eta^{c}\i_{b}\i_{a} + \frac{1}{3!}\fF^{abc}\i_{c}\i_{b}\i_{a}
\eeq
While the R-R flux multiplets do not lend themselves to a concise manner of expression involving the $\eta^{\tau}$ and $\i_{\sigma}$ we can use the matrix representation of the derivative to construct an expression for $\Dp$ in terms of $\iota_{\BasisA}$ and $\BasisC$.
\beq
\Dp (\BasisA) = \DpoddA \cdot h_{\Hbasis} \cdot \BasisC \quad \Rightarrow \quad \Dp = \BasisC^{\top} \cdot h_{\Hbasis}^{\top} \cdot \DpoddA^{\top} \cdot \iota_{\BasisA}
\eeq
The action of $\Dp$ on elements of $\Hnt{3}$ can be constructed from its action on elements of $\Hn{3}$.
\beq
\ba {cccccc}
\Dp(\syma{I}) &=& \fFh_{(I)A}\v{A} &+& \fFh^{\phantom{(I)}B}_{(I)}\vt{B}  \\
\Dp(\symb{J}) &=& \fFh_{\phantom{(J)}A}^{(J)}\v{A} &+& \fFh^{(J)B}\vt{B} 
\ea  \quad \Leftrightarrow \quad 
\ba {cccccc}
\Dp(\v{A}) &=& \fm\fFh^{(I)A}\syma{I} &-& \fFh^{\phantom{(J)}A}_{(J)}\symb{J}  \\
\Dp(\vt{B}) &=& -\fFh_{\phantom{(I)}B}^{(I)}\syma{I} &+& \fFh_{(J)B}\symb{J} 
\ea 
\eeq 

\subsection{Flux Interdependency}

The explicit relationship between the $\Fh$ and the $\sFh$, or more specifically $\DpevenA$ and $\DpoddB$, is found in the same way as the NS-NS sector case, except this time we can use moduli duality directly, without having to change the argument of the derivative.
\begin{itemize}
\item Type IIB R-R superpotential.
\beq
\label{eqn:WRR1}
\int_{\W} \Omega \wedge \sDp(\mho_{c}') = \sg\Big( \Omega,\sDp(\mho_{c}') \Big)  = \u{\sT}^{\top} \cdot \sh_{\HtbasisB}  \cdot \bCp \cdot \DpevenB \cdot \sg_{\Hbasis} \cdot \u{\sU} 
\eeq
\item Type IIA mirror and moduli dual to the Type IIB R-R superpotential.
\beq
\label{eqn:WRR2}
\int_{\M} \mho \wedge \Dp(\Omega_{c}') = g\Big( \mho, \Dp(\Omega_{c}') \Big) = \u{\fU}^{\top} \cdot h_{\Hbasis} \cdot \bCp \cdot \DpoddA \cdot g_{\HtbasisB} \cdot \u{\fT}
\eeq
\end{itemize}
Comparing these two expressions we can determine $\DpevenB$ in terms of $\DpoddB$.
\beq
\label{eqn: RR IIB to IIA}
\sh_{\HtbasisB} \cdot \DpevenB \cdot \sg_{\Hbasis} = h_{\Hbasis} \cdot \DpoddA \cdot g_{\HtbasisB}  \quad \Rightarrow \quad \DpoddA = h_{\Hbasis} \cdot \DpevenB \cdot \sh_{\Hbasis}
\eeq
As a result of this we can express $\sFh$ in terms of the $\Fh$ flux matrices.
\beq
\DpoddA \equiv \bpm \fFh_{(I)A} & \fFh^{\phantom{(I)}B}_{(I)} \\ \fFh_{\phantom{(J)}A}^{(J)} & \fFh^{(J)B}\epm = \bpm \fm\sFh_{(I)A} & -\sFh^{\phantom{(I)}B}_{(I)} \\ -\sFh_{\phantom{(J)}A}^{(J)} & \fm\sFh^{(J)B}\epm   \nn
\eeq
Putting these results together we obtain the R-R version of (\ref{eqn:AlternateD3}) but in this case it is more convenient to express all fluxes in terms of Type IIB fluxes.
\beq
\label{eqn:AlternateD4}
\ba {ccccccccccccccccccccc}
\sDp(\v{I}) &=& \sFh_{(I)A}\syma{A} &-& \sFh_{(I)}^{\phantom{(I)}B}\symb{B} &\quad&&\quad& \sDp(\syma{A}) &=& -\sFh^{(I)A}\v{I} &+& \sFh_{(J)}^{\phantom{(J)}A}\vt{J} \\
\sDp(\vt{J}) &=& \sFh_{\phantom{(J)}A}^{(J)}\syma{A} &-& \sFh^{(J)B}\symb{B}&\quad&&\quad& \sDp(\symb{B}) &=& -\sFh_{\phantom{(I)}B}^{(I)}\v{I} &+& \sFh_{(J)B}\vt{J} \\\\
\Dp(\syma{I}) &=& \fm\sFh_{(I)A}\v{A} &-& \sFh_{(I)}^{\phantom{(I)}B}\vt{B}  &\quad&&\quad& \Dp(\v{A}) &=& \fm\sFh^{(I)B}\syma{I} &+& \sFh_{(J)}^{\phantom{(J)}A}\symb{J}  \\
\Dp(\symb{J}) &=& -\sFh_{\phantom{(J)}A}^{(J)}\v{A} &+& \sFh^{(J)B}\vt{B}  &\quad&&\quad& \Dp(\vt{B}) &=& \fm\sFh_{\phantom{(I)}B}^{(I)}\syma{I} &+& \sFh_{(J)B}\symb{J}
\ea   
\eeq 

\section{Flux constraints}

Thus far we have only considered what fluxes might in principle be required for a flux compactification which is invariant under T and S duality transformations, we have not addressed the issue of their consistency constraints. 

\subsection{T Duality Constraints}

Though T duality alone does not induce additional fluxes in the Type IIA R-R sector we assume that all fluxes which might be required for full T and S duality invariance are turned on. However, we initially restrict our analysis to the conditions which T duality induces. This approach allows us to regard the NS-NS sector and the R-R sector as disjoint from one another, even though the R-R sector is considered to include fluxes which require S duality transformations to be induced. Once we have examined the T duality induced structures in terms of these fluxes we shall extend our analysis to include $\SL_{S}$ transformations, thus providing the proper description of the fluxes.

\subsubsection{Bianchi Constraints}

$\D$ must satisfy Bianchi constraints in order for consistency, which in absense of any fluxes are equivalent to $\d^{2}=0$ and so is trivially satisfied. Turning on the fluxes induced by T duality is equivalent to removing the closure properties of the cohomology base. As previously discussed the action of $\F_{1}$ is closely related to $\d$ itself, as they both map $p$-forms to $p+1$-forms and while the non-geometric fluxes do not have the same equivalent behaviour they too contribute to $\D^{2}$.
\beq
\label{eqn:Bianchi}
\D^{2} = \Big(\d + \F_{0}\wedge \Big)^{2} \sim \Big( \F_{0}\wedge + \F_{1}\cdot + \F_{2}\cdot + \F_{3}\cdot \Big)^{2}  = 0 
\eeq
As previously commented, using the component expressions in (\ref{eqn:Dform1}) the constraints arising from acting such a $\D^{2}$ on elements of each $\Ln{p}(\M)$ are derived in full in \cite{Ihl:2007ah}. Considering only $\Ha$ defined fluxes is insufficient for full nilpotency but it has a number of advantages over the $\La$ analysis; the equivalence of the Bianchi constraints due to $\mathbb{G}$ and $\sD$ in Type IIB is manifest, $\SL_{S}$ multiplets are constructable in both Type IIB and Type IIA and the analysis of tadpoles can be subsumed into the analysis of Bianchi constraints. All of these will be derived and examined in this section. If such a restricted analysis is to be valid then there needs to be a clearcut distinction between light and heavy modes. The light modes are those whose mass is zero in the case where all fluxes are turned off, while the heavy are the converse, they are not harmonic even in the fluxless limit. The expression for the Laplacian in terms of the fluxes is given in Appendix Section \ref{Laplacian section} and since there are two flux sectors there are two Laplacians, ignoring the issue of S duality induced mixing. The construction of the Laplacians follows the same method as the construction of the Bianchi constraints and we first consider the Type IIA NS-NS constraints. These can be expressed in two ways, in terms of the flux components or in terms of flux matrices, the latter of which we consider first.
\beq
\D^{2}(\Basis) = \u{\u{\D}}\cdot h \cdot  \u{\u{\D}} \cdot h \cdot \Basis = \bpm \DevenA \cdot h_{\Hbasis} \cdot \DoddA \cdot h_{\HtbasisB} & 0 \\ 0& \DoddA \cdot h_{\HtbasisB} \cdot \DevenA \cdot h_{\Hbasis} \epm \bpm \BasisC \\ \BasisA \epm \eeq
Using the flux component definition of $\DevenA$ and its relationship to $\DoddA$ we can expand these two expressions into four sets of flux polynomials.
\beq
\label{eqn:Bianchi in Type IIA}
\ba{ccccccccccccccc}
\D^{2}(\v{A}) &=& \Big( & \F_{(A)I}\F^{(B)I} &-& \F_{(A)}^{\phantom{(A)}J}\F_{\phantom{(B)}J}^{(B)} & \Big)\v{B} &+& \Big( & \F_{(A)}^{\phantom{(A)}J}\F_{(B)J} &-& \F_{(A)I}\F^{\phantom{(B)}I}_{(B)} & \Big)\vt{B}\\
\D^{2}(\vt{B}) &=& \Big( & \F_{\phantom{(B)}I}^{(B)}\F^{(A)I} &-& \F^{(B)J}\F_{\phantom{(A)}J}^{(A)} & \Big)\v{A} &+& \Big( & \F^{(B)J}\F_{(A)J} &-& \F_{\phantom{(B)}I}^{(B)}\F_{(A)}^{\phantom{(A)}I} & \Big)\vt{A} \\
\D^{2}(\syma{I}) &=& \Big( & \F^{(A)I}\F_{(A)J} &-& \F_{(B)}^{\phantom{(B)}I}\F_{\phantom{(B)}J}^{(B)} & \Big)\syma{J} &+& \Big( & \F_{(B)}^{\phantom{(B)}I}\F^{(B)J} &-& \F^{(B)I}\F_{(B)}^{\phantom{(B)}J} & \Big)\symb{J}  \\
\D^{2}(\symb{J}) &=& \Big( & \F_{\phantom{(A)}J}^{(A)}\F_{(A)I} &-& \F_{(B)J}\F_{\phantom{(B)}I}^{(B)} & \Big)\syma{I} &+& \Big( & \F_{(B)J}\F^{(B)I} &-& \F_{\phantom{(A)}J}^{(A)}\F_{(A)}^{\phantom{(A)}I} & \Big)\symb{I}  
\ea 
\eeq 
The nilpotency of $\D$ is therefore expressible in terms of an ideal whose generating functions are these components of $(\u{\u{\D}}\cdot h)^{2}$.
\beq
\la \D^{2} \ra = \la \DevenA \cdot h_{\Hbasis} \cdot \DoddA,\DoddA \cdot h_{\HtbasisB} \cdot \DevenA\ra \nn
\eeq
We have neglected the `external' factor of $h$ as it is non-degenerate and therefore does not alter the ideal the constraints generate. The Type IIB case for $\sD$ follows in the same manner, defining four sets of flux component constraints.
\beq
\label{eqn:Bianchi in Type IIB}
\ba{ccccccccccccccc}
\sD^{2}(\v{I}) &=& \Big( & \sF_{(I)A}\sF^{(J)A} &-& \sF_{(I)}^{\phantom{(I)}B}\sF_{\phantom{(J)}B}^{(J)} & \Big)\v{J} &+& \Big( & \sF_{(I)}^{\phantom{(I)}B}\sF_{(J)B} &-& \sF_{(I)A}\sF^{\phantom{(J)}A}_{(J)} & \Big)\vt{J}\\
\sD^{2}(\vt{J}) &=& \Big( & \sF_{\phantom{(J)}A}^{(J)}\sF^{(I)A} &-& \sF^{(J)B}\sF_{\phantom{(I)}B}^{(I)} & \Big)\v{I} &+& \Big( & \sF^{(J)B}\sF_{(I)B} &-& \sF_{\phantom{(J)}A}^{(J)}\sF_{(I)}^{\phantom{(I)}A} & \Big)\vt{I} \\
\sD^{2}(\syma{A}) &=& \Big( & \sF^{(I)A}\sF_{(I)B} &-& \sF_{(J)}^{\phantom{(J)}A}\sF_{\phantom{(J)}B}^{(J)} & \Big)\syma{B} &+& \Big( & \sF_{(J)}^{\phantom{(J)}A}\sF^{(J)B} &-& \sF^{(J)A}\sF_{(J)}^{\phantom{(J)}B} & \Big)\symb{B}  \\
\sD^{2}(\symb{B}) &=& \Big( & \sF_{\phantom{(I)}B}^{(I)}\sF_{(I)A} &-& \sF_{(J)B}\sF_{\phantom{(J)}A}^{(J)} & \Big)\syma{A} &+& \Big( & \sF_{(J)B}\sF^{(J)A} &-& \sF_{\phantom{(I)}B}^{(I)}\sF_{(I)}^{\phantom{(I)}A} & \Big)\symb{A}  
\ea 
\eeq 
Likewise we have a pair of flux matrix dependent expressions.
\beq
\la \sD^{2} \ra = \la \DevenB \cdot \sh_{\Hbasis} \cdot \DoddB,\DoddB \cdot \sh_{\HtbasisB} \cdot \DevenB\ra \nn
\eeq
Due to the manner in which the Type IIB fluxes couple to the K\"{a}hler moduli in a way different to that of the Type IIA fluxes we had to define two different flux dependent expressions, $\mathbb{G}$ and $\sD$. By construation $\mathbb{G}$'s nilpotency constraints were equal to the constraints of the Type IIB gauge algebra but this is not automatically true for $\sD$. By comparing the superpotentials of (\ref{eqn : Different IIB}), which followed from these two different expressions, we obtained a flux matrix of $\sD$ in terms of a flux matrix of $\mathbb{G}$.
\beq
\DevenB = L^{\top} \cdot \mathsf{G} \cdot \sg_{\Hbasis}\cdot K \cdot \sg_{\Hbasis}^{\top}
\eeq
Rather than construct their flux matrix partners it is more convenient to express the Bianchi constraints in terms of single flux matrix by using (\ref{eqn:ZetaofM}), thus allowing us to construct the ideals of $\la \sD^{2} \ra$ and $\la \mathbb{G} \ra$ in terms of $\DevenB$ and $\mathsf{G}$ respectively. 
\beq
\label{eqn : quadratic derivs}
\la \sD^{2} \ra = \la \DevenB \cdot \sg_{\Hbasis} \cdot \DevenB^{\top},\DevenB^{\top} \cdot g_{\Hbasis} \cdot \DevenB\ra  \quad,\quad
\la \mathbb{G}^{2} \ra = \la \mathsf{G} \cdot \sg_{\Hbasis} \cdot \mathsf{G}^{\top},\mathsf{G}^{\top} \cdot g_{\Hbasis} \cdot \mathsf{G}\ra \nn
\eeq
We have once again droped irrelevant factors of non-degenerate matrices since they do not change the ideal. Considering each case in turn and not yet making any assumption about the specific form of $L$ and $K$ we can derive the constraints they must satisfy for $\sD$ and $\mathbb{G}$ to have equivalent Bianchi constraints on the $\Ha$.
\beq
\ba{cccccc}
\la \DevenB \cdot \sg_{\Hbasis} \cdot \DevenB^{\top} \ra &=& \la \mathsf{G} \cdot \sg_{\Hbasis}\cdot K \cdot \sg_{\Hbasis} \cdot K^{\top} \cdot \sg_{\Hbasis} \cdot \mathsf{G}^{\top} \ra &\quad\Rightarrow\quad& K \cdot \sg_{\Hbasis} \cdot K^{\top} = \sg_{\Hbasis} \\
\la \DevenB^{\top} \cdot g_{\Hbasis} \cdot \DevenB \ra &=& \la \mathsf{G}^{\top} \cdot L \cdot g_{\Hbasis} \cdot L^{\top} \cdot \mathsf{G} \ra &\quad\Rightarrow\quad& L \cdot g_{\Hbasis} \cdot L^{\top} = g_{\Hbasis}
\ea
\eeq
Therefore both $L$ and $K$ are sympletic matrices as they preserve the canonical sympletic form. This is further restricted by recalling that we required $\bC$ to commute with $L$ and in order to treat the moduli spaces in the same manner we require $K$ to commute with $\t{\bC}$. Since the existence of $L$ is motivated by exchanging the $\cT_{i}$ and $\cT^{j}$ moduli in the construction of the superpotential we set the action $L$ to not alter the superpotential's $\cT_{0}$ and $\cT^{0}$ terms. This is further justified by the fact $\bC$ treats those moduli differently, coupling them to the dilaton. Thus we can make a general ansatz for $L$ and $K$.
\beq
L = \bpm 1 & 0 & 0 & 0 \\ 0 & 0 & 0 & L_{1} \\ 0 & 0 & 1 & 0 \\ 0 & L_{2} & 0 & 0 \epm \quad,\quad K = \bpm 1 & 0 & 0 & 0 \\ 0 & 0 & 0 & K_{1} \\ 0 & 0 & 1 & 0 \\ 0 & K_{2} & 0 & 0 \epm 
\eeq
The sympletic constraints now reduce to skew-orthogonality of the submatrixes, $L_{2}\cdot L_{1}^{\top} = -\mathbb{I}$ and $K_{2}\cdot K_{1}^{\top} = -\mathbb{I}$. The simplest specific solution is $L_{1} = -L_{2} = \mathbb{I}$ which has the effect $\Jn{1} = \cT_{i}\w{i} \to -\cT^{j}\w{j} = \J_{c}$, the standard definition used for the K\"{a}hler $4$-form. Provided we frame our analysis of the fluxes and their duality induced structures in terms of the flux matrices of $\sD$ we are not required to use specific $L_{i}$ or $K_{j}$ as we can construct the polynomial expressions for the Type IIB superpotential via $\sD$ in the same manner as we did for the Type IIA case. This greatly simplifies the analysis of the Type IIB theory, as everything can be written in terms of $\DevenB$ rather than $\mathsf{G}$. With this knowledge we now turn to comparing the nilpotency constraints of the fluxes of Type IIA to those of Type IIB.
\\

The generic problem of comparing the equivalence of the Type IIA and Type IIB nilpotency constraints reduces to comparing the ideals defined by the associated polynomials.
\beq
\ba{cccccccccccccc}
\textrm{Type IIA} &\;:\;& \cI_{A,N} &=& \la \D^{2} \ra &\quad& \cI_{A,R} &=& \la \Dp^{2} \ra \\
\textrm{Type IIB} &\;:\;& \cI_{B,N} &=& \la \sD^{2} \ra &\quad& \cI_{B,R} &=& \la \sDp^{2} \ra 
\ea
\eeq
To confirm that the constraints are equivalent we use (\ref{eqn: NSNS IIB to IIA}) and (\ref{eqn: RR IIB to IIA}) to convert the flux matrix expressions between the Type II constructions.
\beq
\label{eqn:BtoAsub}
\ba {cccccccccccccccc}
\cI_{B,N} &=& \Big\la & \ba{c}\DoddB \cdot \sh_{\HtbasisB} \cdot \DevenB \\ \DevenB \cdot \sh_{\Hbasis} \cdot \DoddB \ea &\Big\ra &\to& \Big\la & \ba{c}\phantom{-}\sh_{\Hbasis} \cdot \Big(\DevenA \cdot h_{\Hbasis} \cdot \DoddA\Big) \cdot h_{\HtbasisB} \\ -\sh_{\Hbasis} \cdot \Big( \DoddA \cdot h_{\HtbasisB} \cdot \DevenA \Big) \cdot h_{\HtbasisB} \ea &\Big\ra &=& \cI_{A,N} \\
\cI_{B,R} &=& \Big\la & \ba{c}\DpoddB \cdot \sh_{\HtbasisB} \cdot \DpevenB \\ \DpevenB \cdot \sh_{\Hbasis} \cdot \DpoddB \ea &\Big\ra &\to& \Big\la & \ba{c} \phantom{-}h_{\Hbasis} \cdot \Big( \DpevenA \cdot h_{\Hbasis} \cdot \DpoddA \Big) \cdot \sh_{\HtbasisB} \\ -\sh_{\HtbasisB}\cdot \Big( \DpoddA \cdot h_{\HtbasisB} \cdot \DpevenA \Big) \cdot h_{\Hbasis} \ea &\Big\ra &=& \cI_{A,R} 
\ea  
\eeq
In (\ref{eqn : quadratic derivs}) we used the relationship between $\DoddB$ and $\DevenB$ to express the Bianchi constraints in terms of a single flux matrix and this can be done for the other flux sectors and in both Type IIA or Type IIB.
\beq
\label{eqn:Quadconstraints}
\la \sD^{2} \ra \;\Leftrightarrow\; \Bigg\{\ba{ccccc}
\la \DevenB \cdot \sh_{\Hbasis} \cdot \DoddB \ra &\;\Rightarrow\;& \Big\{ \quad 
\ba {rclcrcl}
\la \DevenB \cdot \sg_{\Hbasis} \cdot \DevenB^{\top} \ra \gapA \la \DoddB^{\top} \cdot \sg_{\Hbasis} \cdot \DoddB \ra \\
\la \DoddA \cdot \sg_{\Hbasis} \cdot \DoddA^{\top} \ra \gapA \la \DevenA^{\top} \cdot \sg_{\Hbasis} \cdot \DevenA \ra 
\ea  \\
\la \DoddB \cdot \sh_{\HtbasisB} \cdot \DevenB \ra &\;\Rightarrow\;& \Big\{ \quad 
\ba {rclcrcl}
\la \DevenB^{\top} \cdot g_{\Hbasis} \cdot \DevenB \ra \gapA \la \DoddB \cdot g_{\Hbasis} \cdot \DoddB^{\top} \ra\\
\la \DoddA^{\top} \cdot g_{\Hbasis} \cdot \DoddA \ra \gapA \la \DevenA \cdot g_{\Hbasis} \cdot \DevenA^{\top} \ra 
\ea  
\ea 
\eeq
The expression $\DevenA \cdot g_{\Hbasis} \cdot \DevenA^{\top}$ suggests the nilpotency constraints can be rephrased to be in terms of an integrand defined by a pair of exact $3$-forms, with $g_{\Hbasis}$ implying the forms are defined in the $\Hn{3}$ of Type IIA. To examine this we use a pair of vectors $\u{\chi}$ and $\u{\varphi}$ of dimension $2h^{1,1}+2$ and a pair of vectors $\u{\phi}$ and $\u{\psi}$ of dimension $2h^{2,1}+2$, which then allow us to define sets of forms in either Type II theory.
\beq
\label{table:fluxcomponents}
\textrm{IIA} \quad \Big\{ \quad 
\ba {cccccccccccc}
\chi &\equiv& \u{\chi}^{\top} \cdot h_{\HtbasisB} \cdot \BasisC &\quad \leftrightarrow \quad & \t{\chi} &\equiv& \u{\chi}^{\top} \cdot \sh_{\Hbasis} \cdot \BasisY \\
\phi &\equiv& \u{\phi}^{\top} \cdot h_{\Hbasis} \cdot \BasisA &\quad \leftrightarrow \quad & \t{\phi} &\equiv& \u{\phi}^{\top} \cdot \sh_{\HtbasisB} \cdot \BasisZ
\ea  \quad \Big\} \quad \textrm{IIB} 
\eeq
In order to obtain expressions which are quadratic in $\DevenA$ we consider the pair of exact $3$-forms $\D(\phi)$ and $\D(\varphi)$ and their skew-inner product in Type IIA.
\beq
\label{eqn:DD1}
\int_{\M} \D(\chi) \wedge \D(\varphi) = g_{\Hbasis}\Big( \D(\chi),\D(\varphi)\Big) = \u{\varphi}^{\top} \cdot h_{\HtbasisB} \cdot \DevenA \cdot g_{\Hbasis} \cdot \DevenA^{\top} \cdot h_{\HtbasisB} \cdot \u{\chi} 
\eeq
Therefore the first set of constraints on the fluxes can be reexpressed as the vanishing of the skew-inner product of any two $\D$ exact elements of the Type IIA $\Hn{3}$, with the R-R sector following the same method, but with $\Dp$ in the place of $\D$.
\beq
g_{\HtbasisB}\Big( \chi ,\D\big(\D(\varphi)\big)\Big) = 0 \quad \Leftrightarrow \quad g_{\Hbasis}\Big( \D(\chi),\D(\varphi)\Big) = 0 \nn
\eeq
Since the bilinear form in each equation of (\ref{eqn:Quadconstraints}) is $g_{\Hbasis}$ or $\sg_{\Hbasis}$ it is clear that the constraints can only be rephrased in terms of exact forms belonging to $\Hn{3}$ in either Type IIA or Type IIB. Therefore the only other combination we need consider is that of the exact forms defined by the $\sD$ images of the Type IIB $\Hnt{3}$ elements $\t{\phi}$ and $\t{\psi}$.
\beq
\label{eqn:DD2}
\int_{\W} \sD(\t{\phi}) \wedge \sD(\t{\psi}) = \sg_{\Hbasis}\Big( \sD(\t{\phi}) , \sD(\t{\psi})\Big) = \u{\t{\psi}}^{\top} \cdot \sh_{\HtbasisB} \cdot \DevenB \cdot \sg_{\Hbasis} \cdot \DevenB^{\top} \cdot \sh_{\HtbasisB} \cdot \u{\t{\phi}}
\eeq
Therefore the first set of constraints on the fluxes can be reexpressed as the vanishing of the skew-inner product of any two $\sD$ exact elements of the Type IIB $\Hn{3}$, with the R-R sector following the same method, but with $\sDp$ in the place of $\sD$.
\beq
\sg_{\HtbasisB}\Big( \t{\phi} ,\sD\big(\sD(\t{\psi})\big)\Big) = 0 \quad \Leftrightarrow \quad \sg_{\Hbasis}\Big( \sD(\t{\phi}),\sD(\t{\psi})\Big) = 0
\eeq
Therefore we can write nilpotency defined ideals in terms of orthogonal exact form defined ideals.
\beq
\ba{ccccccccccccccc}
\la \D^{2} \ra &=& \left\la g_{\Hbasis}\Big( \D(\,\cdot\,),\D(\,\cdot\,)\Big) , \sg_{\Hbasis}\Big( \sD(\,\cdot\,),\sD(\,\cdot\,)\Big) \right\ra  &=& 
\la \sD^{2} \ra \\
\la \Dp^{2} \ra &=& \left\la g_{\Hbasis}\Big( \Dp(\,\cdot\,),\Dp(\,\cdot\,)\Big) , \sg_{\Hbasis}\Big( \sDp(\,\cdot\,),\sDp(\,\cdot\,)\Big) \right\ra  &=& 
\la \sDp^{2} \ra 
\ea
\eeq

\subsubsection{Tadpoles}

Given $\M$ is six dimensional the Type IIA models can have non-zero potentials of the form $C_{5}$, $C_{7}$ and $C_{9}$ but due to the emptiness of $\Hn{1}$ and $\Hn{5}$ for $\M$ the only non-trivial case can be $C_{7}$, whose tadpole contributions arise from $3$-form flux combinations. It is known \cite{Grana:2006hr,Ihl:2007ah} that the tadpole expression for spaces with only $F_{RR}$ R-R fluxes is, up to proportionality factors, measured by $\D(F_{RR})$. In our analysis of the Type IIA flux sector we noted that $F_{RR}$ could be written as an exact derivative in terms of $\Dp$ and so the tadpole term can be written as a quadratic derivative.
\beq
F_{RR} \equiv \fF_{0}\cdot\a{0} = \Dp(\a{0}) = \Dp(\syma{0}) \quad \Rightarrow \quad \D(F_{RR}) = \D\Dp(\syma{0})
\eeq
The fact $F_{RR}$ could be written in this way provided a natural extension of the sector to include other $\fF$ fluxes and as a result we are able to form additional $3$-forms which have the schematic form of being a $\D$ derivative of a $\Dp$ exact form.
\beq
\label{eqn:TadpolesA}
\ba{ccccccccccccc}
\D(\fF_{0}\cdot \a{0}) &=& \D\Dp(\a{0}) &=& +\D\Dp(\syma{0}) &\quad,\quad& \D(\fFh_{1}\cdot \a{a}) &=& \D\Dp(\a{a}) &=& -\D\Dp(\symb{a}) \\
\D(\fF_{3}\cdot \b{0}) &=& \D\Dp(\b{0}) &=& +\D\Dp(\symb{0}) &\quad,\quad& \D(\fFh_{2}\cdot \b{b}) &=& \D\Dp(\b{b}) &=& +\D\Dp(\syma{b}) \\
\ea
\eeq
Using (\ref{eqn:AlternateD3}) and (\ref{eqn:AlternateD4}) we can determine the component form of these expressions in terms of the Type IIB fluxes.
\beq
\label{eqn:TadpoleT2}
\ba{ccccccccccccccc}
\D\Dp(\syma{I}) &=& - &\Big( & \sFh_{(I)A}\sF^{(J)A} &+& \sFh_{(I)}^{\phantom{(I)}B}\sF_{\phantom{(J)}B}^{(J)} & \Big)\syma{J} &-& \Big( & \sFh_{(I)A}\sF_{(J)}^{\phantom{(J)}A} &+& \sFh_{(I)}^{\phantom{(I)}B}\sF_{(J)B} & \Big)\symb{J}  \\
\D\Dp(\symb{I}) &=&   &\Big( & \sFh^{(I)A}\sF_{\phantom{(J)}A}^{(J)} &+& \sFh_{\phantom{(I)}A}^{(I)}\sF^{(J)A} & \Big)\syma{J} &+& \Big( & \sFh^{(I)A}\sF_{(J)A} &+& \sFh_{\phantom{(I)}B}^{(I)}\sF_{(J)}^{\phantom{(J)}B} & \Big)\symb{J}  
\ea \qquad
\eeq 
All of these expressions are constructed from applying $\D\Dp$ to $3$-forms and since they are also expanded in terms of the $\Hn{3}$ basis it follows that all of these expressions can contribute tadpoles. The fluxes of $F_{RR}$ have a straightforward definition in terms of the D-brane content of Type IIA but the remaining cases, $\D\Dp(\syma{i})$ and $\D\Dp(\symb{J})$, do not. The $F_{RR}$ tadpole contributions can be decomposed as $\syma{I}\wedge \D\Dp(\syma{0})$ and $\symb{J}\wedge \D\Dp(\syma{0})$ and the physical interpretation is the D$6$-brane wrapping particular $3$-cycles in $\W$. In order to justify considering these expressions as new tadpole contributions we also construct the equivalent expressions for $\Dp\D$.
\beq
\label{eqn:TadpoleT3}
\ba{ccccccccccccccc}
\Dp\D(\syma{I}) &=&  &\Big( & \sF_{(I)A}\sFh^{(J)A} &+& \sF_{(I)}^{\phantom{(I)}B}\sFh_{\phantom{(J)}B}^{(J)} & \Big)\syma{J} &+& \Big( & \sF_{(I)A}\sFh_{(J)}^{\phantom{(J)}A} &+& \sF_{(I)}^{\phantom{(I)}B}\sFh_{(J)B} & \Big)\symb{J}  \\
\Dp\D(\symb{I}) &=& - &\Big( & \sF^{(I)A}\sFh_{\phantom{(J)}A}^{(J)} &+& \sF_{\phantom{(I)}A}^{(I)}\sFh^{(J)A} & \Big)\syma{J} &-& \Big( & \sF^{(I)A}\sFh_{(J)A} &+& \sF_{\phantom{(I)}B}^{(I)}\sFh_{(J)}^{\phantom{(J)}B} & \Big)\symb{J}  
\ea \qquad
\eeq 
By considering (\ref{eqn:TadpoleT2}) and (\ref{eqn:TadpoleT3}) we observe a number of identities relating the individual coefficients which allows us to link this standard tadpole to the new expressions. We use the relabellings $\D = \D_{1}$ and $\Dp = \D_{2}$ for simplicity.
\beq
\label{eqn:IIAtadpoles}
\ba{ccr}
\syma{I} \wedge \D_{n}\D_{m}(\syma{J}) &=& -\syma{J} \wedge \D_{m}\D_{n}(\syma{I}) \\
\syma{I} \wedge \D_{n}\D_{m}(\symb{J}) &=& -\symb{J} \wedge \D_{m}\D_{n}(\syma{I}) \\ 
\symb{I} \wedge \D_{n}\D_{m}(\symb{J}) &=& -\symb{J} \wedge \D_{m}\D_{n}(\symb{I}) 
\ea 
\eeq 
In order to examine the $\D(F_{RR})$ expression further it is convenient to use the projection-like properties of $\i_{\syma{I}}$ and $\i_{\symb{J}}$ because they project out a $0$-form rather than a $6$-form.
\beq
\label{eqn:IIAtadpoles}
\ba{ccrcccr}
\syma{I} \wedge \D_{n}\D_{m}(\syma{J}) &=& -\syma{J} \wedge \D_{m}\D_{n}(\syma{I}) &\quad\to\quad& \i_{\symb{I}}\D_{n}\D_{m}(\syma{J}) &=& -\i_{\symb{J}}\D_{m}\D_{n}(\syma{I}) \\
\syma{I} \wedge \D_{n}\D_{m}(\symb{J}) &=& -\symb{J} \wedge \D_{m}\D_{n}(\syma{I}) &\quad\to\quad& \i_{\symb{I}}\D_{n}\D_{m}(\symb{J}) &=& +\i_{\syma{J}}\D_{m}\D_{n}(\syma{I}) \\ 
\symb{I} \wedge \D_{n}\D_{m}(\symb{J}) &=& -\symb{J} \wedge \D_{m}\D_{n}(\symb{I}) &\quad\to\quad& \i_{\syma{I}}\D_{n}\D_{m}(\symb{J}) &=& -\i_{\syma{J}}\D_{m}\D_{n}(\symb{I}) 
\ea 
\eeq 
Using these and the fact that for $3$-forms $1 = \syma{I}\i_{\syma{I}} + \symb{J}\i_{\symb{J}}$ we can reexpress the $F_{RR}$ tadpole in a new way.
\beq
\D(F_{RR}) = \D\Dp(\syma{0}) = \syma{I}\i_{\syma{I}}\D\Dp(\syma{0}) + \symb{J}\i_{\symb{J}}\D\Dp(\syma{0}) = -\syma{I}\i_{\syma{0}}\Dp\D(\syma{I}) + \symb{J}\i_{\syma{0}}\Dp\D(\symb{J})
\eeq
Though we are not explicitly considering S duality transformations yet it follows from the dilaton inversion that $\D$ and $\Dp$ should be treated in the same manner, though not the flux multiplets which define their components. With the fluxes of $\D\Dp(\syma{0})$ also contributing to particular $\Dp\D$ images of all $\Hn{3}$ basis forms it follows that all coefficients of the $\Hn{3}$ basis elements in (\ref{eqn:TadpoleT2}) are tadpoles, not just those corresponding to $\D\Dp(\syma{0})$. The physical interpretation of these tadpoles is not clear as the fluxes which they are defined in terms of are induced by the inclusion of $\SL_{S}$ symmetry to the Type IIA superpotential. Furthermore, we have not considered a fully consistent picture of these tadpoles since we are assuming the non-zero nature of fluxes which are not required for only T duality invariance; under modular transformations of the dilaton we would expect the tadpoles to form $\SL_{S}$ multiplets, mixing $\D\Dp$ and $\Dp\D$ terms. We defer such an analysis for the time being and instead consider the Type IIB tadpoles.
\\

The Type IIB tadpoles are those flux combinations which couple to the $C_{2n}$ potentials living on the D-branes of Type IIB. Following the same principle as the Type IIA case possible contributions to such tadpoles can be found by considering how a $10-2n$ form can be contructed to couple to $C_{2n}$. $C_{4}$ has a tadpole contribution $H_{3} \wedge F_{3} \to \sFh_{0} \wedge \sF_{0}$ which we immediately note can be written in a number of ways.
\beq
\sFh_{0} \wedge \sF_{0} \quad=\quad \bigg\{ \quad \ba{rcr}\sD(\sF_{0}) &\quad=\quad& \sD\sDp(\v{0}) \\ -\sDp(\sFh_{0}) &\quad=\quad& -\sDp\sD(\v{0}) \ea \quad \bigg\} \quad =\quad \frac{1}{2}(\sD\sDp-\sDp\sD)(\v{0})
\eeq
The flux component version of this tadpole follows from the known actions of $\sD$ and $\sDp$.
\beq
\label{eqn:TadpoleT}
\ba{ccccccccccccccc}
\sD\sDp(\v{I}) &=& \Big( & \sFh_{(I)A}\sF^{(J)A} &-& \sFh_{(I)}^{\phantom{(I)}B}\sF_{\phantom{(J)}B}^{(J)} & \Big)\v{J} &+& \Big( & \sFh_{(I)}^{\phantom{(I)}B}\sF_{(J)B} &-& \sFh_{(I)A}\sF^{\phantom{(J)}A}_{(J)} & \Big)\vt{J}\\
\sD\sDp(\vt{J}) &=& \Big( & \sFh_{\phantom{(J)}A}^{(J)}\sF^{(I)A} &-& \sFh^{(J)B}\sF_{\phantom{(I)}B}^{(I)} & \Big)\v{I} &+& \Big( & \sFh^{(J)B}\sF_{(I)B} &-& \sFh_{\phantom{(J)}A}^{(J)}\sF_{(I)}^{\phantom{(I)}A} & \Big)\vt{I}
\ea 
\eeq 
The $C_{4}$ tadpole's flux polynomial expression is the $\vt{0}$ component of $\sD\sDp(\v{I})$, which can be projected out by $\v{0}\wedge$ or $\i_{\vt{0}}$.
\beq
C_{4} \;:\; \sFh_{(0)}^{\phantom{(0)}B}\sF_{(0)B} &-& \sFh_{(0)A}\sF^{\phantom{(0)}A}_{(0)}
\eeq
The $C_{6}$ case requires a $4$-form, which is $\sF_{1}\cdot \sF_{0}$ for the T duality only case. Since the fluxes of $\sF_{1}$ couple to $\cT^{j} = \sT_{i}$ the $\sF^{\phantom{(i)}B}_{(i)}$ and $\sF_{(i)A}$ are the components of $\sF_{1}$ and so the corresponding terms in (\ref{eqn:TadpoleT}) are seen to be the coefficients of $\vt{j}$ in $\sD\sDp(\v{0})$.
\beq
C_{6} \;:\; \sFh_{(0)}^{\phantom{(0)}B}\sF_{(j)B} &-& \sFh_{(0)A}\sF^{\phantom{(j)}A}_{(j)}
\eeq
Due to the different way in which $\sF_{1}$ couples in a tadpole compared too a Bianchi constraint these terms are not the coefficients of a $4$-form, but a $2$-form. However, the physical interpretation can still be viewed in terms of D$5$-branes wrapping cycles; the D$5$ wraps a $2$-cycle $\A_{i}$ and the associated tadpole is obtained by integrating $\sD\sDp(\v{0})$ over its dual $4$-cycle $\B^{i}$. The $C_{8}$ and $C_{10}$ follow in the same manner, with $C_{8}$ tadpoles being defined on the dual cycles to the wrapped D$7$. 
\\

As in the Type IIA case these expressions only account for a small number of the possible flux polynomials of (\ref{eqn:TadpoleT}). Each of the $C_{2n}$ we have considered has had tadpoles of the form $\sD(\sF_{0})$ but the inclusion of other R-R fluxes allows for other possibilities. They do not naturally form $p$-forms alone, such as $\sF_{2}\cdot \sFh_{2}\cdot $ not being a $p$-form unless it acts on an element of $\Ln{p\geq 2}$. Such a flux combination appears in (\ref{eqn:TadpoleT}), formed by acting $\sD\sDp$ on $\vt{i}$. The physical interpretation of these fluxes and the objects they live on we shall not address, instead we restrict our discussion to how these flux polynomials are constructed and transform under the dualities we are considering. In that regard by constructing the $\sDp\sD$ expressions and comparing with those of $\sD\sDp$ we obtain a set of identities which are the Type IIB version of (\ref{eqn:IIAtadpoles}).
\beq
\label{eqn:Tadpole T again}
\ba{ccccccccccccccc}
\sDp\sD(\v{I}) &=& \Big( & \sF_{(I)A}\sFh^{(J)A} &-& \sF_{(I)}^{\phantom{(I)}B}\sFh_{\phantom{(J)}B}^{(J)} & \Big)\v{J} &+& \Big( & \sF_{(I)}^{\phantom{(I)}B}\sFh_{(J)B} &-& \sF_{(I)A}\sFh^{\phantom{(J)}A}_{(J)} & \Big)\vt{J}\\
\sDp\sD(\vt{J}) &=& \Big( & \sF_{\phantom{(J)}A}^{(J)}\sFh^{(I)A} &-& \sF^{(J)B}\sFh_{\phantom{(I)}B}^{(I)} & \Big)\v{I} &+& \Big( & \sF^{(J)B}\sFh_{(I)B} &-& \sF_{\phantom{(J)}A}^{(J)}\sFh_{(I)}^{\phantom{(I)}A} & \Big)\vt{I}
\ea 
\eeq 
Relabelling the derivatives as $\sD = \sD_{1}$ and $\sDp = \sD_{2}$ we can generalise the results to consider the $\sD_{i}\sD_{j}$ actions on all elements of $\Hnt{3}$.
\beq
\label{eqn:IIAtadpoles}
\ba{ccrcccr}
\v{I} \wedge \sD_{n}\sD_{m}(\v{J}) &=& -\v{J} \wedge \sD_{m}\sD_{n}(\v{I}) &\quad\to\quad& \i_{\vt{I}}\sD_{n}\sD_{m}(\v{J}) &=& -\i_{\vt{J}}\sD_{m}\sD_{n}(\v{I}) \\
\v{I} \wedge \sD_{n}\sD_{m}(\vt{J}) &=& +\vt{J} \wedge \sD_{m}\sD_{n}(\v{I}) &\quad\to\quad& \i_{\vt{I}}\sD_{n}\sD_{m}(\vt{J}) &=& +\i_{\v{J}}\sD_{m}\sD_{n}(\v{I}) \\ 
\vt{I} \wedge \sD_{n}\sD_{m}(\vt{J}) &=& -\vt{J} \wedge \sD_{m}\sD_{n}(\vt{I}) &\quad\to\quad& \i_{\v{I}}\sD_{n}\sD_{m}(\vt{J}) &=& -\i_{\v{J}}\sD_{m}\sD_{n}(\vt{I}) 
\ea 
\eeq 

\subsection{S duality}
\label{sec:Sduality}

Type IIB supergravity is known to have a self dual $\textrm{SL}(2,\mathbb{R})$ symmetry at the ten dimensional level which is broken by compactification down to the quantised $\SL$ subgroup but this is not shared by Type IIA, as we have seen in its asymmetric treatment of the two flux sectors. However, we are able to do much of our analysis of S duality in the Type IIB side and then convert the results, using the relationships between the IIA and IIB fluxes summarised in (\ref{table:fluxmatrices}), into the equivalent Type IIA results. As a result, unless otherwise stated this section is entirely within Type IIB. We shall denote a general $\SL_{S}$ transformation on $S$ by $\Gamma_{S}$, which dependent on the context we will take to be either a matrix on some flux doublet or a M\"{o}bius transformation on $S$ itself.
\beq
\label{eqn:ModularS}
S \to S^{\prime} = \Gamma_{S}(S) \equiv \frac{aS+b}{cS+d} \quad \Leftrightarrow \quad \Gamma_{S} \equiv \bpm a & b \\ c & d \epm 
\eeq
$\SL$ has two generators, which we choose to be those related to the moduli transformations $\Gamma_{1} : S \to -\frac{1}{S}$ and $\Gamma_{2} : S \to S+1$.
\beq
\Gamma_{1} = \bpm 0 & -1 \\ 1 & 0\epm \quad , \quad \Gamma_{2} = \bpm 1 & 1 \\ 0 & 1\epm \nn
\eeq

\subsubsection{Flux matrix transformations}

To examine the flux structures induced by S duality we define a pair of matrices, $\uwave{\F}$ and $\uwave{\Fh}$, in terms of matrices $\A_{n}$ and $\B_{m}$ previously defined in relation to complexification matrices.
\beq
\label{eqn:FFhMN}
\uwave{\F} = \A \cdot \DevenB + \B \cdot \DpevenB\qquad \uwave{\Fh} = \A \cdot \DpevenB+ \B \cdot \DevenB 
\eeq
These definitions can be written in terms of flux matrix doublets.
\beq
\label{eqn:FtoM}
\bpm \uwave{\F} \\ \uwave{\Fh}\epm = \bpm \A & \B \\ \B & \A \epm \bpm \DevenB \\ \DpevenB\epm \quad \Rightarrow \quad \bpm \DevenB \\ \DpevenB\epm = \bpm \A & \B \\ \B & \A \epm\bpm \uwave{\F} \\ \uwave{\Fh} \epm
\eeq
The superpotential can then be rewritten in a way such that its dilaton dependence is manifest.
\beq
W = \u{\fT}^{\top} \cdot \uwave{\mho} \cdot \Big( \bC \cdot \DevenB + \bCp \cdot \DpevenB \Big) \cdot g \cdot \uwave{\Omega} \cdot \u{\fU} = \u{\fT}^{\top} \cdot \uwave{\mho} \cdot \Big( \uwave{\F} - S \uwave{\Fh} \Big) \cdot g \cdot \uwave{\Omega} \cdot \u{\fU}  \nn
\eeq
If the superpotential is to be invariant then $\uwave{\F} - S \uwave{\Fh}$ must be invariant, up to a gauge choice, and as with the individual flux multiplets $\F_{n}$ and $\Fh_{n}$ these two matrices transform as an $\SL_{S}$ doublet.
\beq
\label{eqn:SdoubletF}
S \to \frac{aS+b}{cS+d} \quad \Rightarrow \quad \bpm \uwave{\F} \\ \uwave{\Fh} \epm \to \bpm a & b \\ c & d\epm \bpm \uwave{\F} \\ \uwave{\Fh} \epm \nn
\eeq
The $\SL_{S}$ transformation properties of the $\D$ flux matrices follow from this.
\beq
\label{eqn:MunderS}
\bpm \DevenB \\ \DpevenB\epm &\to& \bpm \A & \B \\ \B & \A \epm\bpm a & b \\ c & d \epm\bpm \A & \B \\ \B & \A \epm\bpm \DevenB \\ \DpevenB\epm  \\
&=& \left[ \bpm a & b \\ c & d \epm \otimes \A + \bpm d & c \\ b & a \epm\otimes \B \right] \bpm \DevenB \\ \DpevenB\epm\nn \\
\label{eqn:Seffect1}
&=& \Big( \Gamma_{S}\otimes \A + \big(\sigma \cdot \Gamma_{S} \cdot \sigma\big)\otimes \B \Big)\bpm \DevenB \\ \DpevenB\epm \nn 
\eeq
If $\DevenB$ transforms as $\DevenB \to m\cdot \DevenB$, where $m$ is a matrix that commutes with both $\sg_{\Hbasis}$ and $\sg_{\HtbasisB}$ then the corresponding transformation on $\DoddB$ is $\DoddB \to \DoddB \cdot m^{\top}$. Both $\A$ and $\B$ satisfy this and are also symmetric, allowing us to express the $\SL_{S}$ transformations on each flux matrix in similar ways.
\beq
\bpm \DoddB & \DpoddB\epm &\to& \bpm \DoddB & \DpoddB\epm\Big( \Gamma_{S}^{\top}\otimes \A + \big(\sigma \cdot \Gamma_{S}^{\top} \cdot \sigma\big)\otimes \B \Big) \nn \\
\label{eqn:Seffect2}
&=& \bpm \DoddB & \DpoddB\epm \bpm \A & \B \\ \B & \A \epm\bpm a & b \\ c & d \epm^{\top}\bpm \A & \B \\ \B & \A \epm
\eeq
Thus far we have only considered the constraints which arise from the nilpotency of the two flux sectors seperately, as without S duality the two sectors can be viewed as disjoint. Under S duality transformations the flux matrices of the derivatives mix and thus the four quadratic derivative expressions used to define Bianchi and tadpole constraints also mix.

\subsubsection{Flux matrix representation}

The analysis of the two Type II theories is much the same when working on the level of the flux matrices, as opposed to working with the flux multiplets, and so we shall restrict our discussion to the Type IIB case, $\sD$ and $\sDp$.
\beq
\ba{ccccccc}
\la \sD^{2} \ra &=& \la \DevenB \cdot \sh_{\Hbasis} \cdot \DoddB ,\DoddB \cdot \sh_{\HtbasisB} \cdot \DevenB  \ra &\qquad& \la \sD\sDp \ra &=& \la \DpevenB \cdot \sh_{\Hbasis} \cdot \DoddB ,\DpoddB \cdot \sh_{\HtbasisB} \cdot \DevenB \ra \\ 
\la \sDp^{2}\ra &=& \la \DpevenB \cdot \sh_{\Hbasis} \cdot \DpoddB ,\DpoddB \cdot \sh_{\HtbasisB} \cdot \DpevenB \ra &\qquad& \la \sDp\sD \ra &=& \la \DevenB \cdot \sh_{\Hbasis} \cdot \DpoddB ,\DoddB \cdot \sh_{\HtbasisB} \cdot \DpevenB \ra
\ea
\eeq
It is noteworthy that due to their linear independence, relationship $\A+\B = \mathbb{I}$ and projection-like multiplicative action we can use $\A$ and $\B$ to decompose any matrix into four disjoint submatrices.
\begin{eqnarray}
\label{eqn:LinDep}
X = (\A+\B)\cdot X \cdot (\A+\B) = \A \cdot X \cdot \A + \A \cdot X \cdot \B + \B \cdot X \cdot \A + \B \cdot X \cdot \B 
\end{eqnarray}

We first consider quadratic derivative actions of the same schematic form as $\D^{2} : \Hnt{3} \to \Hnt{3}$ and as in previous sections we are using the $\ZZ$ orientifold as an explicit example for our analysis. Previous work \cite{Aldazabal:2006up,Guarino:2008ik} on the behaviour of the fluxes in this space under S duality has discussed combinations of geometric and non-geometric fluxes, in both sectors, which transform as either triplets or singlets of $\SL_{S}$. An example of a triplet is seen in the case of the non-geometric fluxes, $Q^{ab}_{c} \sim Q$ and $P^{ab}_{c} \sim P$, by considering the Bianchi constraints due to T duality under $\SL_{S}$. Under T duality only, the NS-NS non-geometric flux satisfies the Lie algebra inspired Jacobi constraint $Q\cdot Q=0$ and the corresponding R-R flux $P$ satisfies the same kind of constraint $P \cdot P = 0$. Separately these expressions form ideals $\la Q\cdot Q\ra$ and $\la P\cdot P\ra$ but since these two fluxes transform as an $\SL_{S}$ doublet these ideals are not closed under S duality. Under the inversion $S \to -\frac{1}{S}$ the generating functions of these two ideals are exchanged and therefore under S duality they must belong to the same ideal, along with a third flux combination $Q\cdot P+P\cdot Q$ in order to make the ideal closed under $\SL_{S}$.
\beq
\label{eqn:QPtriplet}
\bpm Q \\ P \epm \to \bpm a & b \\ c & d \epm\bpm Q \\ P \epm = \bpm aQ+bP \\ cQ+dP \epm \quad \Rightarrow \quad \ba{c} \la Q\cdot Q\ra \\ \la P \cdot P \ra \ea \to \left\la \ba{ccc} Q \cdot Q &,& P\cdot P \\ Q\cdot P&+&P\cdot Q \ea \right\ra  
\eeq
To begin we consider the flux matrix expressions associated to $Q\cdot Q \sim \sF_{2} \cdot \sF_{2}$; the NS-NS sector constraints $\DevenB \cdot \sh_{\Hbasis} \cdot \DoddB$.
\beq
\label{eqn:MN1}
\ba{ccc} 
\DevenB \cdot \sh_{\Hbasis} \cdot \DoddB &=&  \bpm \DevenB & \DpevenB\epm  \bpm \mathbb{I} & 0 \\ 0 & 0\epm \bpm \sh_{\Hbasis} & 0 \\ 0 & \sh_{\Hbasis}\epm \bpm \DoddB \\ \DpoddB\epm
\ea 
\eeq
This scalar product notation is not required to consider the $\SL_{S}$ image of $\DevenB \cdot h_{\Hbasis} \cdot \DoddB$ but it is convenient for more general expressions encountered later. The $\SL_{S}$ image we arrange in accordance with the $\A$, $\B$ inspired decomposition of (\ref{eqn:LinDep}) and for less cluttered notation use $\hdot \equiv \cdot \sh_{\Hbasis} \cdot$.
\beq 
\label{eqn:Sduality1}
\ba{ccccc}
\DevenB \cdot \sh_{\Hbasis} \cdot \DoddB &\to& \left(
\ba{crcl}
& \A \cdot ( a\DevenB+b\DpevenB ) & \hdot & ( a\DoddB+b\DpoddB ) \cdot \A \\
+&\A \cdot ( a\DevenB+b\DpevenB ) & \hdot & ( d\DoddB+c\DpoddB ) \cdot \B \\
+&\B \cdot ( d\DevenB+c\DpevenB ) & \hdot & ( a\DoddB+b\DpoddB ) \cdot \A \\
+&\B \cdot ( d\DevenB+c\DpevenB ) & \hdot & ( d\DoddB+c\DpoddB ) \cdot \B 
\ea \right)
\ea
\eeq
The corresponding $\SL_{S}$ image of $\DpevenB \hdot \DpoddB$, $\DevenB \hdot \DpoddB$ and $\DpevenB \hdot \DoddB$ is given in the Appendix as (\ref{eqn:Sduality1b}-\ref{eqn:Sduality1d}). Two transformations of particular note are $\Gamma_{S} = \mathbb{I}_{2}$ and $\Gamma_{S}=\Gamma_{1}$. The former implies that the NS-NS derivative is still nilpotent and the latter implies that the R-R derivative is also nilpotent.
\beq 
\Gamma_{1}(\DevenB \hdot \DoddB) &=& (\A-\B) \cdot \DpevenB \hdot \DpoddB \cdot (\A-\B) \nn
\eeq
Though $\Gamma_{1}$ has not mapped the NS-NS constraint matrix exactly into the R-R version due to the change of sign on $\B$ the linear independence of the submatrices make this irrelevant and with these two conditions (\ref{eqn:Sduality1}) can be reduced down to only terms which mix the two sectors.
\beq 
\label{eqn:Sduality2}
\ba{ccccc}
\DevenB \cdot h_{\Hbasis} \cdot \DoddB &\to& \left(
\ba{cccccc}
 & \A \cdot \Big( & ab\,\DevenB \hdot \DpoddB &+& ab\, \DpevenB \hdot \DoddB & \Big) \cdot \A \\
+& \A \cdot \Big( & ac\,\DevenB \hdot \DpoddB &+& bd\, \DpevenB \hdot \DoddB & \Big) \cdot \B \\
+& \B \cdot \Big( & bd\,\DevenB \hdot \DpoddB &+& ac\, \DpevenB \hdot \DoddB & \Big) \cdot \A \\
+& \B \cdot \Big( & cd\,\DevenB \hdot \DpoddB &+& cd\, \DpevenB \hdot \DoddB & \Big) \cdot \B
\ea \right)
\ea
\eeq
Comparing this expression with the similarly reduced form of (\ref{eqn:Sduality1b}) we note that in each case the $\SL_{S}$ integers factorise out as overall factors in both the $\A\cdot X \cdot \A$ and $\B \cdot X \cdot \B$ terms. Using this pre- and post-multiplication by $\A$ or $\B$ we can project out particular parts of (\ref{eqn:Sduality1}) to form $\SL_{S}$ multiplets.
\beq
\label{eqn:Triplet1}
\ba{cccccccccccccc}
\mathbf{3}_{\A\A} & \equiv \la & \A \cdot \DevenB \hdot \DoddB \cdot \A & \;,\; & \A \cdot \DpevenB \hdot \DpoddB \cdot \A & \;,\; & \A \cdot ( \DpevenB  \hdot \DoddB + \DevenB \hdot \DpoddB) \cdot \A & \ra \\
\mathbf{3}_{\B\B} & \equiv \la & \B \cdot \DevenB \hdot \DoddB \cdot \B & \;,\; & \B \cdot \DpevenB \hdot \DpoddB \cdot \B & \;,\; & \B \cdot ( \DpevenB  \hdot \DoddB + \DevenB \hdot \DpoddB) \cdot \B & \ra 
\ea \quad
\eeq
Schematically the action of $\A$ is to project out the components associated with $\sF_{1}$ and $\sF_{2}$, while $\B$ does the opposite, projecting out the $\sFh_{0}$ and $\sFh_{3}$ fluxes. As such we would expect the triplet of (\ref{eqn:QPtriplet}) to be associated to a triplet of the form $\mathbf{3}_{\A\A}$. The components of (\ref{eqn:Sduality1}) yet to be put into a multiplet are the $\B\cdot X \cdot \A$ and $\B \cdot X \cdot \A$ terms and by considering the $\SL_{S}$ integers for these parts we can construct another pair of triplets associated to these components.
\beq
\label{eqn:Triplet2}
\ba{cccccccccccccc}
\mathbf{3}_{\A\B} & \equiv \la& \A \cdot \DpevenB \hdot \DoddB \cdot \B & \;,\; & \A \cdot \DevenB \hdot \DpoddB \cdot \B & \;,\; & \A \cdot ( \DevenB \hdot \DoddB + \DpevenB \hdot \DpoddB) \cdot \B & \ra \\
\mathbf{3}_{\B\A} & \equiv \la& \B \cdot \DpevenB \hdot \DoddB \cdot \A & \;,\; & \B \cdot \DevenB \hdot \DpoddB \cdot \A & \;,\; & \B \cdot ( \DevenB \hdot \DoddB + \DpevenB \hdot \DpoddB) \cdot \A & \ra 
\ea 
\eeq 
This is not sufficient for full $\SL_{S}$ invariance, for each of the terms in (\ref{eqn:LinDep}) we have constructed our triplets from the components of four terms, $\DevenB \hdot \DoddB$, $\DpevenB \hdot \DoddB$, $\DevenB \hdot \DpoddB$, $\DpevenB \hdot \DpoddB$. We have seen how a linear combination of these terms makes a triplet and so we would expect it to be possible to build a singlet to go along with each of these triplets. An explicit example in the form of an $\SL_{S}$ singlet of the $\ZZ$ orientifold discussed in \cite{Aldazabal:2006up,Guarino:2008ik} has the schematic form $Q\cdot H - P\cdot F$. This we would associate it with the terms of the form $\A \cdot X \cdot \B$. By considering the triplets of (\ref{eqn:Triplet1}) and (\ref{eqn:Triplet2}) and using (\ref{eqn:Sduality1a}-\ref{eqn:Sduality1d}) we can straightforwardly construct the four singlets associated to the terms of (\ref{eqn:LinDep}).
\beq
\label{eqn:Singlet1}
\ba{cccccccccccccc}
\mathbf{1}_{\A\A} & \equiv \la& \A \cdot ( \DevenB \hdot \DpoddB - \DpevenB \hdot \DoddB ) \cdot \A & \ra \\
\mathbf{1}_{\B\B} & \equiv \la& \B \cdot ( \DevenB \hdot \DpoddB - \DpevenB \hdot \DoddB ) \cdot \B & \ra \\
\mathbf{1}_{\A\B} & \equiv \la& \A \cdot ( \DevenB \hdot \DoddB - \DpevenB \hdot \DpoddB ) \cdot \B & \ra \\
\mathbf{1}_{\B\A} & \equiv \la& \B \cdot ( \DevenB \hdot \DoddB - \DpevenB \hdot \DpoddB ) \cdot \A & \ra 
\ea 
\eeq
Given the four possible combinations of the flux matrices and the four terms arising from the $\A$, $\B$ induced decomposition the multiplets of (\ref{eqn:Triplet1}-\ref{eqn:Singlet1}) make up all possible $\SL_{S}$ multiplets. However, not all of these expressions are independent nor are they all Bianchi constraints. Only those expressions which are an $\SL_{S}$ image of a T duality Bianchi constraint are S duality constraints and not all expressions defining the multiplets of (\ref{eqn:Triplet1}-\ref{eqn:Singlet1}) are of this form. To use a more explicit example, the $\ZZ$ triplet of (\ref{eqn:QPtriplet}) is formed from particular linear combinations of the four terms $Q\cdot Q$, $P\cdot Q$, $Q\cdot P$ and $P\cdot P$ and so there exists an $\SL_{S}$ singlet which is the complement of this triplet and which does not arise as a Bianchi constraint, namely $Q\cdot P-P\cdot Q$. In the $(\Ha,\Ha\dual)$ basis this would belong to the singlet $\mathbf{1}_{\A\A}$. The triplet can be regarded as being the $\SL_{S}$ image of the term $Q \cdot Q$, a pairing of two NS-NS fluxes and only pairwise combinations of two NS-NS fluxes will give T duality constraints if the R-R sector is turned off. In the flux matrix definition of $\mathbf{1}_{\A\A}$ there is no term which can be decomposed into a pair of NS-NS flux terms, all of the terms are a mixture of NS-NS and R-R fluxes. Therefore $\mathbf{1}_{\A\A}$ does not arise as an $\SL_{S}$ image of a T duality Bianchi constraint and we have no apriori reason to expect it to be zero. The case of the singlet $Q\cdot H - P\cdot F \in \mathbf{1}_{\A\B}$ does not follow in precisely the same manner, as $\mathbf{3}_{\A\B}$ has a term with $\DevenB \hdot \DoddB$ in it also. In this case we note that $QF \in\A \cdot \DevenB \hdot \DpoddB \cdot \B$, a known tadpole condition in the T duality only case and thus $\mathbf{3}_{\A\B}$ cannot be set to zero. Also following on from our previous discussion of the tadpole expressions we can deduce that $\mathbf{n}_{\A\B} \cong \mathbf{n}_{\B\A}$ and so we can split these multiplets into two different categories; those whose generating functions are set to zero by Bianchi contraints and those whose generating functions are equal to integers which define the number and type of charged extended objects in the space.
\beq
\label{eqn:BianchiTadpole}
\ba{ccc|ccc}\hline\hline
\quad& \textrm{Bianchi} &\quad&\quad& \textrm{Tadpole} &\quad \\\hline\hline
& \mathbf{3}_{\A\A} &&& \mathbf{1}_{\A\A} \\
& \mathbf{3}_{\B\B} &&& \mathbf{1}_{\B\B} \\
& \mathbf{1}_{\A\B}=\mathbf{1}_{\B\A} &&& \mathbf{3}_{\B\A}=\mathbf{3}_{\A\B} \\\hline\hline
\ea
\eeq 
We now turn our attention to the action of the derivatives of general form $\sD^{2} : \Hnt{3} \to \Hnt{3}$ where unlike the quadratic action on $\Hn{3}$ the $\A$ and $\B$ are `internal' to the flux matrix expressions, rather than projecting out linearly independent sections of the constraints. A result of this is that the induced transformations are not as straightforward and as in the $\DevenB \cdot \sh_{\Hbasis} \cdot \DoddB$ case we find it convenient to express the flux matrix nilpotency expressions as scalar products.
\beq
\label{eqn:NM1}
\ba{ccc} 
\DoddB \cdot h_{\HtbasisB} \cdot \DevenB &=& \bpm \DoddB & \DpoddB \epm \bpm \mathbb{I} & 0 \\ 0 & 0\epm\bpm \sh_{\HtbasisB} & 0 \\ 0 & h_{\HtbasisB}\epm\bpm \DevenB \\ \DpevenB \epm  \\
\DpoddB \cdot h_{\HtbasisB} \cdot \DpevenB &=& \bpm \DoddB & \DpoddB \epm \bpm 0 & 0 \\ 0 & \mathbb{I}\epm\bpm \sh_{\HtbasisB} & 0 \\ 0 & h_{\HtbasisB}\epm\bpm \DevenB \\ \DpevenB \epm
\ea 
\eeq
In order to simplify our expressions we note that $h_{\HtbasisB} = \mathbb{I}$ and so no longer explictly write it and in what follows factors of $\mathbb{I}$ are surpressed. For a general combination of pairs of flux matrices we can define a related quadratic form, $\cX$, and we will consider how this transforms under $\SL_{S}$.
\beq
p\,\DoddB \cdot \DevenB + q\,\DoddB \cdot \DpevenB + r\,\DpoddB \cdot \DevenB + s\,\DpoddB \cdot \DpevenB \equiv \bpm \DoddB & \DpoddB\epm \cdot\cX\cdot  \bpm \DevenB \\ \DpevenB\epm
\eeq 
We can construct the transformation properties of $\cX$ by using (\ref{eqn:Seffect1}) and (\ref{eqn:Seffect2}) and since the entries are scalar multiples of $\mathbb{I}$ the $\A$ and $\B$ terms decouple.
\beq
\label{eqn:Simage}
\cX \equiv \bpm p & q \\ r & s\epm&\to& \Big( \Gamma_{S}^{\top}\otimes \A + \big(\sigma \cdot \Gamma_{S}^{\top} \cdot \sigma\big)\otimes \B \Big)\bpm p & q \\ r & s \epm\Big( \Gamma_{S}\otimes \A + \big(\sigma \cdot \Gamma_{S} \cdot \sigma\big)\otimes \B \Big) \nn \\
&=& \underbrace{\left(\Gamma_{S}^{\top}\bpm p & q \\ r & s \epm\Gamma_{S}\right)}_{\Xi_{\A}}\otimes \A + \underbrace{\left(\big(\sigma \cdot \Gamma_{S}^{\top} \cdot \sigma\big)\bpm p & q \\ r & s \epm\big(\sigma \cdot \Gamma_{S} \cdot \sigma\big)\right)}_{\Xi_{\B}}\otimes \B \qquad
\eeq
Proceeding as before we wish to obtain an $\SL_{S}$ triplet by considering the image of the T duality constraints and also combination of terms which form a singlet. Due to the splitting of $\cX$ by $\A$ and $\B$ the equations which are satisfied by a singlet reduce to $\Xi_{\A} = \cX = \Xi_{\B}$, so by using the fact that any element of $\textrm{SL}(2,\mathbb{R})$ is a symplectic matrix and both $\Xi_{\A}$ and $\Xi_{B}$ are of the form $m^{\top}\cdot \cX \cdot m$ it follows that if $\cX$ is the canonical symplectic form then the equations $\Xi_{\A} = \cX = \Xi_{\B}$ are automatically satisfied for any $\SL_{S}$ transformation and we obtain a singlet.
\beq
\Gamma_{S}\Big( \DoddB \cdot \DpevenB - \DpoddB \cdot\DevenB \Big) = \DoddB \cdot \DpevenB - \DpoddB \cdot \DevenB \nn 
\eeq
This can be taken a step further by noting that due to the linear independence of $\A$ and $\B$ $\cX$ can be written as two independent terms, which transform separately.
\beq
\cX = \bpm p & q \\ r & s\epm= \bpm a_{1} & a_{2} \\ a_{3} & a_{4} \epm\otimes \A + \bpm b_{1} & b_{2} \\ b_{3} & b_{4} \epm\otimes \B \equiv \cX_{\A} \otimes \A + \cX_{\B} \otimes \B
\eeq
With the decomposition of the $\SL_{S}$ image of $\cX$ in (\ref{eqn:Simage}) we have that $\Xi_{\A}$ depends on the $a_{i}$ only and $\Xi_{\B}$ depends on the $b_{j}$ only. Therefore the necessary and sufficient conditions for a singlet become the pair of conditions $\Xi_{\A} = \cX_{\A}$, $\Xi_{\B} = \cX_{\B}$ and we can construct two seperate non-trivial singlets by setting one of $\cX_{\A}$ or $\cX_{\B}$ to zero and the other to the canonical sympletic form. 
\beq
\label{eqn:SingletAB}
\mathbf{1}_{\A} \equiv \la \, \DoddB \cdot \A \cdot \DpevenB - \DpoddB \cdot \A \cdot \DevenB \, \ra \quad , \quad \mathbf{1}_{\B} \equiv \la \,\DoddB \cdot \B \cdot \DpevenB - \DpoddB \cdot \B \cdot \DevenB \, \ra
\eeq
For the triplet we begin with the known NS-NS sector T duality constraint $\DoddB \cdot \DevenB=0$ and consider its images under particular elements of $\SL_{S}$, which in the case of $\Gamma = \mathbb{I}_{2}$ and $\Gamma = \Gamma_{1}$ we obtain the T duality constraints of both the NS-NS sector and the R-R sector.
\beq
\label{eqn:Strans1}
\ba{ccccccl}
\Gamma_{S} &=& \bpm 0 & -1 \\ 1 & 0\epm & \quad:\quad & \bpm p & 0 \\ 0 & q\epm&\to& \bpm q & 0 \\ 0 & p\epm
\ea
\eeq
Therefore we have that given S duality $\DoddB \cdot \DevenB=0$ implies $\DpoddB \cdot \DpevenB=0$. The other generator of $\SL_{S}$ on a general linear combination of these two  terms leads to different transformations in the $\A$ and $\B$ terms.
\beq
\label{eqn:Strans2}
\ba{ccccccl}
\Gamma_{S} &=& \bpm 1 & 1 \\ 0 & 1\epm & \quad:\quad & \bpm p & 0 \\ 0 & q\epm&\to& \bpm p & 0 \\ 0 & q\epm+ p\, \bpm 0 & \A \\ \A & \A\epm+ q\, \bpm \B & \B \\ \B & 0\epm\\
\ea
\eeq
Although setting the expressions associated with the second and third terms in the above expression is a necessary condition for joint T and S duality invariance, it is not sufficient. This can be seen by considering another $\SL_{S}$ transformation, those which is associated to the negative integer shift, $S \to S-1$. 
\beq
\label{eqn:Strans3}
\ba{ccccccl}
\Gamma_{S} &=& \bpm 1 & -1 \\ 0 & \phantom{-}1\epm & \;:\; & \bpm p & 0 \\ 0 & q\epm&\to& \bpm p & 0 \\ 0 & q\epm+ p\, \bpm \phantom{-}0 & -\A \\ -\A & \phantom{-}\A\epm+ q\, \bpm \phantom{-}\B & -\B \\ -\B & \phantom{-}0\epm
\ea
\eeq
Using (\ref{eqn:Strans1}) the requirement that both (\ref{eqn:Strans2}) and (\ref{eqn:Strans3}) are also zero leads to stronger constraints, $\DoddB \cdot \DevenB=0$ is true by virtue of the $\mathbb{I} = \A+\B$ decomposition terms both vanishing seperately. Apriori we could not assume that the $\A$ and $\B$ related terms form two separate, independent, systems but given the singlet structures we would expect the triplets to follow with the same splittings.
\beq
\label{eqn:TripletAB}
\ba{cccccccccccccccccc}
\mathbf{3}_{\A} & \equiv \la& \DoddB \cdot \A \cdot \DevenB & \;,\; & \DpoddB \cdot \A \cdot \DpevenB & \;,\; & \DoddB \cdot \A \cdot \DpevenB &+& \DpoddB \cdot \A \cdot \DevenB & \ra\\
\mathbf{3}_{\B} & \equiv \la& \DoddB \cdot \B \cdot \DevenB & \;,\; & \DpoddB \cdot \B \cdot \DpevenB & \;,\; & \DoddB \cdot \B \cdot \DpevenB &+& \DpoddB \cdot \B \cdot \DevenB & \ra
\ea 
\eeq
As with the quadratic derivative actions of the form $\sD^{2} : \Hnt{3} \to \Hnt{3}$ it is possible that not all of these expressions automatically give Bianchi constraints. However, unlike the previous case we cannot express these flux matrix expressions in terms of the natural Type IIB flux multiplets since they are not defined by derivative actions on $\Hn{3}$. The $\mathbf{3}$ triplets contain expressions which arise in the T duality only case while the $\mathbf{1}$ singlets do not, but they cannot be Type IIB tadpoles as they are expanded in the $\Hn{3}$ basis and so they can only couple to the Type IIA $C_{7}$ form. However, the use of flux matrices to examine the effect of $\SL_{S}$ transformations on the superpotential is independent of how we might label the flux multiplets of the particular Type II theory they are defined in. Hence $\SL_{S}$ transformations in Type IIA will result in its flux matrices forming the same set of multiplets. Therefore while the $\mathbf{1}$s we have constructed here do not represent tadpole constraints in Type IIB they would in Type IIA and conversely the tadpoles found in the $\Hnt{3}$ case would not be Type IIA tadpoles.

\section{Moduli space exchange symmetry}

Thus far we have observed a great deal of symmetry in how the different moduli spaces of $\M$ (and $\W$) can be described, though distinct differences exist. For example, the manner in which the Type IIB superpotential is defined, with the derivatives acting upon the K\"{a}hler moduli dependent terms rather than the complex structure. Af we wish to make our descriptions of the moduli spaces as symmetric as possible we would postulate that the superpotential and fluxes can be reformulated such that the roles of the moduli spaces are exchanged even within the Type IIB framework. However, because this implies the two moduli spaces of $\M$ are equivalent we would not expect it to be possible for all $\M$, only a particular set of spaces.
\\

To lend some justification for this hypothesis we again turn to the explict case of the $\ZZ$ orientifold. At has been shown \cite{Font:2008vd,Guarino:2008ik} that the Bacobi-like set of constraints of non-geometric fluxes possess an invariance under the subset of $\textrm{GL}(6,\bZ)$ which satisfies the orbifold symmetries. An the case of $\Mz$ this manifests itself as the complex structure $U_{i}$ moduli possessing modular invariance due to the way the orientifold factorises into three similar sub-tori, each of which possess $\SL$ symmetries. The modular invariance is only in the complex structure moduli, which can be seen to follow from the fact that a redefinition of the complex structure moduli does not mix different Type IIB fluxes and is related to a redefinition of the sympletic basis $(\syma{A},\symb{B})$ and thus how the fluxes are expanded in terms of components.  The orientifold has also been examined \cite{Ildazabal:2008aa} from the point of view of enforcing, by construction, such modular invariance on the flux Lie algebras and while it does not result in the same constraints as T duality the methodology of their analysis is qualitatively the same.
\\

An \cite{Font:2008vd,Guarino:2008ik} the change of basis on the non-geometric fluxes was viewed in terms of the subset of $\textrm{GL}(6,\bZ)$ which satisfies the orbifold symmetries but this can be reinterpreted as the set of sympletic transformations for the $\Hn{3}$ basis of the orientifold. More generally this is a reflection of how sympletic transformations on $\Hn{3}$ are a symmetry if the complex structure moduli are appropriately redefined, precisely what we have observed in our analysis of the Type IIB theory. Conversely, if the K\"{a}hler moduli can be redefined then we can regard this as being related to the redefinition of the $\Hnt{3}$ basis elements and in the case of the $\Mz$ precisely the same kind of transformations occur for the K\"{a}hler moduli as the complex structure moduli, another result of the properties of the three sub-tori. Overall we have two kinds of symmetry associated with the complex structure moduli space; T duality and redefinitions of the sympletic basis of $\Hn{3}$, but only one kind of symmetry associated to the K\"{a}hler moduli space; redefinitions of the $\Hnt{3}$ basis and the redefinition of the $\Hn{3}$ and $\Hnt{3}$ is a symmetry inherent to $\M$ and $\W$ outside of their string theory contexts \cite{ModuliSpace}. To illustrate this for the $\Mz$ more explicitly we recall its general polynomial form, where the moduli are grouped in terms of their K\"{a}hler moduli dependence.
\beq
\label{eqn:W(U,T)}
W &=& \int_{\W} \Omega \wedge \Big( \sD(\mho_{c}) +\sDp(\mho_{c}^{\prime}) \Big) \nn \\
&=& \left(\ba{ccccc}&\sT_{0}\Big( \cP_{0}(\sU) - S\,\cPh_{0}(\sU) \Big) &+& \sT_{a}\Big( \cP_{1}^{(a)}(\sU) - S\,\cPh_{1}^{(a)}(\sU) \Big) &+\\
+& \sT^{0}\Big( \cP_{3}(\sU) - S\,\cPh_{3}(\sU) \Big) &+& \sT^{b}\Big( \cP_{2}^{(b)}(\sU) - S\,\cPh_{2}^{(b)}(\sU) \Big)
\ea \right)
\eeq
Due to its factorisation in terms of three two dimensional sub-tori the moduli pair off, with $(T_{a},U_{a})$ being those moduli which describe the $a$'th sub-torus. Under $\Upsilon$ the moduli are exchanged $(T_{a},U_{a}) \to (\t{T}_{a},\t{U}_{a}) = (U_{a},T_{a})$ but the underlying structure of the space remains a two dimensional torus, suggesting that the dynamics of the complex structure moduli space is equivalent to the dynamics of the K\"{a}hler moduli space, in line with the definition of mirror symmetry; the complex structure moduli space of $\M$ being related to the K\"{a}hler moduli space of $\W$. We have seen this in the way the different superpotentials are defined, but in cases where $\M=\W$ this would imply an enhanced symmetry on $\M$ itself and the superpotential defined by its fluxes, with the two moduli spaces interchangable and structures existing on one should appear in the other. To illustrate this on the $\Mz$ rather than use T duality or mirror symmetry to alter the superpotential we instead can simply rearrange the polynomial superpotential to exchange the roles of the two moduli types.
\beq
\label{eqn:W(T,U)}
W &\to& \left(\ba{ccccc}&\sU_{0}\Big( \fP_{0}(\sT) - S\,\fPh_{0}(\sT) \Big) &+& \sU_{a}\Big( \fP_{1}^{(a)}(\sT) - S\,\fPh_{1}^{(a)}(\sT) \Big) &+\\
+& \sU^{0}\Big( \fP_{3}(\sT) - S\,\fPh_{3}(\sT) \Big) &+& \sU^{b}\Big( \fP_{2}^{(b)}(\sT) - S\,\fPh_{2}^{(b)}(\sT) \Big)
\ea \right)
\eeq
T duality defines a sequence of fluxes induced by the non-zero nature of $\Fh_{0}$ and for each of these fluxes there is a corresponding polynomial in the complex structure moduli, which in the case of $\Fh_{0}$ we have denoted as $\cPh_{0}(\sU)$. With the reformulation of the superpotential in (\ref{eqn:W(T,U)}) due to the symmetry in the moduli we now have K\"{a}hler moduli dependent polynomials such as $\fPh_{0}$. We can postulate these polynomials are built from some new set of fluxes but precisely what those might be we will consider in the next section. However, if the K\"{a}hler moduli space for the $\Mz$ is equivalent to the complex structure moduli space of $\Mz$ then this leads us to argue that there is a duality which has the same effect on the K\"{a}hler moduli as T duality has on the complex structure moduli, which we will refer to as T$^{\prime}$ duality. To examine this more quantitatively and for spaces other than $\Mz$ we are required to consider the many different ways we have of constructing superpotential-like expressions from objects we have been examining.

\subsection{Alternate superpotential}

Since we will be discussing how various derivatives and their matrix representations relate to one another we will dispense with the different D's used for different derivatives in previous sections and simply label them with an index, $\sD_{i}$, and likewise with their associated matrix representations, which in the case of Type IIB has $\DevenB_{i}$ representing the action on the $\Hnt{3}$ basis and $\DoddB_{j}$ on the $\Hn{3}$ basis and the superpotentials of Type IIA are distinguished from those of Type IIB by $W$ compared to $\mathsf{W}$. In Type IIB we can construct objects which have a superpotential-like form in two different ways; one of which is the Type IIB superpotential and the second resembles its $\ModB$ moduli dual form, except it is defined upon $\W$ rather than $\M$, which is a critical difference.
\beq 
\label{eqn:TypeIIB1}
\mathsf{W}_{1} &=& \int_{\W} \Omega \wedge \Big( \sD_{1}(\mho_{c}) +\sDp_{1}(\mho_{c}^{\prime}) \Big) = \u{\sT}^{\top}\cdot \sh_{\HtbasisB}\cdot\Big( \bC\cdot \DevenB_{1} + \bC^{\prime}\cdot \DevenB_{1}^{\prime} \Big)\cdot \sg_{\Hbasis} \cdot \u{\sU}  \\
\label{eqn:TypeIIB2}
\mathsf{W}_{2} &=& \int_{\W} \mho \wedge \Big( \sD_{2}(\Omega_{c}) +\sDp_{2}(\Omega_{c}^{\prime}) \Big) = \u{\sU}^{\top}\cdot \sh_{\Hbasis}\cdot\Big( \t{\bC}\cdot \DoddB_{2} + \t{\bC}^{\prime}\cdot \DoddB_{2}^{\prime} \Big)\cdot \sg_{\HtbasisB} \cdot \u{\sT} 
\eeq
In general these are the only\footnote{We do not consider $\Omega \wedge \sD(\mho_{c})$ and $\sD(\Omega) \wedge \mho_{c}$ as different for reasons which will shortly be given.} two expressions which can be formed of integrals and from pairs of elements of either $\Hn{3}$ or $\Hnt{3}$. It is possible, however, to construct Type IIB scalar products which are dependent upon the bilinear forms $g$ and $h$ defined in Type IIA. 
\beq 
\label{eqn:TypeIIB3}
\mathsf{W}_{3} &=& \u{\sT}^{\top}\cdot h_{\Hbasis}\cdot\Big( \bC\cdot \DoddB_{3} + \bC^{\prime}\cdot \DoddB_{3}^{\prime} \Big)\cdot g_{\HtbasisB} \cdot \u{\sU} \\
\label{eqn:TypeIIB4}
\mathsf{W}_{4} &=& \u{\sU}^{\top}\cdot h_{\HtbasisB}\cdot\Big( \t{\bC}\cdot \DevenB_{4}+ \t{\bC}^{\prime}\cdot \DevenB_{4}^{\prime} \Big)\cdot g_{\Hbasis} \cdot \u{\sT} 
\eeq
These two expressions are constructable using matrices because the dimensions of such pairs as $h_{\HtbasisB}$ and $\sh_{\Hbasis}$ are equal by $h^{1,1}(\M) = h^{2,1}(\W)$. This allows us to build forms such as $\u{\sT}^{\top} \cdot h_{\Hbasis} \cdot \BasisZ$, hybrids of terms defined in different spaces and in different Type II theories but this fact means that generally such constructs are ill defined. The expression $\u{\sT}^{\top} \cdot h_{\Hbasis} \cdot \BasisZ$ can be built in $\W$ if $h^{1,1} = h^{2,1}$ and it is possible to choose $\Hn{3}$ bases in $\M$ and $\W$ such that $h_{\Hbasis} = \sh_{\Hbasis}$. This is a reflection of the link between the K\"{a}hler moduli space of $\W$ and the complex structure moduli space of $\M$, $\sT_{I} \leftrightarrow \fU_{I}$. Af the link is to be between the two moduli spaces of $\W$ itself then we instead wish to consider the equivalence $\sT_{I} \leftrightarrow \fU_{I}$ which is possible if, given $h^{1,1} = h^{2,1}$, $\fT_{A} \leftrightarrow \sU_{A}$ and as a result we narrow our considerations to those spaces which satisfy $\M = \W$, those spaces which are self mirror dual. Such a restriction automatically allows us to make the equivalence $\sg_{\Hbasis} = g_{\Hbasis}$ and likewise with the other bilinear forms because of the equality of the Hodge numbers\footnote{At should be noted that although the IIA complex structure indices $I,J,\ldots$ and the IIA K\"{a}hler indices $A,B,\ldots$ range over the same values we retain their distinction for the purposes of clarity.} and that a redefinition of the basis of $\Hn{3}(\M)$ is also a redefinition of the basis of $\Hn{3}(\W)$ since they are one and the same. Is a result it is possible to construct the Type IIB form $\t{\Omega} \equiv \Omega \Big|_{\sU_{A} \to \sT_{A}} = \sT^{\top} \cdot \sh_{\Hbasis} \cdot \BasisY = \u{\sT}^{\top} \cdot h_{\Hbasis} \cdot \BasisY$ on $\W$. With this equality between the Type IIA and Type IIB bilinear forms on $\M=\W$ both (\ref{eqn:TypeIIB3}) and (\ref{eqn:TypeIIB4}) therefore obtain an integral representation, in terms of $\t{\Omega}$ and $\t{\mho} \equiv \mho \Big|_{\sT_{I} \to \sU_{I}}$.
\beq 
\label{eqn:TypeIIB5}
\mathsf{W}_{3} = \int_{\W} \t{\mho} \wedge \Big( \sD_{3}(\t{\Omega}_{c}) +\sDp_{3}(\t{\Omega}_{c}^{\prime}) \Big) \qquad \label{eqn:TypeIIB6}
\mathsf{W}_{4} = \int_{\W} \t{\Omega} \wedge \Big( \sD_{4}(\t{\mho}_{c}) +\sDp_{4}(\t{\mho}_{c}^{\prime}) \Big)
\eeq
To illustrate this more explicitly we consider an integral similar to that of (\ref{eqn:TypeIIB1}), namely using non-complexified holomorphic forms and use the properties of the symplectic basis and the equality of the Hodge number to convert it into something similar to (\ref{eqn:TypeIIB5}). Thus illustrating a rearrangement of the superpotential akin to that between (\ref{eqn:W(T,U)}) and (\ref{eqn:W(U,T)}). For the purpose of simplification we neglect dilaton complexifications.
\beq
\int_{\W} \Omega \wedge \sD_{1}(\mho) &=& \int_{\W} \Big(\sU_{A}\syma{A} - \sU^{B}\symb{B} \Big) \wedge \left[ \ba {l}\sT_{I}\Big(\sF_{(I)A}\syma{A} - \sF_{(I)}^{\phantom{(I)}B}\symb{B} \Big)  \\ \qquad - \sT^{J}\Big(\sF_{\phantom{(I)}A}^{(I)}\syma{A} - \sF^{(J)B}\symb{B} \Big)\ea  \right] \nn \\
&=& \int_{\W} \Big(\sT_{I}\syma{I} - \sT^{J}\symb{J} \Big) \wedge \left[ \ba {l} \sU_{A} \Big( \sF^{(I)A}\syma{I} - \sF_{(J)}^{\phantom{(J)}B}\symb{J} \Big) \\ \qquad - \sU^{B} \Big( \sF^{(I)}_{\phantom{(I)}A}\syma{I} - \sF_{(J)B}\symb{J} \Big) \ea  \right] \nn \\
\label{eqn:SwapUT}
&=& \int_{\W} \t{\Omega} \wedge \sD_{4}(\t{\mho}) 
\eeq
We have had to make the assumption that $A,B$ and $I,J$ range over the same indices and that the sympletic structure of $\M$ is equivalent to that of $\W$, as such expressions as $\sT_{I}\syma{I} - \sT^{J}\symb{J}$ are the Type IIB holomorphic $3$-form $\Omega$ but with the moduli labelled in the Type IIA manner. The general fact that these expressions bear a striking resemblence to the Type IIA superpotential integrals prompts us to now turn our attention to those superpotential-like integrals defined in Type IIA on a generic $\M$. As with Type IIB, there are two expressions which can be written as integrals and two which, in general, cannot. We label the Type IIA derivatives with a tilde and an index, $\D_{i}$ and likewise with their associated matrix representations, which in the case of Type IIA has $\DevenA_{i}$ representing the action on the $\Hnt{3}$ basis and $\DoddA_{j}$ on the $\Hn{3}$ basis.
\beq 
\label{eqn:IIA1}
W_{1} &=& \int_{\M} \mho \wedge \Big( \D_{1}(\Omega_{c}) +\Dp_{1}(\Omega_{c}^{\prime}) \Big) = \u{\fU}^{\top}\cdot h_{\Hbasis}\cdot\Big( \bC\cdot \DoddA_{1} + \bC^{\prime}\cdot \DoddA_{1}^{\prime} \Big)\cdot g_{\HtbasisB} \cdot \u{\fT}\\
\label{eqn:IIA2}
W_{2} &=& \int_{\M} \Omega \wedge \Big( \D_{2}(\mho_{c}) +\Dp_{2}(\mho_{c}^{\prime}) \Big) = \u{\fT}^{\top}\cdot h_{\HtbasisB}\cdot\Big( \t{\bC}\cdot \DevenA_{2}+ \t{\bC}^{\prime}\cdot \DevenA_{2}^{\prime} \Big)\cdot g_{\Hbasis} \cdot \u{\fU} 
\eeq
As in Type IIB we can construct a pair of superpotential-like scalar products which are a mixture of Type IIA and Type IIB defined objects.
\beq 
\label{eqn:IIA3}
W_{3} &=& \u{\fU}^{\top}\cdot \sh_{\HtbasisB}\cdot\Big( \bC\cdot \DevenA_{3} + \bC^{\prime}\cdot \DevenA_{3}^{\prime} \Big)\cdot \sg_{\Hbasis} \cdot \u{\fT} \\
\label{eqn:IIA4}
W_{4} &=& \u{\fT}^{\top}\cdot \sh_{\Hbasis}\cdot\Big( \t{\bC}\cdot \DoddA_{4} + \t{\bC}^{\prime}\cdot \DoddA_{4}^{\prime} \Big)\cdot \sg_{\HtbasisB} \cdot \u{\fU}
\eeq
For the case of $\M=\W$ it is possible to construct integral representations in the same manner as the Type IIB case and we again take $\t{\mho}$ and $\t{\Omega}$ to represent the holomorphic forms which have had their moduli spaces exchanged. 
\beq 
\label{eqn:TypeIIA5}
W_{3} = \int_{\M} \t{\Omega} \wedge \Big( \D_{3}(\t{\mho}_{c}) +\t{\sDp_{3}}(\t{\mho}_{c}^{\prime}) \Big) \qquad \label{eqn:TypeIIB6}
W_{4} = \int_{\M} \t{\mho} \wedge \Big( \D_{4}(\t{\Omega}_{c}) +\t{\sDp_{4}}(\t{\Omega}_{c}^{\prime}) \Big)
\eeq
These eight ways of constructing a superpotential can split into two subsets by considering how the moduli are coupled to the dilaton by the complexification matrices. The standard Type IIB superpotential has its dilaton dependence defined in the same manner as $\mathsf{W}_{1}$, it is the $\sT$ which are complexified. The Type IIA superpotential has terms of the form seen in both $W_{1}$ and $W_{3}$ but in both cases it is the $\fU$ which are complexified. In these three cases, as well as $\mathsf{W}_{3}$ it is the same degrees of freedom which is dilaton complexified, the $\sT$ and $\fU$ are different labellings for the same moduli. 
\beq
\ba{cccccccccccccc}
\textrm{Type IIA} &\;:\;& \Omega_{c} & \;=\; &-& S\,\fU_{0}\syma{0} &+& \fU_{i}\syma{i} &-& \fU^{j}\symb{j} &+& S\,\fU^{0}\symb{0} \\
\textrm{Type IIB} &\;:\;& \mho_{c} & \;=\; &-& S\,\sT_{0}\v{0} &+& \sT_{i}\v{i} &+& \sT^{j}\vt{j} &-& S\,\sT^{0}\vt{0}
\ea
\eeq
When comparing the Type IIA and IIB fluxes for each flux sector we noted that the complexification matrices could be neglected due to each Type II superpotential being dependent on the same complexification matrices. By considering all of these possible superpotential constructions it can be seen that all elements of $W_{-} = \{ \mathsf{W}_{1},\mathsf{W}_{3},W_{1},W_{3}\}$ and, seperately, $W_{+} = \{ \mathsf{W}_{2},\mathsf{W}_{4},W_{2},W_{4}\}$ have dilaton complexification of the same form. The elements of $W_{\pm}$ are related to one another by moduli relabelling and $\ModA$ and these lead to simple expressions for the interdependency of their associated flux matrices. Comparing $\mathsf{W}_{1}$ with $W_{1}$ gives the flux matrix relations previously observed in the NS-NS sector in (\ref{eqn:Mirrorrelation}), but now in both sectors.
\beq 
\sh_{\HtbasisB}\cdot \DevenB_{1}\cdot \sg_{\Hbasis} = h_{\Hbasis}\cdot \DoddA_{1} \cdot g_{\HtbasisB} \quad \Rightarrow \quad 
\ba{ccc}
\DevenB_{1} &=& \fm h_{\Hbasis} \cdot \DoddA_{1} \cdot \sh_{\Hbasis} \\
\DoddB_{1} &=&    -\sh_{\HtbasisB} \cdot \DevenA_{1} \cdot h_{\HtbasisB}
\ea \nn
\eeq
Provided $\M=\W$ the expressions for $W_{3}$ and $\mathsf{W}_{3}$ are well defined and their flux matrices are related in a similar way. Their interdependency with the flux matrices of $W_{1}$ and $\mathsf{W}_{1}$ are trivial since $\mathsf{W}_{1}$ is $W_{3}$ under a moduli relabelling and likewise for $\mathsf{W}_{3}$ and $W_{1}$.
\beq 
\sh_{\HtbasisB}\cdot \DevenA_{3} \cdot \sg_{\Hbasis} = h_{\Hbasis}\cdot \DoddB_{3} \cdot g_{\HtbasisB} \quad \Rightarrow \quad 
\ba{ccccc}
\DevenA_{3} &=& \fm  h_{\Hbasis} \cdot \DoddB_{3} \cdot \sh_{\Hbasis} &=& \DevenB_{1} \\
\DoddA_{3} &=&    -\sh_{\HtbasisB} \cdot \DevenB_{3} \cdot h_{\HtbasisB} &=& \DoddB_{1}
\ea \nn
\eeq
We previously saw that the Type IIA nilpotency constraints are equal to the Type IIB nilpotency constraints, in either flux sector, and it therefore follows that the nilpotency constraints of the derivatives used to construct the superpotentials of $W_{-}$ are all equivalent. Repeating this analysis for $W_{+} = \{ \mathsf{W}_{2},\mathsf{W}_{4},W_{2},W_{4}\}$ we observe they too are linked by moduli relabelling and moduli duality. The dilaton couples to the same degrees of freedom in each case and therefore the complexification matrices can be factorised out when comparing the expressions. From this it is straightforward to obtain the relationship between the different flux matrices and to show the nilpotency conditions to be equal. As an example we consider $\mathsf{W}_{2}$ and $W_{2}$ where the complexification matrices combine with either $\fT$ or $\sU$, as is the case for any other pairwise comparision of $W_{+}$ elements.
\beq 
\sh_{\Hbasis} \cdot \DoddB_{2} \cdot \sg_{\HtbasisB} = h_{\HtbasisB}\cdot \DevenA_{2} \cdot g_{\Hbasis} \quad \Rightarrow \quad 
\ba{ccccc}
\DevenA_{2} &=& \fm  \sh_{\Hbasis} \cdot \DoddB_{2} \cdot h_{\Hbasis} &=& \DevenB_{4} \\
\DoddA_{2} &=&    -h_{\HtbasisB} \cdot \DevenB_{2} \cdot \sh_{\HtbasisB} &=& \DoddB_{4}
\ea \quad \Leftarrow \quad \sh_{\Hbasis} \cdot \DevenA_{4} \cdot \sg_{\HtbasisB} = h_{\HtbasisB}\cdot \DoddB_{4} \cdot g_{\Hbasis}  \nn
\eeq
By the same reasoning as $W_{-}$ the Bianchi constraints for each flux sector of each superpotential are equivalent to one another. However, if we are to equate an element of $W_{-}$ with an element of $W_{+}$ it is no longer the case that the Bianchi constraints are equivalent. The Bianchi constraints of $\sD_{1}$ and $\D_{1}$ are equivalent and those of $\sD_{2}$ and $\D_{2}$ are equivalent but $\sD_{1}$ and $\sD_{2}$ are inequivalent, the change in dilaton couplings between the expressions is a non-trivial effect. To examine this more indepth we shall consider the specific cases of the Bianchi constraints of $\mathsf{W}_{1}$ and $\mathsf{W}_{2}$, a choice motivated by the results of (\ref{eqn:symW}). We shall denote the map which converts the standard Type IIB fluxes and derivatives of $\mathsf{W}_{1}$ into those of $\mathsf{W}_{2}$ by $\merge$. In terms of the polynomial form of the superpotential this is equivalent to converting (\ref{eqn:W(U,T)}) into (\ref{eqn:W(T,U)}) and it is this which we wish to express in terms of derivatives and holomorphic forms.

\subsection{Alternate fluxes}

The K\"{a}hler moduli in (\ref{eqn:TypeIIB1}) arise due to the K\"{a}hler forms $\Jn{n}$ and we have previously defined their complex structure counterparts $\fJn{n}$. The $\mathsf{W}_{2}$ of (\ref{eqn:TypeIIB2}) can be broken down into simpler expressions by defining a set of flux multiplets, $\fF_{n}$ and $\fFh_{m}$, as the images of these $\fJ$ under the derivatives. The hatted and unhatted fluxes choosen to follow the same layout as the $\sF$ and $\sFh$ fluxes.
\beq 
\int_{\M} \mho \wedge \sD_{2}(\Omega_{c}) &=& \int_{\M} \mho \wedge \sD_{2}\left(-S\,\fJn{0} + \fJn{1} - \fJn{2} + S\,\fJn{3}\right) \nn \\
\label{eqn:NewW1}
&=& \int_{\M} \mho \wedge \left(-S\,\fFh_{0}\cdot\fJn{0} + \fF_{1}\cdot\fJn{1} - \fF_{2}\cdot\fJn{2} + S\,\fFh_{3}\cdot\fJn{3}\right) \\
\label{eqn:NewW2}
\int_{\M} \mho \wedge \sDp_{2}(\Omega_{c}^{\prime}) 
&=& \int_{\M} \mho \wedge \left(\fF_{0}\cdot\fJn{0} -S\,\fFh_{1}\cdot\fJn{1} + S\,\fFh_{2}\cdot\fJn{2} - \fF_{3}\cdot\fJn{3}\right)
\eeq
We can define the components of the $\fF$ fluxes in the same manner as we have for the $\sF$ but due to their definition in mapping $\cHn{3-n,n}$ to $\Hnt{3}$ their indices will run over different ranges. 
\beq
\ba{cccccccccccccccc}
\fFh_{0} &\quad:\quad& \Big( & \fF_{(0)I}\v{I} &+& \fF_{(0)}^{\phantom{(0)}J}\vt{J} & \Big) & \iota_{\syma{0}} &\quad:\quad& \cHn{3,0} &\to& \Hnt{3} \\
\fF_{1}  &\quad:\quad& \Big( & \fF_{(a)I}\v{I} &+& \fF_{(i)}^{\phantom{(i)}J}\vt{J} & \Big) & \iota_{\syma{a}} &\quad:\quad& \cHn{2,1} &\to& \Hnt{3} \\
\fFh_{3} &\quad:\quad& \Big( & \fF^{(0)}_{\phantom{(0)}I}\v{I} &+& \fF^{(0)J}\vt{J} & \Big) & \iota_{\symb{0}} &\quad:\quad& \cHn{0,3} &\to& \Hnt{3} \\
\fF_{2}  &\quad:\quad& \Big( & \fF^{(b)}_{\phantom{(j)}I}\v{I} &+& \fF^{(j)J}\vt{J} & \Big) & \iota_{\symb{b}} &\quad:\quad& \cHn{1,2} &\to& \Hnt{3} 
\ea
\eeq
The superpotential is then straightforward to express in terms of these fluxes, in the same manner as (\ref{eqn:W1components}).
\beq 
\int_{\M} \mho \wedge \sD_{2}(\Omega_{c}) &=& \left(
\ba{cccccccccccc}
-S\,\sU_{0} \Big( & \fF_{(0)I}\sT^{I} &+& \fF_{(0)}^{\phantom{(0)}J}\sT_{J} & \Big) &+& \sU_{a} \Big( & \fF_{(a)I}\sT^{I} &+& \fF_{(i)}^{\phantom{(i)}J}\sT_{J} & \Big)\\
+S\,\sU^{0} \Big( & \fF^{(0)}_{\phantom{(0)}I}\sT^{I} &+& \fF^{(0)J}\sT_{J} & \Big) &-& \sU^{b} \Big( & \fF^{(b)}_{\phantom{(j)}I}\sT^{I} &+& \fF^{(j)J}\sT_{J} & \Big) 
\ea \right) \nn \\ \label{eqn:W2components} \\
\int_{\M} \mho \wedge \sDp_{2}(\Omega_{c}^{\prime}) &=& \left(
\ba{cccccccccccc}
\phantom{+}\,\sU_{0} \Big( & \fFh_{(0)I}\sT^{I} &+& \fFh_{(0)}^{\phantom{(0)}J}\sT_{J} & \Big) &-& S\,\sU_{a} \Big( & \fFh_{(a)I}\sT^{I} &+& \fFh_{(i)}^{\phantom{(i)}J}\sT_{J} & \Big)\\
-\,\sU^{0} \Big( & \fFh^{(0)}_{\phantom{(0)}I}\sT^{I} &+& \fFh^{(0)J}\sT_{J} & \Big) &+& S\,\sU^{b} \Big( & \fFh^{(b)}_{\phantom{(j)}I}\sT^{I} &+& \fFh^{(j)J}\sT_{J} & \Big) 
\ea \right) \nn
\eeq
The action of $\merge$ on the various objects of Type IIB theory can now be written in a more explicit manner, one which bears close resemblence to the action of $\zeta_{2}$ in (\ref{eqn:ModMirror}) but without $\M \leftrightarrow \W$ or Type IIA $\leftrightarrow$ Type IIB.
\beq
\label{eqn:Mergeeffects}
\merge \quad:\quad \ba{ccccccc}
\sD_{1}^{(\prime)} &\to& \sD_{2}^{(\prime)} &\quad,\quad& (\sF,\sFh) &\to& (\fF,\fFh) \\
\DevenB_{1}^{(\prime)} &\to& \DevenB_{2}^{(\prime)} &\quad,\quad& \Omega &\to& \mho\\
\DoddB_{1}^{(\prime)} &\to& \DoddB_{2}^{(\prime)} &\quad,\quad& \mho &\to& \Omega \ea 
\eeq
These actions are such that the superpotential is left invariant by $\merge$ but the fluxes and derivatives are redefined. At is noteworthy also that $\merge$ satisfies $\merge^{2} = \textrm{Ad}$, where $\textrm{Id}$ is the identity map which leaves all objects in (\ref{eqn:Mergeeffects}) unchanged.

\subsubsection{T$^{\prime}$ duality constraints}

By comparing these two ways of writing the superpotential we can obtain the components of the $\fF$ and $\fFh$ in terms of the usual fluxes $\sF$ and $\sFh$, which are given in Table \ref{table:Alternatefluxes1}. The global factor of $-1$ in the definition of the signs could be removed by defining the superpotential integrands in the opposite order and so the $\Hn{3}$ defined $\mathsf{W}_{1}$ would pick up a sign change, but this is purely a matter of definition. %
\begin{table}
\beq
\setL{2}
\ba{lcllllllllllllllllllllll}
\hline\hline
\,& \fF &\,:\,& \fm\fF_{(0)0} && \fm\fF_{(0)i} && \fm\fF_{(0)}^{\phantom{(0)}0} && \fm\fF_{(0)}^{\phantom{(0)}j} &&& \fm\fF^{(0)}_{\phantom{(0)}0} && \fm\fF^{(0)}_{\phantom{(0)}i} && \fm\fF^{(0)0} && \fm\fF^{(0)j} & \,\\\hline
 & \in \mathsf{W} &:& -S\,\sU_{0}\sT^{0} &\,& -S\,\sU_{0}\sT^{i} &\,& -S\,\sU_{0}\sT_{0} &\,& -S\,\sU_{0}\sT_{j} &\,& \, & \fm S\,\sU^{0}\sT^{0} &\,& \fm S\,\sU^{0}\sT^{i} &\,& \fm S\,\sU^{0}\sT_{0} &\,& \fm S\,\sU^{0}\sT_{j} & \\\hline
& \sF &:& -\sF^{(0)0} && -\sFh^{(i)0} && -\sF_{(0)}^{\phantom{(0)}0} && -\sFh_{(j)}^{\phantom{(j)}0} &&& -\sF^{(0)}_{\phantom{(0)}0} && -\sFh^{(i)}_{\phantom{(i)}0} && -\sF_{(0)0} && -\sFh_{(j)0} & \\\hline\hline 

\,& \fF &:& \fm\fF_{(a)0} && \fm\fF_{(a)i} && \fm\fF_{(a)}^{\phantom{(0)}0} && \fm\fF_{(a)}^{\phantom{(0)}j} &&& \fm\fF^{(b)}_{\phantom{(b)}0} && \fm\fF^{(b)}_{\phantom{(b)}i} && \fm\fF^{(b)0} && \fm\fF^{(b)j} & \,\\\hline
 & \in \mathsf{W} &:& \fm\sU_{a}\sT^{0} &\,& \fm\sU_{a}\sT^{i} &\,& \fm\sU_{a}\sT_{0} &\,& \fm\sU_{a}\sT_{j} &\,& \, & -\sU^{b}\sT^{0} &\,& -\sU^{b}\sT^{i} &\,& -\sU^{b}\sT_{0} &\,& -\sU^{b}\sT_{j} & \\\hline
& \sF &:& -\sFh^{(0)a} && -\sF^{(i)a} && -\sFh_{(0)}^{\phantom{(0)}a} && -\sF_{(j)}^{\phantom{(j)}a} &&& -\sFh^{(0)}_{\phantom{(0)}b} && -\sF^{(i)}_{\phantom{(i)}b} && -\sFh_{(0)b} && -\sF_{(j)b} & \\\hline\hline

\,& \fFh &\,:\,& \fm\fFh_{(0)0} && \fm\fFh_{(0)i} && \fm\fFh_{(0)}^{\phantom{(0)}0} && \fm\fFh_{(0)}^{\phantom{(0)}j} &&& \fm\fFh^{(0)}_{\phantom{(0)}0} && \fm\fFh^{(0)}_{\phantom{(0)}i} && \fm\fFh^{(0)0} && \fm\fFh^{(0)j} & \,\\\hline
 & \in \mathsf{W} &:& \fm\sU_{0}\sT^{0} &\,& \fm\sU_{0}\sT^{i} &\,& \fm\sU_{0}\sT_{0} &\,& \fm\sU_{0}\sT_{j} &\,& \, & -\sU^{0}\sT^{0} &\,& -\sU^{0}\sT^{i} &\,& -\sU^{0}\sT_{0} &\,& -\sU^{0}\sT_{j} & \\\hline
& \sF &:& -\sFh^{(0)0} && -\sF^{(i)0} && -\sFh_{(0)}^{\phantom{(0)}0} && -\sF_{(j)}^{\phantom{(j)}0} &&& -\sFh^{(0)}_{\phantom{(0)}0} && -\sF^{(i)}_{\phantom{(i)}0} && -\sFh_{(0)0} && -\sF_{(j)0} & \\\hline\hline 

\,& \fFh &:& \fm\fFh_{(a)0} && \fm\fFh_{(a)i} && \fm\fFh_{(a)}^{\phantom{(0)}0} && \fm\fFh_{(a)}^{\phantom{(0)}j} &&& \fm\fFh^{(b)}_{\phantom{(b)}0} && \fm\fFh^{(b)}_{\phantom{(b)}i} && \fm\fFh^{(b)0} && \fm\fFh^{(b)j} & \,\\\hline
 & \in \mathsf{W} &:& -S\,\sU_{a}\sT^{0} &\,& -S\,\sU_{a}\sT^{i} &\,& -S\,\sU_{a}\sT_{0} &\,& -S\,\sU_{a}\sT_{j} &\,& \, & \fm\sU^{b}\sT^{0} &\,& \fm\sU^{b}\sT^{i} &\,& \fm\sU^{b}\sT_{0} &\,& \fm\sU^{b}\sT_{j} & \\\hline
& \sF &:& -\sF^{(0)a} && -\sFh^{(i)a} && -\sF_{(0)}^{\phantom{(0)}a} && -\sFh_{(j)}^{\phantom{(j)}a} &&& -\sF^{(0)}_{\phantom{(0)}b} && -\sFh^{(i)}_{\phantom{(i)}b} && -\sF_{(0)b} && -\sFh_{(j)b} & \\\hline\hline
\ea \nn
\eeq
\caption{$\merge$ defined components of $\fF$ and $\fFh$ in terms of the components of $\fF$ and $\fFh$ and associated superpotential coefficients.}
\label{table:Alternatefluxes1}
\end{table}
As a result of different dilaton couplings the fluxes of $\sD_{2}$ are a non-trivial mixture of the fluxes from both $\sD_{1}$ and $\sDp_{1}$. The $I,J=0$ terms being in a different flux sector to those of the $I,J>0$ terms and similarly the $A,B=0$ terms are in a different flux sector to the $A,B>0$ terms. As a result $\mathsf{W}_{2}$ has no clear distinction between the usual notion of NS-NS fluxes and R-R fluxes in $\mathsf{W}_{1}$ and the immediate corrollary of this is that the Bianchi constraints due to $\sD_{2}$ and $\sDp_{2}$ will not be the same as those due to $\sD_{1}$ and $\sDp_{1}$. We can construct the Bianchi constraints for $\sD_{2}$ in terms of the $\fF$ components in the same manner as was done for $\sD_{1}$; converting the flux actions on $\Hn{3}$ into the equivalent actions on $\Hnt{3}$.
\beq
\ba{cccccccccccccccccccc}
\fFh_{0} &\;:\;& \fm\symb{0} & \Big( &   & \fF_{(0)I}\iota_{\vt{I}} &-& \fF_{(0)}^{\phantom{(0)}J}\iota_{\v{J}} & \Big) &\quad& \fF_{1}  &\;:\;& \fm\symb{a} & \Big( &   & \fF_{(a)I}\iota_{\vt{I}} &-& \fF_{(a)}^{\phantom{(0)}J}\iota_{\v{J}} & \Big)  \\
\fFh_{3} &\;:\;& -\syma{0} & \Big( & & \fF^{(0)}_{\phantom{(0)}I}\iota_{\vt{I}} &-& \fF^{(0)J}\iota_{\v{J}} & \Big) &\quad& \fF_{2}  &\;:\;& -\syma{b} & \Big( & & \fF^{(b)}_{\phantom{(b)}I}\iota_{\vt{I}} &-& \fF^{(b)J}\iota_{\v{J}} & \Big)
\ea
\eeq
Therefore we have the two equivalent actions of $\sD_{2}$, in the same manner as (\ref{eqn:AlternateD}), with the actions of $\fFh_{0}$ and $\fFh_{3}$ once again able to be subsumed into the other fluxes.
\beq
\label{eqn:AlternateD2}
\ba {cccccc}
\sD_{2}(\syma{A}) &=& \fF_{(A)I}\v{I} &+& \fF_{(A)}^{\phantom{(A)}J}\vt{J}  \\
\sD_{2}(\symb{B}) &=& \fF_{\phantom{(B)}I}^{(B)}\v{I} &+& \fF^{(B)J}\vt{J} 
\ea \quad \Leftrightarrow \quad
\ba {cccccc}
\sD_{2}(\v{I})  &=& \fm\fF^{(A)I}\syma{A} &-& \fF_{(B)}^{\phantom{(B)}I}\symb{B}  \\
\sD_{2}(\vt{J}) &=&   -\fF_{\phantom{(A)}J}^{(A)}\syma{A} &+& \F_{(B)J}\symb{B} 
\ea
\eeq 
Combining these two actions of each flux we obtain their Bianchi constraints.
\beq
\label{eqn:BianchiT2}
\ba{ccccccccccccccc}
\sD_{2}^{2}(\syma{A}) &=& \Big( & \fF_{(A)J}\fF^{(B)J} &-& \fF_{(A)}^{\phantom{(A)}I}\fF_{\phantom{(B)}I}^{(B)} & \Big)\syma{B} &+& \Big( & \fF_{(A)}^{\phantom{(A)}J}\fF_{(B)J} &-& \fF_{(A)I}\fF_{(B)}^{\phantom{(B)}I} & \Big)\symb{B} \\
\sD_{2}^{2}(\symb{A}) &=& \Big( & \fF_{\phantom{(A)}J}^{(A)}\fF^{(B)J} &-& \fF^{(A)I}\fF_{\phantom{(B)}I}^{(B)} & \Big)\syma{B} &+& \Big( & \fF^{(A)J}\fF_{(B)J} &-& \fF_{\phantom{(A)}I}^{(A)}\fF_{(B)}^{\phantom{(B)}I} & \Big)\symb{B} \\
\sD_{2}^{2}(\v{I}) &=& \Big( & \fF^{(B)I}\fF_{(B)J} &-& \fF_{(A)}^{\phantom{(A)}I}\fF_{\phantom{(A)}J}^{(A)} & \Big) \v{J} &+& \Big( & \fF^{(B)I}\fF_{(B)}^{\phantom{(B)}J} &-& \fF_{(A)}^{\phantom{(A)}I}\fF^{(A)J} & \Big)\vt{J}\\
\sD_{2}^{2}(\vt{I})&=& \Big( & \fF_{(B)I}\fF_{\phantom{(B)}J}^{(B)} &-& \fF_{\phantom{(A)}I}^{(A)}\fF_{(A)J} & \Big)\v{J} &+& \Big( & \fF_{(A)I}\fF^{(A)J} &-& \fF_{\phantom{(B)}I}^{(B)}\fF_{(B)}^{\phantom{(B)}J} & \Big)\vt{J}
\ea \qquad
\eeq 
Although we can use Table \ref{table:Alternatefluxes1} to convert these expressions into the $\sF$ and $\sFh$ components, it is more convenient to work with flux matrices, as the generalisation to the S duality case is more forthcoming in that formulation; a fact we have previously seen in Section \ref{sec:Sduality}.

\subsection{Alternate flux matrices}

Before considering the constraints induced on the fluxes of $\sD_{2}^{(\prime)}$ we shall derive the dependence of those fluxes on the usual $\sD_{1}^{(\prime)}$ fluxes by equating the two ways of writing the superpotential in terms of flux matrices in (\ref{eqn:TypeIIB1}) and (\ref{eqn:TypeIIB2}).
\beq 
\label{eqn:DefIlt}
\u{\sT}^{\top}\cdot \sh_{\HtbasisB}\cdot\Big( \bC\cdot \DevenB_{1} + \bC^{\prime}\cdot \DevenB_{1}^{\prime} \Big)\cdot \sg_{\Hbasis} \cdot \u{\sU} = \u{\sU}^{\top}\cdot \sh_{\Hbasis}\cdot\Big( \t{\bC}\cdot \DoddB_{2} + \t{\bC}^{\prime}\cdot \DoddB_{2}^{\prime} \Big)\cdot \sg_{\HtbasisB} \cdot \u{\sT}
\eeq
Since the Type IIB superpotential is naturally written in terms of $\DevenB_{1}$ and $\DevenB_{1}^{\prime}$ we wish to express $\DoddB_{2}$ and $\DoddB_{2}^{\prime}$ in terms of them. Hence, because of the non-trivial dilaton coupling caused by the inability to neglect the complexification matrices we must consider the NS-NS and R-R sector simultaneously.
\beq 
\bC\cdot \DevenB_{1} + \bC^{\prime}\cdot \DevenB_{1}^{\prime}  = \sh_{\HtbasisB}\cdot \sg_{\HtbasisB}^{\top} \cdot\Big( \DoddB_{2}^{\top}\cdot \t{\bC} +(\DoddB_{2}^{\prime})^{\top} \cdot \t{\bC}^{\prime} \Big)\cdot \sh_{\Hbasis}\cdot \sg_{\Hbasis}^{\top}
\eeq
The complexification matrices, tilded and not, are all diagonal and commute with the $g$ and $h$ bilinear forms in both Type IIA and Type IIB and though we are assuming $h^{1,1}=h^{2,1}$ we retain the distinction between $\bC^{(\prime)}$ and $\t{\bC}^{(\prime)}$ and their definition in terms of other matrices.
\beq
\ba{ccccccccccc}
\bC &=& \A -S\, \B &=& \A_{h^{1,1}} -S\, \B_{h^{1,1}} &\qquad & \bC^{\prime} &=& \B -S\, \A &=& \B_{h^{1,1}} -S\, \A_{h^{1,1}} \\
\t{\bC} &=& \t{\A} -S\, \t{\B} &=& \A_{h^{2,1}} -S\, \B_{h^{2,1}} &\quad& \t{\bC}^{\prime} &=& \t{\B} -S\, \t{\A} &=&  \B_{h^{2,1}} -S\, \A_{h^{2,1}}
\ea
\eeq
Commuting the $\bC^{(\prime)}$ through the bilinear forms we can reexpress the $\DoddB_{2}^{(\prime)}$ in terms of $\DevenB_{2}^{(\prime)}$ so that all the transposed matrices are removed and the result is an expression which can be written with manifest dilaton dependence and $\uwave{\sF}$ and $\uwave{\sFh}$ matrices.
\beq 
\bC\cdot \DevenB_{1} + \bC^{\prime}\cdot \DevenB_{1}^{\prime}  &=& -\sh_{\Hbasis} \cdot \Big( \DevenB_{2}  \cdot \t{\bC} + \DevenB_{2}^{\prime}  \cdot \t{\bC}^{\prime} \Big) \nn \\
\Rightarrow \quad \uwave{\sF_{1}} - S \, \uwave{\sFh_{1}} &=& -\sh_{\Hbasis} \cdot \Big( \uwave{\sF_{2}} - S \, \uwave{\sFh_{2}} \Big)
\eeq
By considering dilaton couplings this decomposes into a pair of equations, each involving all of the flux matrices, which can be written in terms of the $\uwave{\F_{n}}$ and $\uwave{\Fh_{m}}$ matrices.
\beq 
\ba{cccccccccccccccccc}
\uwave{\F_{2}} &=& \DevenB_{2}\cdot \t{\A} &+& \DevenB_{2}^{\prime}\cdot \t{\B} &=& -\sh_{\Hbasis} \cdot \Big( & \A \cdot \DevenB_{1} &+& \B \cdot \DevenB_{1}^{\prime} & \Big) &=& -\sh_{\Hbasis} \cdot \uwave{\F_{1}} \\
\uwave{\Fh_{2}}&=& \DevenB_{2}\cdot \t{\B} &+& \DevenB_{2}^{\prime}\cdot \t{\A} &=& -\sh_{\Hbasis} \cdot \Big( & \B \cdot \DevenB_{1} &+& \A \cdot \DevenB_{1}^{\prime} & \Big) &=& -\sh_{\Hbasis} \cdot \uwave{\Fh_{1}} 
\ea
\eeq
Using the properties of $\t{\A}$ and $\t{\B}$ these simultaneous equations allow us to express $\DevenB_{2}^{(\prime)}$ entirely in terms of $\DevenB_{1}^{(\prime)}$.
\beq
\label{eqn:Swappedmoduli1} 
\ba{cccccc}
\DevenB_{2} &=& -\sh_{\Hbasis} \cdot \left( \Big( \A \cdot \DevenB_{1} + \B \cdot \DevenB_{1}^{\prime} \Big)\cdot \t{\A} + \Big( \B \cdot \DevenB_{1} + \A \cdot \DevenB_{1}^{\prime} \Big)\cdot \t{\B} \right) &=& -\sh_{\Hbasis} \cdot \Big( \uwave{\F_{1}}\cdot \t{\A} + \uwave{\Fh_{1}}\cdot \t{\B} \Big)\\
\DevenB_{2}^{\prime} &=& -\sh_{\Hbasis} \cdot\left( \Big( \A \cdot \DevenB_{1} + \B \cdot \DevenB_{1}^{\prime} \Big)\cdot \t{\B} + \Big( \B \cdot \DevenB_{1} + \A \cdot \DevenB_{1}^{\prime} \Big)\cdot \t{\A} \right) &=& -\sh_{\Hbasis} \cdot \Big( \uwave{\F_{1}}\cdot \t{\B} + \uwave{\Fh_{1}}\cdot \t{\A} \Big)
\ea  \nn
\eeq
It was previously noted that Table \ref{table:Alternatefluxes1} shows the $A,B=0$ cases of the flux components of $\fF$ and $\fFh$ are treated differently to the $A,B\not=0$ case, the reverse of the behaviour seen in our examination of the standard Type IIB fluxes, where $I,J=0$ cases were different from those with $I,J\not=0$. An the case of the K\"{a}hler indices it is related to the left action of $\A$ and $\B$ on $\DevenB_{1}^{(\prime)}$, with the dimensions of $\A$ and $\B$ being defined in terms of $h^{1,1}$. Here, however, it is due to the right multiplication of $\t{\A}$ and $\t{\B}$, whose dimensions are defined in terms of $h^{2,1}$ instead. The combined left and right actions of these matrices is that which provides this alteration and so  (\ref{eqn:Swappedmoduli1}) is of the general form we might have apriori expected, given the results of Table \ref{table:Alternatefluxes1}. Previously, when discussing S duality transformations in Type IIB, it was convenient to view the two flux matrices as doublet partners due to their relationship with the $\SL_{S}$ doublets and the same is true here; we can express the relationship between the $\DevenB_{2}^{(\prime)}$ and the $\DevenB_{1}^{(\prime)}$ in terms of transformations on a two component vector using the same transformation matrices that relate the S duality flux doublet with the flux matrix doublet, as in (\ref{eqn:FtoM}).
\beq 
\label{eqn:SwappedmoduliM}
\bpm \DevenB_{2} \\ \DevenB_{2}^{\prime}\epm = -\sh_{\Hbasis} \bpm \A & \B \\ \B & \A\epm_{L}\bpm \DevenB_{1} \\ \DevenB_{1}^{\prime}\epm\bpm \t{\A} & \t{\B} \\ \t{\B} & \t{\A}\epm_{R} = -\sh_{\Hbasis} \bpm \uwave{\F_{1}} \\ \uwave{\Fh_{1}}\epm\bpm \t{\A} & \t{\B} \\ \t{\B} & \t{\A}\epm_{R} 
\eeq
The $\sh_{\Hbasis}$ is understood to be a common factor, hence its factorisation out of the matrices in a way not entirely consistent with its matrix definition. The $L$ and $R$ subscripts define the direction of multiplication.
\beq 
\bpm \A & \B \\ \B & \A\epm_{L}\bpm X \\ Y\epm\equiv \bpm \A\cdot X + \B\cdot Y \\ \A\cdot X + \B\cdot Y\epm\quad , \quad 
\bpm X \\ Y \epm\bpm \t{\A} & \t{\B} \\ \t{\B} & \t{\A}\epm_{R} \equiv \bpm X\cdot \t{\A} + Y\cdot \t{\B} \\ X\cdot \t{\A} + Y\cdot \t{\B}\epm\nn
\eeq
Rather than repeating the entire method just used, the $\DoddB_{i}^{(\prime)}$ forms of these expressions are straightforward to construct from (\ref{eqn:Swappedmoduli1}) but this time removing the $\DevenB_{j}^{(\prime)}$ by using the fact $\sh_{\Hbasis}$ anticommutes with the $\sg_{\Hbasis/\HtbasisB}$ and so picks up a factor of $-1$.
\beq 
\label{eqn:Swappedmoduli2} 
\ba{ccc}
\DoddB_{2} &=&  \left( \t{\A}\cdot \Big(  \DoddB_{1}\cdot \A +  \DoddB_{1}^{\prime}\cdot \B \Big) + \t{\B}\cdot \Big(  \DoddB_{1}\cdot \B +  \DoddB_{1}^{\prime}\cdot \A \Big) \right)(\sh_{\Hbasis})  \\
\DoddB_{2}^{\prime} &=&  \left( \t{\B}\cdot \Big(  \DoddB_{1}\cdot \A +  \DoddB_{1}^{\prime}\cdot \B \Big) + \t{\A}\cdot \Big(  \DoddB_{1}\cdot \B +  \DoddB_{1}^{\prime}\cdot \A \Big) \right)(\sh_{\Hbasis}) 
\ea
\eeq
These form the same kind of tranformed doublet structure as in (\ref{eqn:SwappedmoduliM}), but with the $\sh_{\Hbasis}$ now an overall factor on the right.
\beq 
\label{eqn:SwappedmoduliN}
\bpm \DoddB_{2} \\ \DoddB_{2}^{\prime}\epm&=&  \bpm \t{\A} & \t{\B} \\ \t{\B} & \t{\A}\epm_{L}\bpm \DoddB_{1} \\ \DoddB_{1}^{\prime}\epm \bpm \A & \B \\ \B & \A\epm_{R} (\sh_{\Hbasis})
\eeq

\subsubsection{T$^{\prime}$ duality constraints}

The constraints on the fluxes as a result of the nilpotency of $\sD_{2}$ are not equivalent to the $\sD_{1}$ nilpotency constraints, due to the existence and placement of the projection-like matrices $\A$ and $\B$. To examine this we redefine our notation for each of the flux matrices such that the expressions relating to $\DevenB_{i}\cdot \sh_{\Hbasis} \cdot \DoddB_{i}=0$ simplify and we again use $\cdot \sh_{\Hbasis} \cdot  = \hdot$.\footnote{The case of $\DoddB_{i}\cdot \sh_{\HtbasisB} \cdot \DevenB_{i}=0$ follows in the same manner if we did a different redefinition in which we factorised out the matrices $\bpm \A & \B \\ \B & \A\epm$.}
\beq 
\bpm \DevenB_{2} \\ \DevenB_{2}^{\prime}\epm= \bpm m_{2} \\ m_{2}^{\prime}\epm\bpm \t{\A} & \t{\B} \\ \t{\B} & \t{\A}\epm_{R} \quad , \quad \bpm \DoddB_{2} \\ \DoddB_{2}^{\prime}\epm= \bpm \t{\A} & \t{\B} \\ \t{\B} & \t{\A}\epm_{L}\bpm n_{2} \\ n_{2}^{\prime}\epm
\eeq
Due to the orthogonality of $\t{\A}$ and $\t{\B}$ half of the terms in the expansion of $\DevenB_{2}^{(\prime)} \hdot \DoddB_{2}^{(\prime)}$ as linear  combinations of $\DevenB_{1}^{(\prime)} \hdot \DoddB_{1}^{(\prime)}$ are identically zero, as was seen when considering S duality constraints. Using $\t{\A}^{2} = \t{\A}$, and likewise for $\t{\B}$, the constraints can be put into a particular format which was seen previously in (\ref{eqn:MN1}) except that the vectors of flux matrices are redefined.
\beq 
\label{list:W+matrices}
\ba{cccccccc}
\DevenB_{2} \hdot \DoddB_{2} &=& \bpm m_{2} & m_{2}^{\prime}\epm\bpm \t{\A} & 0 \\ 0 & \t{\B}\epm \bpm \sh_{\Hbasis} & 0 \\ 0 & \sh_{\Hbasis}\epm \bpm n_{2} \\ n_{2}^{\prime}\epm \\
\DevenB_{2}^{\prime} \hdot \DoddB_{2}^{\prime} &=& \bpm m_{2} & m_{2}^{\prime}\epm\bpm \t{\B} & 0 \\ 0 & \t{\A}\epm \bpm \sh_{\Hbasis} & 0 \\ 0 & \sh_{\Hbasis}\epm \bpm n_{2} \\ n_{2}^{\prime}\epm \\
\DevenB_{2} \hdot \DoddB_{2}^{\prime} &=& \bpm m_{2} & m_{2}^{\prime}\epm\bpm 0 & \t{\A} \\ \t{\B} & 0\epm\bpm \sh_{\Hbasis} & 0 \\ 0 & \sh_{\Hbasis}\epm \bpm n_{2} \\ n_{2}^{\prime}\epm \\
\DevenB_{2}^{\prime} \hdot \DoddB_{2} &=& \bpm m_{2} & m_{2}^{\prime}\epm\bpm 0 & \t{\B} \\ \t{\A} & 0\epm\bpm \sh_{\Hbasis} & 0 \\ 0 & \sh_{\Hbasis}\epm \bpm n_{2} \\ n_{2}^{\prime}\epm
\ea
\eeq
With each of the four cases being of the same format, only differing by location and number of primed flux matrices, without much loss of generality we explicitly consider the first case.
\beq 
\label{eqn:M2N2}
\ba{ccccc}
-\sh_{\Hbasis} \cdot \DevenB_{2} \hdot \DoddB_{2} \cdot \sh_{\Hbasis} &=& -\sh_{\Hbasis} \cdot \Big( m_{2}\cdot \t{\A} \hdot \t{\A} \cdot n_{2} + m_{2}^{\prime}\cdot \t{\B} \hdot \t{\B} \cdot n_{2}^{\prime} \Big) \cdot \sh_{\Hbasis} \\
&=& \left(
\ba{crcl}
&\A \cdot \Big( \DevenB_{1} \cdot \t{\A} \hdot \DoddB_{1} &+& \DevenB_{1}^{\prime} \cdot \t{\B} \hdot \DoddB_{1}^{\prime}  \Big) \cdot \A \\
+&\B \cdot \Big( \DevenB_{1}^{\prime} \cdot \t{\A} \hdot \DoddB_{1} &+& \DevenB_{1} \cdot \t{\B} \hdot \DoddB_{1}^{\prime} \Big) \cdot \A \\
+&\A \cdot \Big( \DevenB_{1} \cdot \t{\A} \hdot \DoddB_{1}^{\prime} &+& \DevenB_{1}^{\prime} \cdot \t{\B} \hdot \DoddB_{1} \Big) \cdot \B \\
+&\B \cdot \Big( \DevenB_{1}^{\prime} \cdot \t{\A} \hdot \DoddB_{1}^{\prime} &+& \DevenB_{1}\cdot \t{\B} \hdot \DoddB_{1}  \Big) \cdot \B 
\ea \right)
\ea
\eeq
This bears a strong resemblence to (\ref{eqn:Sduality1}), except that there are $\A$ and $\B$ factors between the two flux matrices as well as being external to each term. By using the projection properties of the external $\A$ and $\B$ we can compare the components of $\DevenB_{2} \hdot \DoddB_{2}$ with those of $\DevenB_{1} \hdot \DoddB_{1}$ and $\DevenB_{1}^{\prime} \hdot \DoddB_{1}^{\prime}$ from (\ref{eqn:MN1}), as well as $\DevenB_{2}^{\prime} \hdot \DoddB_{2}^{\prime}$. In order to drop the non-degenerate factors of $\sh_{\Hbasis}$ we consider the ideals generated by the components of the flux matrices instead.
\beq 
\label{eqn:IXI}
\ba{cccrclcc}
\Big\la \A \cdot \DevenB_{2} \hdot \DoddB_{2} \cdot \A \Big\ra &=& \Big\la \A \cdot \Big( & \DevenB_{1} \cdot \t{\A} \hdot \DoddB_{1} &+& \DevenB_{1}^{\prime} \cdot \t{\B} \hdot \DoddB_{1}^{\prime} & \Big) \cdot \A \Big\ra \\
\Big\la \A \cdot \DevenB_{2}^{\prime} \hdot \DoddB_{2}^{\prime} \cdot \A \Big\ra &=& \Big\la \A \cdot \Big( & \DevenB_{1} \cdot \t{\B} \hdot \DoddB_{1} &+& \DevenB_{1}^{\prime} \cdot \t{\A} \hdot \DoddB_{1}^{\prime} & \Big) \cdot \A \Big\ra \\
\Big\la \A \cdot \DevenB_{1} \hdot \DoddB_{1} \cdot \A \Big\ra &=& \Big\la \A \cdot \Big( & \DevenB_{1} \cdot \mathbb{I} \hdot \DoddB_{1} &+& \DevenB_{1}^{\prime} \cdot 0 \hdot \DoddB_{1}^{\prime} & \Big) \cdot \A \Big\ra \\
\Big\la \A \cdot \DevenB_{1}^{\prime} \hdot \DoddB_{1}^{\prime} \cdot \A \Big\ra &=& \Big\la \A \cdot \Big( & \DevenB_{1} \cdot 0 \hdot \DoddB_{1} &+& \DevenB_{1}^{\prime} \cdot \mathbb{I} \hdot \DoddB_{1}^{\prime} & \Big) \cdot \A \Big\ra
\ea
\eeq
At is clear from the fact  $\t{\A}$ and $\t{\B}$ are internal to the flux matrix pairings of $\DevenB_{2}^{(\prime)} \hdot \DoddB_{2}^{(\prime)}$ that they cannot be written as some linear combination of the $\DevenB_{1}^{(\prime)} \hdot \DoddB_{1}^{(\prime)}$ and so the T$^{\prime}$ constraints associated with the derivatives defining $\mathsf{W}_{2}$ in (\ref{eqn:TypeIIB2}) provide different constraints to those of $\mathsf{W}_{1}$ in (\ref{eqn:TypeIIB1}). However, it is clear from (\ref{eqn:IXI}) that the constraints are equivalent on a slightly weaker level, in that the sum of the two terms associated with $\mathsf{W}_{1}$ is equal to the sum of the terms associated with $\mathsf{W}_{2}$. This can be obtained by making use of (\ref{eqn:W2fluxes1}-\ref{eqn:W2fluxes4}).
\beq 
\ba{ccccccccccccccc}
\Big\la \A \cdot \Big(& \DevenB_{2} \hdot \DoddB_{2} &+& \DevenB_{2}^{\prime} \hdot \DoddB_{2}^{\prime} &\Big) \cdot \A \Big\ra &=& \Big\la \A \cdot \Big(& \DevenB_{1} \hdot \DoddB_{1} &+& \DevenB_{1}^{\prime} \hdot \DoddB_{1}^{\prime} &\Big) \cdot \A \Big\ra \\
\Big\la \A \cdot \Big(& \DevenB_{2}^{\prime} \hdot \DoddB_{2}&+&\DevenB_{2}^{\prime} \hdot \DoddB_{2} &\Big) \cdot \A \Big\ra &=& \Big\la \A \cdot \Big(& \DevenB_{1} \hdot \DoddB_{1}^{\prime} &+& \DevenB_{1}^{\prime} \hdot \DoddB_{1} &\Big) \cdot \A \Big\ra 
\ea
\eeq
These kinds of flux combinations have been previously seen in our analysis of S duality, forming terms in $\SL_{S}$ multiplets. Since we have explicitly assumed both NS-NS and R-R fluxes are all potentially non-zero we have to consider what kind of flux structures are induced by S duality.

\subsubsection{S duality constraints}

We repeat the method used to examine the S duality of the Type IIB $\mathsf{W}_{1}$ superpotential but now we look at $\mathsf{W}_{2}$, by expressing $\DevenB_{2} \cdot \t{\bC} + \DevenB_{2}^{\prime} \cdot \t{\bC}^{\prime}$ as an inner product.
\beq
\DevenB_{2} \cdot \t{\bC} + \DevenB_{2}^{\prime} \cdot \t{\bCp} = \bpm \DevenB_{2} & \DevenB_{2}^{\prime}\epm \cdot \bpm \t{\bC} \\ \t{\bCp} \epm\nn
\eeq
Using previous results for how the complexification matrices transform under $\SL_{S}$ we have the transformation properties of the doublet formed of the two flux matrices and the transformation on the $\DoddB_{2}^{(\prime)}$ follow or can be obtained directly from the definition of $\mathsf{W}_{2}$.
\beq
\label{eqn:IltS}
\ba{ccclcc}
\bpm \t{\bC} \\ \t{\bC}^{\prime}\epm&\to& &\left( \big(\Gamma_{S}^{\top}\big)^{-1}\otimes \A + \big(\sigma \cdot \Gamma_{S}^{\top} \cdot \sigma \big)^{-1}\otimes \B \right) \bpm \t{\bC} \\ \t{\bC}^{\prime}\epm \\
\bpm \DevenB_{2} & \DevenB_{2}^{\prime}\epm &\to& \bpm \DevenB_{2} & \DevenB_{2}^{\prime}\epm & \left( \Gamma_{S}^{\top}\otimes \A + \big(\sigma \cdot \Gamma_{S}^{\top} \cdot \sigma\big)\otimes \B \right)  \\
\bpm \DoddB_{2} \\ \DoddB_{2}^{\prime}\epm&\to&  & \left( \Gamma_{S}\otimes \A + \big(\sigma \cdot \Gamma_{S} \cdot \sigma\big)\otimes \B \right) \bpm \DoddB_{2} \\ \DoddB_{2}^{\prime} \epm
\ea 
\eeq
These are precisely those transformations seen in our previous analysis S duality in (\ref{eqn:Seffect1}) and (\ref{eqn:Seffect2}) but with the roles of $M$ and $N$ exchanged, a result which could be deduced apriori from our definition (\ref{eqn:DefIlt}). The immediate implication of this fact and that $\A$ and $\B$ commute with $h_{\Hbasis/\HtbasisB}$ is that we can deduce all the $\SL_{S}$ multiplets associated to $\mathsf{W}_{2}$ from the known $\SL_{S}$ multiplets associated to $\mathsf{W}_{1}$ by exchanging $\DevenB_{1,2} \leftrightarrow \DoddB_{2,1}$, $\DevenB_{1,2}' \leftrightarrow \DoddB_{2,1}'$ and $h_{\Hbasis} \leftrightarrow h_{\HtbasisB}$. Applying this to $\mathbf{3}_{\A/\B}$ of (\ref{eqn:Triplet1}) we obtain $\t{\mathbf{3}}_{\A/\B}$ and the pair of singlets $\t{\mathbf{1}}_{\A/\B}$ follow in the same manner from $\mathbf{1}_{\A/\B}$ in (\ref{eqn:Singlet1}).
\beq
\label{eqn:W2MultipletIJ}
\ba{cccccccccccccc}
\t{\mathbf{3}}_{\A} & \equiv \la & \DevenB_{2} \cdot \t{\A} \hdot \DoddB_{2} & \;,\; & \DevenB_{2}^{\prime} \cdot \t{\A} \hdot \DoddB_{2}^{\prime} & \;,\; & \DevenB_{2}^{\prime} \cdot \t{\A} \hdot \DoddB_{2}+\DevenB_{2} \cdot \t{\A} \hdot \DoddB_{2}^{\prime} & \ra \\
\t{\mathbf{3}}_{\B} & \equiv \la & \DevenB_{2} \cdot \t{\B} \hdot \DoddB_{2} & \;,\; & \DevenB_{2}^{\prime} \cdot \t{\B} \hdot \DoddB_{2}^{\prime} & \;,\; & \DevenB_{2}^{\prime} \cdot \t{\B} \hdot \DoddB_{2}+\DevenB_{2} \cdot \t{\B} \hdot \DoddB_{2}^{\prime} & \ra \\
\t{\mathbf{1}}_{\A}  & \equiv \la & \DevenB_{2}^{\prime} \cdot \t{\A} \hdot \DoddB_{2} &-& \DevenB_{2} \cdot \t{\A} \hdot \DoddB_{2}^{\prime} & \ra \\
\t{\mathbf{1}}_{\B}  & \equiv \la & \DevenB_{2}^{\prime} \cdot \t{\B} \hdot \DoddB_{2} &-& \DevenB_{2} \cdot \t{\B} \hdot \DoddB_{2}^{\prime} & \ra
\ea 
\eeq
The introduction of these $\A$ and $\B$ terms inside the flux matrix pairings allows us to make use of (\ref{eqn:W2fluxes1}-\ref{eqn:W2fluxes4}) to compare these $\mathsf{W}_{2}$ multiplets with the $\mathsf{W}_{1}$ multiplets. Due to the linearly independent decomposition (\ref{eqn:LinDep}) $\t{\mathbf{3}}_{\A}$ can be written as a union of ideals defined by this decomposition.
\beq
\ba{cccccccccccccc}
\t{\mathbf{3}}_{\A} & = & \la & \A \cdot \DevenB_{1} \cdot \t{\A} \hdot \DoddB_{1} \cdot \A & \;,\; & \A \cdot \DevenB_{1}^{\prime} \cdot \t{\A} \hdot \DoddB_{1}^{\prime} \cdot \A & \;,\; & \A \cdot ( \DevenB_{1}^{\prime} \cdot \t{\A} \hdot \DoddB_{1} + \DevenB_{1} \cdot \t{\A} \hdot \DoddB_{1}^{\prime} ) \cdot \A & \ra \\
 & \cup & \la & \B \cdot \DevenB_{1}^{\prime} \cdot \t{\A} \hdot \DoddB_{1} \cdot \A & \;,\; & \B \cdot \DevenB_{1} \cdot \t{\A} \hdot \DoddB_{1}^{\prime} \cdot \A & \;,\; & \B \cdot ( \DevenB_{1} \cdot \t{\A} \hdot \DoddB_{1} + \DevenB_{1}^{\prime} \cdot \t{\A} \hdot \DoddB_{1}^{\prime} ) \cdot \A & \ra \\
 & \cup & \la & \A \cdot \DevenB_{1} \cdot \t{\A} \hdot \DoddB_{1}^{\prime} \cdot \B & \;,\; & \A \cdot \DevenB_{1}^{\prime} \cdot \t{\A} \hdot \DoddB_{1} \cdot \B & \;,\; & \A \cdot ( \DevenB_{1}^{\prime} \cdot \t{\A} \hdot \DoddB_{1}^{\prime} + \DevenB_{1} \cdot \t{\A} \hdot \DoddB_{1} ) \cdot \B & \ra \\
 & \cup & \la & \B \cdot \DevenB_{1}^{\prime} \cdot \t{\A} \hdot \DoddB_{1}^{\prime} \cdot \B & \;,\; & \B \cdot \DevenB_{1} \cdot \t{\A} \hdot \DoddB_{1} \cdot \B & \;,\; & \B \cdot ( \DevenB_{1} \cdot \t{\A} \hdot \DoddB_{1}^{\prime} + \DevenB_{1}^{\prime} \cdot \t{\A} \hdot \DoddB_{1} ) \cdot \B & \ra 
\ea \nn
\eeq
By considering the splittings and decompositions due to $\A$, $\B$, $\t{\A}$ and $\t{\B}$ it can be seen that the union of all the $\SL_{S}$ ideals of $\mathsf{W}_{1}$ is equal to the union of all the $\SL_{S}$ ideals of $\mathsf{W}_{2}$ but individually the ideals are not equal to one another.
\beq
\t{\mathbf{3}}_{\A} \cup \t{\mathbf{3}}_{\B} \cup \t{\mathbf{1}}_{\A} \cup \t{\mathbf{1}}_{\B} = \mathbf{3}_{\A} \cup \mathbf{3}_{\B} \cup \mathbf{1}_{\A} \cup \mathbf{1}_{\B}
\eeq
The second set of $\SL_{S}$ triplets on $\mathsf{W}_{2}$ follow (\ref{eqn:Triplet1}) and (\ref{eqn:Triplet2}) by the same relabelling.
\beq
\ba{cccccccccccccc}
\t{\mathbf{3}}_{\t{\A}\t{\A}} & \equiv \la & \t{\A} \cdot \DoddB_{2} \cdot \DevenB_{2} \cdot \t{\A} & \;,\; & \t{\A} \cdot \DoddB_{2}^{\prime} \cdot \DevenB_{2}^{\prime} \cdot \t{\A} & \;,\; & \t{\A} \cdot ( & \DoddB_{2}^{\prime} \cdot \DevenB_{2} + \DoddB_{2} \cdot \DevenB_{2}^{\prime} &) \cdot \t{\A} & \ra \\
\t{\mathbf{3}}_{\t{\A}\t{\B}} & \equiv \la & \t{\A} \cdot \DoddB_{2}^{\prime} \cdot \DevenB_{2} \cdot \t{\B} & \;,\; & \t{\A} \cdot \DoddB_{2} \cdot \DevenB_{2}^{\prime} \cdot \t{\B} & \;,\; & \t{\A} \cdot ( & \DoddB_{2} \cdot \DevenB_{2} + \DoddB_{2}^{\prime} \cdot \DevenB_{2}^{\prime} &) \cdot \t{\B} & \ra \\
\t{\mathbf{3}}_{\t{\B}\t{\A}} & \equiv \la & \t{\B} \cdot \DoddB_{2} \cdot \DevenB_{2}^{\prime} \cdot \t{\A} & \;,\; & \t{\B} \cdot \DoddB_{2}^{\prime} \cdot \DevenB_{2} \cdot \t{\A} & \;,\; & \t{\B} \cdot ( & \DoddB_{2}^{\prime} \cdot \DevenB_{2}^{\prime} + \DoddB_{2} \cdot \DevenB_{2} &) \cdot \t{\A} & \ra \\
\t{\mathbf{3}}_{\t{\B}\t{\B}} & \equiv \la & \t{\B} \cdot \DoddB_{2}^{\prime} \cdot \DevenB_{2}^{\prime} \cdot \t{\B} & \;,\; & \t{\B} \cdot \DoddB_{2} \cdot \DevenB_{2} \cdot \t{\B} & \;,\; & \t{\B} \cdot ( & \DoddB_{2} \cdot \DevenB_{2}^{\prime} + \DoddB_{2}^{\prime} \cdot \DevenB_{2} &) \cdot \t{\B} & \ra 
\ea 
\eeq 
These can then be written in terms of the $\mathsf{W}_{1}$ flux matrices using (\ref{eqn:Swappedmoduli1}) and (\ref{eqn:Swappedmoduli2}), though we only do so explicitly for $\t{\mathbf{3}}_{\t{\A}\t{\A}}$ due to the length of the expressions. The remaining multiplets follow the same general structure but with appropriate (un)priming of the flux matrices.
\beq
\ba{ccccccccccc} \t{\mathbf{3}}_{\t{\A}\t{\A}}  &= 
&       & \la \t{\A} \cdot ( & \DoddB_{1}\cdot \A \cdot \DevenB_{1} &+& \DoddB_{1}^{\prime} \cdot \A \cdot \DevenB_{1}^{\prime} & ) \cdot \t{\A} \ra & \cup \\
&& \cup & \la \t{\A} \cdot ( & \DoddB_{1}\cdot \B \cdot \DevenB_{1} &+& \DoddB_{1}^{\prime} \cdot \A \cdot \DevenB_{1}^{\prime} & ) \cdot \t{\A} \ra & \cup \\
&& \cup & \la \t{\A} \cdot ( & \DoddB_{1}\cdot \DevenB_{1}^{\prime} &+& \DoddB_{1}^{\prime} \cdot \DevenB_{1} & ) \cdot \t{\A} \ra & 
\ea
\eeq 
As before the singlets are the third term of each triplet with a sign change.
\beq
\ba{cccccccccccccc}
\t{\mathbf{1}}_{\t{\A}\t{\A}} & \equiv \la & \t{\A} \cdot ( & \DoddB_{2}^{\prime} \cdot \DevenB_{2} &-& \DoddB_{2} \cdot \DevenB_{2}^{\prime} &) \cdot \t{\A} & \ra \\
\t{\mathbf{1}}_{\t{\A}\t{\B}} & \equiv \la & \t{\A} \cdot ( & \DoddB_{2} \cdot \DevenB_{2} &-& \DoddB_{2}^{\prime} \cdot \DevenB_{2}^{\prime} &) \cdot \t{\B} & \ra \\
\t{\mathbf{1}}_{\t{\B}\t{\A}} & \equiv \la & \t{\B} \cdot ( & \DoddB_{2}^{\prime} \cdot \DevenB_{2}^{\prime} &-& \DoddB_{2} \cdot \DevenB_{2} &) \cdot \t{\A} & \ra \\
\t{\mathbf{1}}_{\t{\B}\t{\B}} & \equiv \la & \t{\B} \cdot ( & \DoddB_{2} \cdot \DevenB_{2}^{\prime} &-& \DoddB_{2}^{\prime} \cdot \DevenB_{2} &) \cdot \t{\B} & \ra 
\ea 
\eeq 
In our examination of the standard formulation of the fluxes and superpotential we noted that not all of these $\SL_{S}$ multiplets are Bianchi constraints, some of them are non-zero and measure tadpole contributions due to branes and their S duality images. Which type of constraint a particular multiplet fell into was given in (\ref{eqn:BianchiTadpole}) and we would expect a similar behaviour in these multiplets. The simplest tadpole considered was the $C_{4}$ potential which coupled to the external space filling D$3$ branes whose flux contribution $H_{3} \wedge F_{3} \sim \Fh_{0} \wedge \F_{0}$ could be written in terms of derivatives as being proportional to $\vt{0} \wedge (\sD_{1}\sDp_{1}-\sDp_{1}\sD_{1})(\vt{0})$. The $\merge$ image of this is obtained by replacing the $\mathsf{W}_{1}$ derivatives with those of $\mathsf{W}_{2}$ and the flux polynomials associated to that appear in the $\t{\mathbf{1}}_{\t{\B}\t{\B}}$ singlet and can be written in terms of derivatives as $\vt{0} \wedge (\sD_{2}\sDp_{2}-\sDp_{2}\sD_{2})(\vt{0})$. This construction does not have a straight forward definition in terms of the $\fF$ and $\fFh$ fluxes of $\mathsf{W}_{2}$, in the same manner that $\SL_{S}$ images of Type IIA fluxes do not have a straight forward expression in terms of fluxes due to the way they are defined. How the tadpole contributions are to be viewed in terms of the action of $\merge$ on the branes of the Type IIB theory is a question we shall not address other than to comment that $\vt{0} \wedge (\sD_{2}\sDp_{2}-\sDp_{2}\sD_{2})(\vt{0})$ contains the fluxes found on branes other than the D$3$s in the formulation of the $\mathsf{W}_{1}$ superpotential including extended objects which are the NS-NS counterparts of the D branes. 

\subsection{Reduced superpotential expression}

If we assume that the formulation of $\mathsf{W}_{2}$ is as valid as that of $\mathsf{W}_{1}$ then we can express the superpotential in a way which is symmetric in its treatment of the moduli spaces. By using (\ref{eqn:symW}) as a guide we have obtained the relationship between the flux matrices of $\mathsf{W}_{1}$ in (\ref{eqn:TypeIIB1}) and those of $\mathsf{W}_{2}$ in (\ref{eqn:TypeIIB2}). To motivate this further we consider a superpotential-like expression $\mathsf{W}_{\fD}$ which is defined as a scalar product involving $\u{\Psi}$, whose entries are the moduli vectors $\u{\sT}$ and $\u{\sU}$, and a matrix $\u{\u{\fD}}$. We do not treat $\u{\u{\fD}}$ as the matrix associated to a derivative, only a linear operator on the cohomology bases so that the associated flux matrices $\DevenB_{\fD}$ and $\DoddB_{\fD}$ are independent but we use notation which follows previous superpotential-like scalar products.
\beq 
\mathsf{W}_{\fD} \equiv \u{\Psi}^{\top} \cdot \sh \cdot \fC \cdot \u{\u{\fD}} \cdot \sg \cdot \u{\Psi}
\qquad \u{\u{\fD}} = \bpm 0 & \DevenB_{\fD} \\ \DoddB_{\fD} & 0 \epm
\qquad \fC = \bpm \bC & 0 \\ 0 & \t{\bC} \epm
\eeq
With $h^{1,1}=h^{2,1}$ the complexification matrices are equal, $\bC=\t{\bC}$, and so $\fC = \mathbb{I}_{2} \otimes \bC$. Expanding $W_{\fD}$ out in terms of the individual moduli sectors results in a pair of terms, one of the form seen in $\mathsf{W}_{1}$ and the other of the form seen in $\mathsf{W}_{2}$.
\beq 
\label{eqn:ID}
\mathsf{W}_{\fD} \;=\; \u{\sT}^{\top} \cdot \sh_{\HtbasisB} \cdot \bC \cdot \DevenB_{\fD} \cdot \sg_{\Hbasis} \cdot \u{\sU} \;+\; \u{\sU}^{\top} \cdot \sh_{\Hbasis} \cdot \bC \cdot \DoddB_{\fD} \cdot \sg_{\HtbasisB} \cdot \u{\sT}
\eeq
This is in contrast to previous superpotential expressions considered, where the matrices $\u{\u{\Omega}}$ and $\u{\u{\mho}}$ are defined with a projection matrix $\cP^{\pm}$ so that one of the two terms is projected out. In general there are two contributions to the superpotential due to the different flux sectors so if the  two moduli spaces are equivalent we would expect it to be possible to express the full superpotential in the same manner as (\ref{eqn:AD}). On the assumption that $\mathsf{W}_{2} \equiv \merge(\mathsf{W}_{1}) = \mathsf{W}_{1}$ the superpotential $\mathsf{W}$, which is normally written as having the form of $\mathsf{W}_{1}$, is proportional to $\mathsf{W}_{1}+\mathsf{W}_{2}$ and the proportionality constant can be gauged to $1$. 
\beq 
\mathsf{W} = \mathsf{W}_{1}+\mathsf{W}_{2} &=& \left( \ba{ccccc} & \sT^{\top}\cdot \sh_{\HtbasisB} \cdot \bC \cdot \DevenB_{1} \cdot \sg_{\Hbasis} \cdot \sU & + & \sU^{\top}\cdot \sh_{\Hbasis}\cdot \t{\bC}\cdot \DoddB_{2} \cdot \sg_{\HtbasisB} \cdot \sT \\
+& \sT^{\top}\cdot \sh_{\HtbasisB} \cdot \bC^{\prime}\cdot \DevenB_{1}^{\prime} \cdot \sg_{\Hbasis} \cdot \sU &+& \sU^{\top}\cdot \sh_{\Hbasis}\cdot \t{\bC}^{\prime}\cdot \DoddB_{2}^{\prime} \cdot \sg_{\HtbasisB} \cdot \sT \ea \right)
\eeq
Given the fact $\merge^{2} = \textrm{Id}$ by construction we have that $\mathsf{W} = \mathsf{W}_{1} + \merge(\mathsf{W}_{1})$ is $\merge$ invariant and therefore the two moduli spaces are treated in the same manner. Comparing the scalar product expression for $\mathsf{W}$ with (\ref{eqn:ID}) we can see that the two pairs of terms, relating to primed and non-primed flux matrices, suggest we consider a pair of matrices where there is no mixing between the primed and non-primed flux matrices.
\beq 
\ba{ccccccccccc}
\u{\u{\fD}} &=& \bpm 0 & \DevenB_{1} \\ \DoddB_{2} & 0\epm&=& \bpm \mathbb{I} & 0 \\ 0 & 0 \epm\bpm 0 & \DevenB_{1} \\ \DoddB_{1} & 0\epm&+& \bpm 0 & 0 \\ 0 & \mathbb{I} \epm\bpm 0 & \DevenB_{2} \\ \DoddB_{2} & 0 \epm &\equiv& P^{+}\cdot\u{\u{\sD_{1}}}+P^{-}\cdot\u{\u{\sD_{2}}} \\
\u{\u{\fDp}} &=& \bpm 0 & \DevenB_{1}^{\prime} \\ \DoddB_{2}^{\prime} & 0 \epm &=& \bpm \mathbb{I} & 0 \\ 0 & 0 \epm\bpm 0 & \DevenB_{1}^{\prime} \\ \DoddB_{1}^{\prime} & 0 \epm &+& \bpm 0 & 0 \\ 0 & \mathbb{I} \epm\bpm 0 & \DevenB_{2}^{\prime} \\ \DoddB_{2}^{\prime} & 0\epm&\equiv& P^{+}\cdot\u{\u{\sDp_{1}}}+P^{-}\cdot\u{\u{\sDp_{2}}}
\ea\nn
\eeq
Due to the non-trivial mixing between the NS-NS and R-R sectors in (\ref{eqn:Swappedmoduli1}) and (\ref{eqn:Swappedmoduli2}) the distinction between the two flux sectors is no longer a simple one but with $\Psi = \u{\Psi}^{\top}\cdot \sh \cdot \BasisX = \mho+\Omega$ we are able to express the superpotential in a way which treats the two moduli spaces in the same manner, using $\fD_{\circ}\fC(\Psi) \equiv \fD\big(\fC(\Psi)\big) = \u{\Psi}^{\top} \cdot \fC \cdot \u{\u{\fD}} \cdot \BasisX$.
\beq 
W &=& \u{\Psi}^{\top} \cdot \sh \cdot \left( \fC \cdot \u{\u{\fD}} + \fC^{\prime} \cdot \u{\u{\fDp}} \right) \cdot \sg \cdot \u{\Psi} \nn \\
&=& \sg\Big( \Psi \,,\, (\fD_{\circ}\fC + \fDp_{\circ}\fC^{\prime})(\Psi) \Big) \nn \\
&=& \int_{\W} \Big( \mho+\Omega \Big) \wedge \Big(\fD_{\circ}\fC + \fDp_{\circ}\fC^{\prime}\Big) \Big( \mho+\Omega \Big)
\eeq
In our examination of S duality we found it convenient to consider the invariance of $\bC \cdot \DevenB + \bCp \cdot \DpevenB = \uwave{\sF} - S\, \uwave{\sFh}$, from which we could deduce the S duality transformation properties of the fluxes and (\ref{eqn:FFhMN}). Now that we have combined the two moduli spaces we can extend this further.
\beq
\fC \cdot \u{\u{\fD}} + \fC^{\prime} \cdot \u{\u{\fDp}} \equiv \u{\u{\F}} - S \, \u{\u{\Fh}} \quad \Rightarrow \quad \bpm \fC \\ \fC^{\prime} \epm \cdot \bpm \u{\u{\fD}} & \u{\u{\fDp}} \epm = \bpm 1 \\  -S \epm \cdot \bpm \u{\u{\sF}} & \u{\u{\sFh}} \epm 
\eeq
With $\fC = \bC \otimes \mathbb{I}_{2}$ and likewise for $\fC^{\prime}$ the same relationship between the matrices as (\ref{eqn:FFhMN}) occurs, which can be rephrased as (\ref{eqn:FtoM}) except that the dimensions of the matrices have increased so we denote $\A \otimes \mathbb{I}_{2}$ by $\sA$ and likewise for $\sB$.
\beq
\bpm \u{\u{\sF}} \\ \u{\u{\sFh}} \epm = \bpm \sA & \sB \\ \sB & \sA \epm \bpm \u{\u{\fD}} \\ \u{\u{\fDp}} \epm \quad \Rightarrow \quad \bpm \u{\u{\fD}} \\ \u{\u{\fDp}} \epm = \bpm \sA & \sB \\ \sB & \sA \epm\bpm \u{\u{\sF}} \\ \u{\u{\sFh}} \epm
\eeq
This allows us to much more succinctly state the S duality transformation properties of the fluxes in this moduli symmetric formulation under $S \to \frac{aS+b}{cS+d}$, in line with (\ref{eqn:MunderS}).
\beq
\bpm \u{\u{\fD}} \\ \u{\u{\fDp}} \epm &\to& \bpm \sA & \sB \\ \sB & \sA \epm\bpm a & b \\ c & d \epm\bpm \sA & \sB \\ \sB & \sA \epm \bpm \u{\u{\fD}} \\ \u{\u{\fDp}} \epm
\eeq
This result combined with the flux matrix definitions of $\u{\u{\fD}}$ and $\u{\u{\fDp}}$ and the $\mathbb{I}_{2}$ term in $\sA$ and $\sB$ again illustrates that the $\DevenB_{1}^{(\prime)}$ and $\DoddB_{2}^{(\prime)}$ have equivalent $\SL_{S}$ transformations, as noted in (\ref{eqn:IltS}) and required by definition (\ref{eqn:DefIlt}). With this definition of $\u{\u{\sF}}$ and $\u{\u{\sFh}}$ we can reduce the superpotential down to a simple form.
\beq 
W &=& \u{\Psi}^{\top} \cdot \sh \cdot \left( \u{\u{\sF}} - S \, \u{\u{\sFh}} \right) \cdot \sg \cdot \u{\Psi}
\eeq
This formulation makes S duality transformation properties and the symmetry in moduli treatment manifest.

\section{The $\ZZ$ orientifold}
\label{sec:ZZ}

In their discussion of non-geometric fluxes induced by T duality \cite{Shelton:2005cf,Shelton:2006fd,Wecht:2007wu} use the isotropic case of the $\ZZ$ orientifold, henceforth $\Mz$, as their explict example for constructing non-geometric vacua and this is extended to include S duality induced constraints in both Type IIA and Type IIB on the non-isotropic space in \cite{Aldazabal:2006up}, with particular S and T duality invariant vacua constructed in the isotropic case. The non-geometric fluxes have been analysed in terms of Lie algebras for cases involving only T duality \cite{Aldazabal:2008aa} and cases where S duality is included \cite{Guarino:2008ik}, where the latter demonstrates how non-geometric flux constraints can be examined through the use of integrability conditions. The two $\bZ_{2}$ orbifold groups factorise $\Mz$ into three two dimensional tori $\bT^{6} \to \otimes_{n=1}^{3} \bT_{n}^{2}$ and thus inherits the property of being its own mirror dual from the two dimensional tori.

\subsection{Definitions}

The orbifold symmetries are such that any $3$-form must have an index on each torus and from this we construct our symplectic basis given in Table \ref{table:H3}. From the definition of the $\Hn{3}$ basis the coefficients of the holomorphic $3$-form $\Omega$ can be written in terms of the period matrix entries.
\beq
\Omega = dz_{1} \wedge dz_{2} \wedge dz_{3} &=& (\eta^{1}+\tau_{1}\eta^{2}) \wedge (\eta^{3}+\tau_{2}\eta^{4}) \wedge (\eta^{5}+\tau_{3}\eta^{6})  \nn \\
&=&  \a{0} + \tau_{i}\a{i} + \frac{\tau_{1}\tau_{2}\tau_{3}}{\tau_{j}}\b{j} - \tau_{1}\tau_{2}\tau_{3}\b{0}  \nn \\
&=&  \syma{0} - \tau_{i}\symb{i} + \frac{\tau_{1}\tau_{2}\tau_{3}}{\tau_{j}}\symb{j} - \tau_{1}\tau_{2}\tau_{3}\symb{0}   \nn
\eeq
The action of the orbifold groups are such that only those $2n$-forms with indices on $n$ tori are allowed and we define the even dimensional cohomology bases in Table \ref{table:Ht3} to be as simple as possible.
\beq
\J \equiv B+iJ = T_{1}\,\eta^{12} + T_{2}\,\eta^{34} + T_{3}\,\eta^{56} = T_{1}\w{1} + T_{2}\w{2} + T_{3}\w{3} = T_{1}\vt{1} + T_{2}\vt{2} + T_{3}\vt{3}\nn
\eeq
The K\"{a}hler moduli holomorphic form can also be defined and takes the same general form as in the complex structure case, but for the sign changes in half the terms.
\beq
\mho = e^{\J} &=& \w{0} + T_{a}\w{a} + \frac{T_{1}T_{2}T_{3}}{T_{b}}\wt{b} + T_{1}T_{2}T_{3}\wt{0} \nn \\
&=& \v{0} + T_{a}\vt{a} + \frac{T_{1}T_{2}T_{3}}{T_{b}}\v{b} + T_{1}T_{2}T_{3}\vt{0} \nn 
\eeq
\begin{table}[htb]
\beq
\ba {ccccccccccc}\hline\hline
\quad \a{0} \gapA \a{1} \gapA \a{2} \gapA \a{3} \gapA -\b{0} \gapA \b{1} \gapA \b{2} \gapA \b{3} \quad \\ \hline 
\quad \eta^{135} \gapA \eta^{235} \gapA \eta^{145} \gapA \eta^{136} \gapA \eta^{246} \gapA \eta^{146} \gapA \eta^{236} \gapA \eta^{245}  \quad \\  \hline
\quad \syma{0} \gapA -\symb{1} \gapA -\symb{2} \gapA -\symb{3} \gapA -\symb{0} \gapA \syma{1} \gapA \syma{2} \gapA \syma{3} \quad \\  \hline\hline
\ea  \nn
\eeq
\caption{Sympletic bases of $\Hn{3}$.}
\label{table:H3}
\end{table}
\begin{table}[htb!]
\beq
\ba {ccccccccccc}\hline\hline
\quad \w{0} \gapA \w{1} \gapA \w{2} \gapA \w{3} \gapA \wt{0} \gapA \wt{1} \gapA \wt{2} \gapA \wt{3} \quad \\ \hline
\quad 1 \gapA \eta^{12} \gapA \eta^{34} \gapA \eta^{56} \gapA \eta^{123456} \gapA \eta^{3456} \gapA \eta^{1256} \gapA \eta^{1234}  \quad \\ \hline
\quad \v{0} \gapA \vt{1} \gapA \vt{2} \gapA \vt{3} \gapA \vt{0} \gapA \v{1} \gapA \v{2} \gapA \v{3} \quad \\ \hline \hline
\ea  \nn
\eeq
\caption{Commutitative bases of $\Hnt{3}$.}
\label{table:Ht3}
\end{table}
\subsection{Alternate derivative actions}

Given an explicit basis we are able to consider the equivalence of the expressions in (\ref{eqn:CliffShort}). In the case of the terms responsible for the sequence $\cHn{1,1} \rightarrow \Hn{3} \rightarrow \cHn{2,2}$ we consider the case of those terms which are responsible for $\w{1} \to \a{0}$ and $\w{1} \to \b{0}$.
\beq
\ba {rrrrrrrrr}
 \eta^{35}\i_{1}  &\quad:\quad& \w{1}&=&\eta^{12} &\quad\to\quad&  \eta^{235} &=& \a{1} \\
-\eta^{46}\i_{2}  &\quad:\quad& \w{1}&=&\eta^{12} &\quad\to\quad&  \eta^{146} &=& \b{1} 
\ea  \nn
\eeq
In general for terms whose action on $\Hnt{3}$ is $\w{a} \to \a{I}$ the corresponding action on $\Hn{3}$ will be to map $\b{I}$ to some element in $\cHn{2,2}$. Hence we apply $-\eta^{46}\i_{1}$ to $\a{1}$ and $-\eta^{35}\i_{2}$ to $\b{1}$.
\beq
\ba {rrrrrrrrr}
 \eta^{35}\i_{1} &\quad:\quad& \b{1} &=& \eta^{146} &\quad\to\quad& -\eta^{3456} &=& -\wt{1} \\
-\eta^{46}\i_{2} &\quad:\quad& \a{1} &=& \eta^{235} &\quad\to\quad&  \eta^{3456} &=&  \wt{1} 
\ea  \nn
\eeq
We therefore have two ways of expressing these $\eta^{ab}\i_{c}$ operators in terms of the cohomology bases and Tables \ref{table:H3} and \ref{table:Ht3} allow for conversion into the alternative basis.
\beq
\ba {rrrrrrrrrrrrrrrr}
 \eta^{35}\i_{1} &=& \a{1}\i_{\w{1}} &=& -\wt{1}\i_{\b{1}} &=& -\symb{1}\i_{\vt{1}} &=& -\v{1}\i_{\syma{1}} \\
-\eta^{46}\i_{2} &=& \b{1}\i_{\w{1}} &=&  \wt{1}\i_{\a{1}} &=& \syma{1}\i_{\vt{1}} &=&  -\v{1}\i_{\symb{1}} 
\ea  \nn
\eeq
The other cases for $\syma{i}$ and $\symb{j}$ follow the same pattern, it is straightforward to see that acting the operators onto other elements of $\cHn{1,1}$ and $\Hn{3}$ are zero and so we obtain half of the results given in (\ref{eqn:CliffShort}). The second half we obtain by considering those terms  which are responsible for $\cHn{2,2} \rightarrow \Hn{3} \rightarrow \cHn{1,1}$.
\beq
\ba {rrrrrrrrrrrrrrrrrrr}
-\eta^{1}\i_{5}\i_{3}  &\quad:\quad& \wt{1}&=&\eta^{3456} &\quad\to\quad& \eta^{146} &=& \b{1} &\quad:\quad& \a{1} &=& \eta^{235} &\quad\to\quad& -\eta^{12} &=& -\w{1} \\
 \eta^{2}\i_{6}\i_{4}  &\quad:\quad& \wt{1}&=&\eta^{3456} &\quad\to\quad& \eta^{235} &=& \a{1} &\quad:\quad& \b{1} &=& \eta^{146} &\quad\to\quad& \eta^{12} &=&  \w{1}  \\
\ea  \nn
\eeq
We therefore have two ways of expressing these $\eta^{ab}\i_{c}$ operators in terms of the cohomology bases and Tables \ref{table:H3} and \ref{table:Ht3} allow for conversion into the alternative basis.
\beq
\ba {rrrrrrrrrrrrrrrr}
-\eta^{1}\i_{5}\i_{3} &=& \b{1}\i_{\wt{1}} &=& -\w{1}\i_{\a{1}} &=&  \syma{1}\i_{\v{1}} &=& \vt{1}\i_{\symb{1}} \\
 \eta^{2}\i_{6}\i_{4} &=& \a{1}\i_{\wt{1}} &=&  \w{1}\i_{\b{1}} &=& -\symb{1}\i_{\v{1}} &=& \vt{1}\i_{\syma{1}} 
\ea  \nn
\eeq
The other cases for $\syma{i}$ and $\symb{j}$ follow the same pattern and give the second half of the results in (\ref{eqn:CliffShort}).

\subsection{Flux components}

There are a number of different flux structures we have considered.
\begin{itemize}
\item The Type IIA NS-NS fluxes $\F$ given in Table \ref{table:Alternatefluxes2} obtained from (\ref{eqn:TypeIIA fluxes}).
\item The Type IIA R-R flux $\fF_{0}$ given in Table \ref{table: IIA RR} obtained from (\ref{eqn:TypeIIARR}).
\begin{table}[h!]
\setlength\arraycolsep{5.0pt}
\beq
\ba{ccc|ccccccccc|ccccccccc|ccc}\hline\hline
& \gapa\gapa \fF_{(0)0} \gapa \fF_{(0)1} \gapa \fF_{(0)2} \gapa \fF_{(0)3} \gapa\gapa \fF_{(0)}^{\phantom{(0)}0} \gapa \fF_{(0)}^{\phantom{(0)}1} \gapa \fF_{(0)}^{\phantom{(0)}2} \gapa \fF_{(0)}^{\phantom{(0)}3} \gapa\gapa\gapa \\
\hline
& \fF_{(0)A} \gapa\gapa -\F^{135} \gapa -\F^{1}_{46} \gapa -\F^{3}_{62} \gapa -\F^{5}_{24} \gapa\gapa \F_{246} \gapa \F^{35}_{2} \gapa \F^{51}_{4} \gapa \F^{13}_{6} \gapa\gapa \fF_{(0)}^{\phantom{(0)}B} \gapa \\
\hline \hline
\ea \nn
\eeq
\caption{Component labels for the fluxes of $F_{RR}$}
\label{table: IIA RR}
\end{table}
\item The Type IIB NS-NS fluxes $\sF_{n}$ and $\sFh_{m}$ given in Table \ref{table:IIBNSNS} from (\ref{eqn : Different IIB}). We set $L$ and $K$ equal to the canonical sympletic forms of their respective dimensions and determine the fluxes of $\sD$ in terms of $\mathbb{G}$ by equating K\"{a}hler moduli coefficients.
\beq
\ba{ccccccccc}
\mathbb{G}(\cT_{0}\w{0}) &=& \sD(\cT_{0}\v{0})   &\qquad& \mathbb{G}(-\cT_{i}\wt{i}) &=& \sD(\cT_{i}\vt{i}) \\
\mathbb{G}(\cT^{0}\wt{0}) &=& \sD(\cT^{0}\vt{0}) &\qquad& \mathbb{G}( \cT^{j} \w{j}) &=& \sD(\cT^{j}\v{j}) \\
\ea
\eeq
This choice results in those fluxes coupling to $\cT_{i}$ acting on $\J_{c} = -\cT_{i}\wt{i}$. The redefinition of the sympletic basis simplifies our analysis and to illustrate this we consider the first case $\mathbb{G}(\cT_{0}\w{0}) = \sD(\cT_{0}\v{0})$ and what it reduces to.
\beq
\frac{1}{3!}\sFh_{pqr}\eta^{pqr} = \sF_{(0)A}\a{A}-\sF_{(0)}^{\phantom{(0)}B}\b{B}
\eeq
The redefinition of the sympletic basis reverts our notation back to the $(\a{A},\b{B})$ basis and the relationship between the $\sFh_{pqr}$ and the $\sF_{(0)A}$ etc follow from comparing coefficients.
\item The Type IIB alternate fluxes in the $\sD_{2}$ of $\mathsf{W}_{2}$ given in Table \ref{table:Alternatefluxes3}. The fluxes associated to $\mathsf{W}_{2}$ do not allow for them to be written in terms of coefficients in the same manner as the $\F$ and $\Fh$. They lack a simple tensor formulation such as $\F_{abc}\eta^{abc}$, a problem also encountered with the R-R fluxes of Type IIA in Table \ref{table: IIA RR}. 
\end{itemize}
\begin{table}
\setlength\arraycolsep{5.0pt}
\beq
\ba{ccc|ccccccccc|ccccccccc|ccc} \hline\hline
& \gapa\gapa \F_{(0)0} \gapa  \F_{(0)1} \gapa \F_{(0)2} \gapa \F_{(0)3} \gapa\gapa \F_{(0)}^{\phantom{(0)}0} \gapa \F_{(0)}^{\phantom{(0)}1} \gapa \F_{(0)}^{\phantom{(0)}2} \gapa \F_{(0)}^{\phantom{(0)}3} \gapa\gapa\gapa \\
\hline
& \F_{(0)I} \gapa\gapa \F_{135} \gapa \F_{146} \gapa \F_{236} \gapa \F_{245} \gapa\gapa  \F_{246} \gapa \F_{235} \gapa \F_{145} \gapa \F_{136} \gapa\gapa \F_{(0)}^{\phantom{(0)}J}\gapa\\
\hline \hline
\\
\hline\hline
& \gapa\gapa \F^{(a)}_{\phantom{(a)}0} \gapa \F^{(a)}_{\phantom{(a)}1} \gapa \F^{(a)}_{\phantom{(a)}2} \gapa \F^{(a)}_{\phantom{(a)}3} \gapa\gapa \F^{(a)0} \gapa \F^{(a)1} \gapa \F^{(a)2} \gapa \F^{(a)3} \gapa\gapa\gapa \\
\hline
& \F^{(1)}_{\phantom{(a)}I} \gapa\gapa -\F^{2}_{35} \gapa -\F^{2}_{46} \gapa +\F^{1}_{36} \gapa +\F^{1}_{45} \gapa\gapa +\F^{1}_{46} \gapa +\F^{1}_{35} \gapa -\F^{2}_{45} \gapa -\F^{2}_{36} \gapa\gapa \F^{(1)J}\gapa \\
& \F^{(2)}_{\phantom{(a)}I} \gapa\gapa -\F^{4}_{51} \gapa +\F^{3}_{61} \gapa -\F^{4}_{62} \gapa +\F^{3}_{52} \gapa\gapa +\F^{3}_{62} \gapa -\F^{4}_{52} \gapa +\F^{3}_{51} \gapa -\F^{4}_{61} \gapa\gapa \F^{(2)J}\gapa\\
& \F^{(3)}_{\phantom{(a)}I} \gapa\gapa -\F^{6}_{13} \gapa +\F^{5}_{14} \gapa +\F^{5}_{23} \gapa -\F^{6}_{24} \gapa\gapa +\F^{5}_{24} \gapa -\F^{6}_{23} \gapa -\F^{6}_{14} \gapa +\F^{5}_{13} \gapa\gapa \F^{(3)J}\gapa\\
\hline \hline
\\
\hline\hline
& \gapa\gapa \F^{(0)}_{\phantom{(0)}0} \gapa \F^{(0)}_{\phantom{(0)}1} \gapa \F^{(0)}_{\phantom{(0)}2} \gapa \F^{(0)}_{\phantom{(0)}3} \gapa\gapa \F^{(0)0} \gapa \F^{(0)1} \gapa \F^{(0)2} \gapa \F^{(0)3} \gapa\gapa\gapa \\
\hline
& \F^{(0)}_{\phantom{(0)}I} \gapa\gapa \F^{246} \gapa \F^{235} \gapa \F^{145} \gapa \F^{136} \gapa\gapa \F^{135} \gapa \F^{146} \gapa \F^{236} \gapa \F^{245} \gapa\gapa \F^{(0)J}\gapa \\
\hline \hline
\\
\hline\hline
& \gapa\gapa \F_{(a)0} \gapa \F_{(a)1} \gapa \F_{(a)2} \gapa \F_{(a)3} \gapa\gapa \F_{(a)}^{\phantom{(a)}0} \gapa \F_{(a)}^{\phantom{(a)}1} \gapa \F_{(a)}^{\phantom{(a)}2} \gapa \F_{(a)}^{\phantom{(a)}3} \gapa\gapa\gapa \\
\hline
& \F_{(1)I} \gapa\gapa -\F_{1}^{46} \gapa -\F_{1}^{35} \gapa +\F_{2}^{45} \gapa +\F_{2}^{36} \gapa\gapa -\F_{2}^{35} \gapa -\F_{2}^{46} \gapa +\F_{1}^{36} \gapa +\F_{1}^{45} \gapa\gapa \F_{(1)}^{\phantom{(1)}J}\gapa\\
& \F_{(2)I} \gapa\gapa -\F_{3}^{62} \gapa +\F_{4}^{52} \gapa -\F_{3}^{51} \gapa +\F_{4}^{61} \gapa\gapa -\F_{4}^{51} \gapa +\F_{3}^{61} \gapa -\F_{4}^{62} \gapa +\F_{3}^{52} \gapa\gapa \F_{(2)}^{\phantom{(2)}J}\gapa\\
& \F_{(3)I} \gapa\gapa -\F_{5}^{24} \gapa +\F_{6}^{23} \gapa +\F_{6}^{14} \gapa -\F_{5}^{13} \gapa\gapa -\F_{6}^{13} \gapa +\F_{5}^{14} \gapa +\F_{5}^{23} \gapa -\F_{6}^{24} \gapa\gapa \F_{(3)}^{\phantom{(3)}J}\gapa\\
\hline \hline
\ea \nn
\eeq
\caption{Alternate component labels for fluxes $\F_{n}$}
\label{table:Alternatefluxes2}
\end{table}
\begin{table}
\setlength\arraycolsep{5.0pt}
\beq
\ba{ccc|ccccccccc|ccccccccc|ccc} \hline\hline
& \gapa\gapa \sF_{(0)0} \gapa  \sF_{(0)1} \gapa \sF_{(0)2} \gapa \sF_{(0)3} \gapa\gapa \sF_{(0)}^{\phantom{(0)}0} \gapa \sF_{(0)}^{\phantom{(0)}1} \gapa \sF_{(0)}^{\phantom{(0)}2} \gapa \sF_{(0)}^{\phantom{(0)}3} \gapa\gapa\gapa \\
\hline
& \sF_{(0)A} \gapa\gapa \sFh_{135} \gapa \sFh_{235} \gapa \sFh_{145} \gapa \sFh_{136} \gapa\gapa  \sFh_{246} \gapa -\sFh_{146} \gapa -\sFh_{236} \gapa -\sFh_{245} \gapa\gapa \sF_{(0)}^{\phantom{(0)}B}\gapa\\
\hline \hline
\\
\hline\hline
& \gapa\gapa \sF_{(i)0} \gapa \sF_{(i)1} \gapa \sF_{(i)2} \gapa \sF_{(i)3} \gapa\gapa \sF_{(i)}^{\phantom{(i)}0} \gapa \sF_{(i)}^{\phantom{(i)}1} \gapa \sF_{(i)}^{\phantom{(i)}2} \gapa \sF_{(i)}^{\phantom{(i)}3} \gapa\gapa\gapa \\
\hline
& \sF_{(1)A} \gapa\gapa -\sF^{2}_{35} \gapa +\sF^{1}_{35} \gapa -\sF^{2}_{45} \gapa -\sF^{2}_{36} \gapa\gapa +\sF^{1}_{46} \gapa +\sF^{2}_{46} \gapa -\sF^{1}_{36} \gapa -\sF^{1}_{45} \gapa\gapa \sF_{(1)}^{\phantom{(1)}B} \gapa \\
& \sF_{(2)A} \gapa\gapa -\sF^{4}_{51} \gapa -\sF^{4}_{52} \gapa +\sF^{3}_{51} \gapa -\sF^{4}_{61} \gapa\gapa +\sF^{3}_{62} \gapa -\sF^{3}_{61} \gapa +\sF^{4}_{62} \gapa -\sF^{3}_{52} \gapa\gapa \sF_{(2)}^{\phantom{(2)}B} \gapa\\
& \sF_{(3)A} \gapa\gapa -\sF^{6}_{13} \gapa -\sF^{6}_{23} \gapa -\sF^{6}_{14} \gapa +\sF^{5}_{13} \gapa\gapa +\sF^{5}_{24} \gapa -\sF^{5}_{14} \gapa -\sF^{5}_{23} \gapa +\sF^{6}_{24} \gapa\gapa \sF_{(3)}^{\phantom{(3)}B} \gapa\\
\hline \hline
\\
\hline\hline
& \gapa\gapa \sF^{(0)}_{\phantom{(0)}0} \gapa \sF^{(0)}_{\phantom{(0)}1} \gapa \sF^{(0)}_{\phantom{(0)}2} \gapa \sF^{(0)}_{\phantom{(0)}3} \gapa\gapa \sF^{(0)0} \gapa \sF^{(0)1} \gapa \sF^{(0)2} \gapa \sF^{(0)3} \gapa\gapa\gapa \\
\hline
& \sF^{(0)}_{\phantom{(0)}A} \gapa\gapa \sFh^{246} \gapa -\sFh^{146} \gapa -\sFh^{236} \gapa -\sFh^{245} \gapa\gapa -\sFh^{135} \gapa -\sFh^{235} \gapa -\sFh^{145} \gapa -\sFh^{136} \gapa\gapa \sF^{(0)B}\gapa \\
\hline \hline
\\
\hline\hline
& \gapa\gapa \sF^{(i)}_{\phantom{(i)}0} \gapa \sF^{(i)}_{\phantom{(i)}1} \gapa \sF^{(i)}_{\phantom{(i)}2} \gapa \sF^{(i)}_{\phantom{(i)}3} \gapa\gapa \sF^{(i)0} \gapa \sF^{(i)1} \gapa \sF^{(i)2} \gapa \sF^{(i)3} \gapa\gapa\gapa \\
\hline
& \sF^{(1)}_{\phantom{(i)}A} \gapa\gapa +\sF_{1}^{46} \gapa +\sF_{2}^{46} \gapa -\sF_{1}^{36} \gapa -\sF_{1}^{45} \gapa\gapa +\sF_{2}^{35} \gapa -\sF_{1}^{35} \gapa +\sF_{2}^{45} \gapa +\sF_{2}^{36} \gapa\gapa \sF^{(1)B} \gapa\\
& \sF^{(2)}_{\phantom{(i)}A} \gapa\gapa +\sF_{3}^{62} \gapa -\sF_{3}^{61} \gapa +\sF_{4}^{62} \gapa -\sF_{3}^{52} \gapa\gapa +\sF_{4}^{51} \gapa +\sF_{4}^{52} \gapa -\sF_{3}^{51} \gapa +\sF_{4}^{61} \gapa\gapa \sF^{(2)B} \gapa\\
& \sF^{(3)}_{\phantom{(i)}A} \gapa\gapa +\sF_{5}^{24} \gapa -\sF_{5}^{14} \gapa -\sF_{5}^{23} \gapa +\sF_{6}^{24} \gapa\gapa +\sF_{6}^{13} \gapa +\sF_{6}^{23} \gapa +\sF_{6}^{14} \gapa -\sF_{5}^{13} \gapa\gapa \sF^{(3)B} \gapa\\
\hline \hline
\ea \nn
\eeq
\caption{Alternate component labels for fluxes $\sF_{n}$ and $\sFh_{m}$.}
\label{table:IIBNSNS}
\end{table}
\begin{table}
\beq
\ba{ccc|ccccccccc|ccccccccc|ccc} \hline\hline
\,& &\,&\,& \fF_{(0)0} && \fF_{(0)1} && \fF_{(0)2} && \fF_{(0)3} &&& \fF_{(0)}^{\phantom{(0)}0} && \fF_{(0)}^{\phantom{(0)}1} && \fF_{(0)}^{\phantom{(0)}2} && \fF_{(0)}^{\phantom{(0)}3} &\,&\,&\,&\, \\
\hline
\,& \fF_{(0)I} &\,&\,& +\sFh^{135} &\,& -\sFh_{2}^{35} &\,& -\sFh_{4}^{51} &\,& -\sFh_{6}^{13} &\,&\,& -\sFh_{246} &\,& 
-\sFh^{1}_{46} &\,& -\sFh^{3}_{62} &\,& -\sFh^{5}_{24} &\,&\,& \fF_{(0)}^{\phantom{(0)}J}\\
\hline \hline
\\
\hline\hline
\,& &\,&\,& \fF_{(a)0} && \fF_{(a)1} && \fF_{(a)2} && \fF_{(a)3} &&& \fF_{(a)}^{\phantom{(a)}0} && \fF_{(a)}^{\phantom{(a)}1} && \fF_{(a)}^{\phantom{(a)}2} && \fF_{(a)}^{\phantom{(a)}3} &\,&\,&\,&\, \\
\hline
\,& \fF_{(1)I} &\,&\,& +\sF^{235} &\,& +\sF_{1}^{35} &\,& -\sF_{4}^{52} &\,& -\sF_{6}^{23} &\,&\,& +\sF_{235} &\,& -\sF^{2}_{46} &\,& +\sF^{3}_{61} &\,& +\sF^{5}_{14} &\,&\,& \fF_{(1)}^{\phantom{(1)}J}\\
\,& \fF_{(2)I} &\,&\,& +\sF^{145} &\,& -\sF_{2}^{45} &\,& +\sF_{3}^{51} &\,& -\sF_{6}^{14} &\,&\,& +\sF_{145} &\,& +\sF^{1}_{36} &\,& -\sF^{4}_{62} &\,& +\sF^{5}_{23} &\,&\,& \fF_{(2)}^{\phantom{(2)}J}\\
\,& \fF_{(3)I} &\,&\,& +\sF^{136} &\,& -\sF_{2}^{36} &\,& -\sF_{4}^{61} &\,& +\sF_{5}^{13} &\,&\,& +\sF_{136} &\,& +\sF^{1}_{45} &\,& +\sF^{3}_{52} &\,& -\sF^{6}_{24} &\,&\,& \fF_{(3)}^{\phantom{(3)}J}\\
\hline \hline
\\
\hline\hline
\,& &\,&\,& \fF^{(0)}_{\phantom{(0)}0} && \fF^{(0)}_{\phantom{(0)}1} && \fF^{(0)}_{\phantom{(0)}2} && \fF^{(0)}_{\phantom{(0)}3} &&& \fF^{(0)0} && \fF^{(0)1} && \fF^{(0)2} && \fF^{(0)3} &\,&\,&\,&\, \\
\hline
\,& \fF^{(0)}_{\phantom{(0)}I} &\,&\,& -\sFh^{246} &\,& -\sFh_{1}^{46} &\,& -\sFh_{2}^{62} &\,& -\sFh_{5}^{24} &\,&\,& -\sFh_{135} &\,& +\sFh^{2}_{35} &\,& +\sFh^{4}_{51} &\,& +\sFh^{6}_{13} &\,&\,& \fF^{(0)J} \\
\hline \hline
\\
\hline\hline
\,& &\,&\,& \fF^{(b)}_{\phantom{(b)}0} && \fF^{(b)}_{\phantom{(b)}1} && \fF^{(b)}_{\phantom{(b)}2} && \fF^{(b)}_{\phantom{(b)}3} &&& \fF^{(b)0} && \fF^{(b)1} && \fF^{(b)2} && \fF^{(b)3} &\,&\,&\,&\, \\
\hline
\,& \fF^{(1)}_{\phantom{(1)}I} &\,&\,& +\sF^{146} &\,& -\sF_{2}^{46} &\,& +\sF_{3}^{61} &\,& +\sF_{5}^{14} &\,&\,& -\sF_{235} &\,& -\sF^{1}_{35} &\,& +\sF^{4}_{52} &\,& +\sF^{6}_{23} &\,&\,& \fF^{(1)J} \\
\,& \fF^{(2)}_{\phantom{(2)}I} &\,&\,& +\sF^{236} &\,& +\sF_{1}^{36} &\,& -\sF_{4}^{62} &\,& +\sF_{5}^{23} &\,&\,& -\sF_{145} &\,& +\sF^{2}_{45} &\,& -\sF^{3}_{51} &\,& +\sF^{6}_{14} &\,&\,& \fF^{(2)J}\\
\,& \fF^{(3)}_{\phantom{(3)}I} &\,&\,& +\sF^{245} &\,& +\sF_{1}^{45} &\,& +\sF_{3}^{52} &\,& -\sF_{6}^{24} &\,&\,& -\sF_{136} &\,& +\sF^{2}_{36} &\,& +\sF^{4}_{61} &\,& -\sF^{5}_{13} &\,&\,& \fF^{(3)J}\\
\hline \hline
\ea \nn
\eeq
\caption{Explicit $\merge$ defined components for fluxes $\fF_{n}$}
\label{table:Alternatefluxes3}
\end{table}

\section{Conclusions}

In this work we have seen that using the cohomology bases for the construction of the fluxes simplifies the analysis of T and S dualities considerably and naturally takes into account the properties of the internal space, rather than working with $\La$ and $\La\dual$. Using $\La$ and $\La\dual$ as the bases for the fluxes has the advantage that no additional work was required to find the action $\F_{n} : \Ln{3} \to \Ln{6-2n}$ if given $\F_{n} : \Ln{2n} \to \Ln{3}$ and the resultant nilpotency conditions are complete. However, this has the disadvantages that it does not make the number of independent fluxes on $\M$ manifest, the superpotential cannot be stated in terms of the components easily and the Type IIA R-R flux sector does not have a simple factorisation for its covariant derivative.  All of these issues were resolved by restricting our attention to the cohomology bases, which were sufficient to entirely describe the independent fluxes and moduli coefficients in the superpotentials of both Type IIA and Type IIB. As seen in the literature the price paid for this was that not all Bianchi constraints are captured by the formulation and due to the fact the fluxes broke the closure of the elements of $\Ha$ the distinction between the lightest modes and the massive modes was lost. The assumption that the cohomology bases are sufficient to describe the space could be justified if the $\Ha$ basis elements are eigenforms of the Laplacians of (\ref{eqn:Laplacian}) with small eigenvalues.
\\

We observed the differences and similarities between the two Type II constructions. The derivatives act on the holomorphic forms of different moduli spaces in Type IIA, thus preventing Type IIA from being self S-dual. Type IIB possesses self S duality but the manner in which the fluxes couple to the K\"{a}hler moduli is different from the Type IIA case. This fact required additional attention to ensure that when the Type IIB superpotential is written in terms of the natural holomorphic forms the Bianchi constraints of the resultant covariant derivatives are equal to those obtained from its Type IIA mirror dual. The tadpole constraints were reformalised into a representation similar to Bianchi constraints but with derivatives from each flux sector. In the case of Type IIB the inclusion of R-R partners to all T duality induced fluxes suggested an extension to the tadpole constraints which was further justified in the Type IIA construction.
\\

Having used the cohomology bases to provide a convenient notation to examine Type II superpotentials under combinations T and S dualities we noted that for particular $\M$ it is possible for reformulate the superpotential such that the roles of the two moduli spaces are exchanged. Such a symmetry is not clear without including all superpotential contributions under T and S dualities. The fluxes, constraints and multiplets of the alternative formulation have the same schematic form as the usual formulation and we have argued that for $\M=\W$ these two formulations are equally valid. In Type IIB the standard flux multiplets are defined as $\sD_{1}(\Jn{n})$, with T duality inducing contributions for each value of $n$, but the alternative set of fluxes were defined as $\sD_{2}(\fJn{n})$. Since the flux contributions for each $n$ are not synonymous with those of $\sD_{1}(\Jn{n})$ the duality which induces the sequence is not T duality, but one which we referred to as T$^{\prime}$ duality. Since the dualities induce different flux multiplet sequences they have different covariant derivatives, $\sD_{1}\neq \sD_{2}$, and we demonstrated that as a result their respective nilpotency constraints are not equivalent. This reformulation was not justified rigorously but rather based on arguments of symmetry. We considered the explicitly case of the $\ZZ$ orientifold whose factorisation into three two dimensional sub-tori results in the complex structure and K\"{a}hler moduli having analogous roles. There are a number choices in how this work might be extended and used :
\begin{itemize}

\item We found a large number of additional terms which could in principle contribute to tadpole constraints but how they do this and what the physical interpretation in terms of D-branes, NS-branes and O-planes was not considered. Whether this can be done on the effective supergravity level or requires the contributions to descend from a full string theory is not clear and in the latter case the fact some of the fluxes are non-geometric is a large obstruction as stringy constructions of spaces with non-geometric fluxes is an open problem at present.

\item We have only used symmetries on the level of the effective moduli theory to motivate the existence of $\merge$, rather than rigorously derived it from a string theory and a stringy proof to its existence (if it indeed exists) might lead to a greater understanding of particular internal spaces.

\item Assuming the existence of $\merge$ the additional constraints restrict the number of independent fluxes on the space and thus narrows down the number of possible vacua a particular $\M=\W$ might be able to have. The implications for this in terms of phenomenology and moduli stablisation would be worth investigating.

\item The enhanced symmetry group of $g = g_{\Hbasis} \oplus g_{\HtbasisB}$ compared to $g_{\Hbasis}$ and $g_{\HtbasisB}$ individually would further connect the NS-NS and R-R flux sectors and the implications this would have for the fluxes and possible vacua constructed from resultant superpotentials are not immediately clear. Constructions of the $\Hnt{3}$ basis which are manifestly symplectic \cite{ModuliSpace} require Grassman valued bases and how this affects flux compactifications is unknown.

\end{itemize}

The fluxes of the $\ZZ$ orientifold given in Section \ref{sec:ZZ} allow for these unresolved problems to be examined in more explicit detail and may lead to a more general understanding of flux compactifications.

\subsection*{Acknowledgements}

The author would like to thank Anthony Ashton and James Gray for useful discussions, as well as Southampton University for the support of a scholarship.

\appendix

\section{Moduli masses}
\label{Laplacian section}
If the approximation to include only `light' modes spanned by $\Ha$ is to be valid then the eigenvalues of the $\Ha$ bases in terms of the Laplacian are to be small. This is complicated by the fact we have constructed a pair of derivatives and so can build two different Laplacians, which is further complicated when S duality is included. To that end we construct the Laplacian operator in terms of the fluxes in $\D$ and $\Dp$, $\Delta_{\D} \equiv \D \, \D\ad + \D\ad \, \D$ using $\phi \in \Hnt{3}$ and $\chi \in \Hn{3}$ in Type IIA on $\M$.
\beq
\ba{ccccccc}
\D(\phi) &=& \u{\phi}^{\top} \cdot h_{\HtbasisB} \cdot \DevenA \cdot h_{\Hbasis} \cdot \BasisA &\qquad& \D(\chi) &=& \u{\chi}^{\top} \cdot h_{\Hbasis} \cdot \DoddA \cdot h_{\HtbasisB} \cdot \BasisC \\
\D\ad(\phi) &=& \u{\phi}^{\top} \cdot h_{\HtbasisB} \cdot \DevenA\ad \cdot h_{\Hbasis} \cdot \BasisA &\qquad& \D\ad(\chi) &=& \u{\chi}^{\top} \cdot h_{\Hbasis} \cdot \DoddA\ad \cdot h_{\HtbasisB} \cdot \BasisC
\ea \nn
\eeq
It should be noted that $\DevenA\ad$ and $\DoddA\ad$ are not the hermitian conjugates of $\DevenA$ and $\DoddA$ but the flux matrices associated to $\D\ad$ instead.
\beq
\ba{cccccccccccc}
\u{\phi}^{\top} \cdot h_{\HtbasisB} \cdot \DevenA \cdot \u{\chi} &=& \IP{ \chi , \D(\phi) } &\equiv& \IP{ \phi , \D\ad(\chi) } &=& \u{\chi}^{\top} \cdot h_{\Hbasis} \cdot \DoddA\ad \cdot \u{\phi} &\Rightarrow& \DoddA\ad &=& h_{\Hbasis} \cdot \DevenA^{\top} \cdot h_{\HtbasisB}\\
\u{\chi}^{\top} \cdot h_{\Hbasis} \cdot \DoddA \cdot \u{\phi} &=& \IP{ \phi , \D(\chi) } &\equiv& \IP{ \chi , \D\ad(\phi) } &=& \u{\phi}^{\top} \cdot h_{\HtbasisB} \cdot \DevenA\ad \cdot \u{\chi} &\Rightarrow& \DevenA\ad &=&  h_{\HtbasisB} \cdot \DoddA^{\top} \cdot h_{\Hbasis}
\ea
\eeq
From this the two terms of the Laplacian can be constructed for both $\Hn{3}$ and $\Hnt{3}$.
\beq
\ba{ccccccccc}
\D\ad \D(\phi) &=& \u{\phi}^{\top} \cdot h_{\HtbasisB} \cdot \DevenA \cdot h_{\Hbasis} \cdot \DoddA\ad \cdot h_{\HtbasisB} \cdot \BasisC &=& \u{\phi}^{\top} \cdot h_{\HtbasisB} \cdot \DevenA \cdot \DevenA^{\top} \cdot \BasisC \\
\D \D\ad(\phi) &=& \u{\phi}^{\top} \cdot h_{\HtbasisB} \cdot \DevenA\ad \cdot h_{\Hbasis} \cdot \DoddA \cdot h_{\HtbasisB} \cdot \BasisC &=& \u{\phi}^{\top} \cdot \DoddA^{\top} \cdot \DoddA \cdot h_{\HtbasisB} \cdot \BasisC \\
\D\ad \D(\chi) &=& \u{\chi}^{\top} \cdot h_{\Hbasis} \cdot \DoddA \cdot h_{\HtbasisB} \cdot \DevenA\ad \cdot h_{\Hbasis} \cdot \BasisA &=& \u{\chi}^{\top} \cdot h_{\Hbasis} \cdot \DoddA \cdot \DoddA^{\top} \cdot \BasisA \\
\D \D\ad(\chi) &=& \u{\chi}^{\top} \cdot h_{\Hbasis} \cdot \DoddA\ad \cdot h_{\HtbasisB} \cdot \DevenA \cdot h_{\Hbasis} \cdot \BasisA &=& \u{\chi}^{\top} \cdot \DevenA^{\top} \cdot \DevenA \cdot h_{\Hbasis} \cdot \BasisA 
\ea
\eeq
Using (\ref{eqn:ZetaofM}) the Laplacian for each flux sector can be written entirely in terms of the $\Hnt{3} \to \Hn{3}$ defined fluxes, in the same manner as the Bianchi constraints for T duality could.
\beq
\label{eqn:Laplacian}
\ba{ccl}
\IP{ \phi , (\D \D\ad+\D\ad \D)(\phi) } &=& \u{\phi}^{\top} \cdot h_{\HtbasisB} \cdot \Big( \DevenA \cdot \DevenA^{\top} + \DoddA^{\top} \cdot \DoddA \Big) \cdot \u{\phi} \\
&=& \u{\phi}^{\top} \cdot h_{\HtbasisB} \cdot \Big( \DevenA \cdot \DevenA^{\top} + \sg_{\Hbasis}\cdot\DevenA \cdot \DevenA^{\top}\cdot \sg_{\Hbasis}^{\top} \Big) \cdot \u{\phi} \\
\IP{ \chi , (\D \D\ad+\D\ad \D)(\chi) } &=& \u{\chi}^{\top} \cdot h_{\Hbasis} \cdot \Big( \DevenA^{\top} \cdot \DevenA + \DoddA \cdot \DoddA^{\top} \Big) \cdot \u{\chi} \\
&=& \u{\chi}^{\top} \cdot h_{\Hbasis} \cdot \Big( \DevenA^{\top} \cdot \DevenA + \sg_{\HtbasisB} \cdot \DevenA^{\top} \cdot \DevenA \cdot \sg_{\HtbasisB}^{\top} \Big) \cdot \u{\chi}
\ea
\eeq
Therefore if $\DevenA \cdot \DevenA^{\top} = 0 = \DevenA^{\top} \cdot \DevenA$ the $\Hn{p}$ are $\D$-harmonic in terms of the NS-NS fluxes of $\D$ being non-zero and the R-R sector has the same result but $\DevenA \to \DpevenA$. If both $\Delta_{\D}$ and $\Delta_{\Dp}$ have the $\Hn{p}$ are harmonic then we would expect the distinction between the moduli and the `heavy' fields to be enough to make the analysis of the fluxes using the cohomology bases consistent. 

\section{Interior form identities}

In order to derive the action of $\DoddA$ given $\DevenA$ we generalise the $\i_{\tau}(\eta^{\sigma}) = \delta^{\sigma}_{\tau}$ to the basis elements of $\Ha$ and its interior form dual.

\subsection{$\Hn{3}$ basis}

The basis of $\Hn{3}$ is defined to be sympletic, $g(\syma{I},\symb{J}) = \delta_{I}^{J}$ and from this basis we define a set of interior forms.
\beq
\Hn{3} = \la \syma{I} , \symb{J} \ra \quad,\quad \int_{\M}\syma{I}\wedge\symb{J} = \delta_{I}^{J} \quad,\quad \iota_{\syma{J}}(\syma{I}) = \delta_{I}^{J} = \iota_{\symb{I}}(\symb{J})
\eeq
The $3$-forms can be rewritten in terms of the general space of $3$-forms and a set of coefficient, $\syma{I} = (\syma{I})_{abc}\eta^{abc}$ and $\symb{J} = (\symb{J})_{ijk}\eta^{ijk}$, with the sympletic definition of the basis defining a set of constraints on these coefficients.
\beq
\delta_{I}^{J} = \int_{\M} \syma{I} \wedge \symb{J} = (\syma{I})_{abc}(\symb{J})_{ijk} \int_{\M}\eta^{abc}\eta^{ijk} = (\syma{I})_{abc}\epsilon^{abcijk}(\symb{J})_{ijk} 
\eeq
By the same methodology we can rewrite the interior forms in terms of $\iota_{abc}$ and sets of coefficients, $\iota_{\syma{I}} = (A^{I})^{abc}\iota_{cba}$ and $\iota_{\symb{J}} = (B_{J})^{ijk}\iota_{kji}$, which are also constrained in terms of the basis coefficients by their relationship with the sympletic forms.
\beq
\ba{ccccccccc}
\delta_{I}^{J} &=& \iota_{\syma{J}}(\syma{I}) &=& (A^{J})^{abc}(\syma{I})_{ijk}\iota_{cba}\eta^{ijk} &=& 3!(A^{J})^{abc}(\syma{I})_{abc} \\
\delta_{I}^{J} &=& \iota_{\symb{I}}(\symb{J}) &=& (B_{I})^{abc}(\symb{J})_{ijk}\iota_{cba}\eta^{ijk} &=& 3!(B_{I})^{abc}(\symb{J})_{abc} 
\ea
\eeq
Comparing the three coefficient expansions for $\delta_{I}^{J}$ we obtain the $(A^{J})^{abc}$ and $(B_{I})^{abc}$ in terms of the $(\syma{I})_{abc}$, $(\symb{J})_{abc} $ and the antisymmetric $\epsilon$.
\beq
(A^{J})^{abc} = \frac{1}{3!}\epsilon^{abcijk}(\symb{J})_{ijk} \quad , \quad (B_{I})^{abc} = \frac{1}{3!}\epsilon^{abcijk}(\syma{I})_{ijk} 
\eeq
With these explicit expressions for the coefficients we can construct $\iota_{\syma{J}}(\symb{I})$ and $\iota_{\symb{I}}(\syma{J})$ in terms of the $\syma{I}$ and $\symb{J}$ components.
\beq
\ba{ccccccccc}
\iota_{\syma{I}}(\symb{J}) &=& \frac{1}{3!}\epsilon^{abcijk}(\symb{I})_{ijk}\iota_{cba}(\symb{J})_{pqr}\eta^{pqr} &=& (\symb{I})_{ijk}\epsilon^{abcijk}(\symb{J})_{abc} \\
\iota_{\symb{J}}(\syma{I}) &=& \frac{1}{3!}\epsilon^{abcijk}(\syma{J})_{ijk}\iota_{cba}(\syma{I})_{pqr}\eta^{pqr} &=& (\syma{J})_{ijk}\epsilon^{abcijk}(\syma{I})_{abc}
\ea
\eeq
Converting the antisymmetric tensor back into an integral over $\M$ of the $6$-form $\eta^{abcdef}$ these expressions can be written entirely in terms of the symplectic basis.
\beq
\iota_{\syma{I}}(\symb{J}) = (\symb{I})_{ijk}\epsilon^{abcijk}(\symb{J})_{abc} = \int_{\M}(\symb{J})_{abc}\eta^{abc}(\symb{I})_{ijk}\eta^{ijk} = \int_{\M}\symb{J}\wedge \symb{I} = 0
\eeq
By the same method we obtain the second expression.
\beq
\iota_{\symb{J}}(\syma{I}) = (\syma{J})_{ijk}\epsilon^{abcijk}(\syma{I})_{abc} = \int_{\M}(\syma{I})_{abc} \eta^{abc}(\syma{J})_{ijk}\eta^{ijk} = \int_{\M}\syma{I}\wedge \syma{J} = 0
\eeq
With these integral expressions we can see the relationship between the $\iota_{\alpha}$ and $\iota_{\beta}$ and integrating over the $3$-cycles of $\M$.

\subsection{$\Hnt{3}$ basis}

In the case of $\Hnt{3}$ the $\w{a}$ and $\wt{b}$ are not in the same cohomologies and so such expressions as $\iota_{\w{a}}(\wt{b})$ do not reduce to scalar quantities. However particular expressions of interest can be obtained by expressing the $\wt{b}$ in terms of the $\w{a}$ as $\wt{a} \equiv f^{abc}\w{b}\wedge\w{c}$, with the $f^{abc}$ coefficients being related to the K\"{a}hler intersection numbers.
\beq
\delta_{a}^{b} = g_{a}^{\phantom{a}b} = \int_{\M}\w{a}\wedge\wt{b} = \int_{\M}\w{a}\wedge\w{c}\wedge\w{d}f^{bcd} = \kappa_{acd}f^{bcd}
\eeq
Using this allows for $\iota_{\w{a}}(\wt{b})$ and $\iota_{\wt{b}}(\w{a})$ to be simplified, once we make use of an additional identity relating interior forms and $p$-forms. Given a general $p$-form $\lambda$ we suppose it has factorisation $\lambda = \xi \wedge \zeta$ and consider their associated interior forms, under the map $\varphi : \lambda \to \iota_{\lambda}$.
\beq
\ba{cccccccccc}
\varphi &\quad : \quad& \lambda &=& \xi \wedge \zeta &\quad \to \quad& \iota_{\lambda} &=& \iota_{\zeta}\iota_{\xi} \\
& \Rightarrow& \wt{a} &=& f^{abc}\w{b} \wedge \w{c} &\quad \to \quad& \iota_{\wt{a}} &=& f^{abc}\iota_{\w{c}}\iota_{\w{b}} 
\ea
\eeq 
With this factorisation of $\wt{b}$ and the associated interior forms we can easily express the remaining combination of interior forms and $p$-forms, where we make use of the fact $f^{abc}=f^{acb}$.
\beq
\ba{ccccccccc}
\iota_{\w{a}}\wt{b} &=& \iota_{\w{a}}(f^{bcd}\w{c} \wedge \w{d}) &=& 2f^{bcd}\w{c} \wedge \iota_{\w{a}}(\w{d}) &=& 2f^{bca}\w{c}\\
\iota_{\wt{b}}\w{a} &=& \iota_{(f^{bcd}\w{c}\wedge\w{d})}\w{a} &=& f^{bcd}\iota_{\w{d}}\iota_{\w{c}}\w{a} &=& f^{bad}\iota_{\w{d}}
\ea
\eeq

\section{Flux matrix identities}

\subsection{Type IIB $\SL_{S}$ action}

Given an $\SL_{S}$ transformation on the flux matrices of Type IIB they transform in the following ways : 
\beq
\label{eqn:Saction}
S \to \frac{aS+b}{cS+d} \quad \Rightarrow \quad \Bigg\{ \quad \ba{ccc}
\bpm \DevenB \\ \DpevenB\epm &\to& \bpm \A \cdot (a\DevenB + b\DpevenB) + \B \cdot (d\DevenB + c\DpevenB) \\ \A \cdot (c\DevenB + d\DpevenB) + \B \cdot (b\DevenB + a\DpevenB)  \epm\\
\bpm \DoddB \\ \DpoddB\epm &\to& \bpm (a\DoddB+b\DpoddB) \cdot \A + (d\DoddB+c\DpoddB) \cdot \B \\ (c\DoddB+d\DpoddB) \cdot \A + (b\DoddB+a\DpoddB) \cdot \B\epm
\ea
\eeq

\subsection{Flux matrix constraints on $\Hn{3}$}

As used in the main body of work we compress notation by replacing $\cdot h_{\Hbasis} \cdot$ by $\hdot$ and consider how the flux matrix expressions associated to $\dD_{i}\dD_{j} : \Hn{3}\to\Hn{3}$ transform under (\ref{eqn:Saction}).

\beq 
\label{eqn:Sduality1a}
\ba{ccccc}
\DevenB \cdot \sh_{\Hbasis} \cdot \DoddB &\to& \Big(\A\,(a\DevenB+b\DpevenB) + \B\,(d\DevenB+c\DpevenB)\Big) \hdot \Big( ( a \DoddB + b\DpoddB) \A + (d\DoddB+c\DpoddB)\B \Big) \\
&=& \left(
\ba{ccccrc}
& \A \cdot \Big( & a^{2}\, \DevenB \hdot \DoddB + ab\, \DevenB \hdot \DpoddB &+& ab\, \DpevenB \hdot \DoddB + b^{2}\, \DpevenB \hdot \DpoddB & \Big) \cdot \A \\
+&\A \cdot \Big( & ad\, \DevenB \hdot \DoddB + ac\, \DevenB \hdot \DpoddB &+& bd\, \DpevenB \hdot \DoddB + bc\, \DpevenB \hdot \DpoddB & \Big) \cdot \B \\
+&\B \cdot \Big( & ad\, \DevenB \hdot \DoddB + bd\, \DevenB \hdot \DpoddB &+& ac\, \DpevenB \hdot \DoddB + bc\, \DpevenB \hdot \DpoddB & \Big) \cdot \A \\
+&\B \cdot \Big( & d^{2}\, \DevenB \hdot \DoddB + cd\, \DevenB \hdot \DpoddB &+& cd\, \DpevenB \hdot \DoddB + c^{2}\, \DpevenB \hdot \DpoddB & \Big) \cdot \B
\ea \right)
\ea
\eeq
\beq 
\label{eqn:Sduality1b}
\ba{ccccc}
\DevenB \cdot \sh_{\Hbasis} \cdot \DpoddB &\to& \Big(\A\,(a\DevenB+b\DpevenB) + \B\,(d\DevenB+c\DpevenB)\Big) \hdot \Big( ( c \DoddB + d\DpoddB) \A + (b\DoddB+a\DpoddB)\B \Big) \\
&=& \left(
\ba{ccccrc}
& \A \cdot \Big( & ac \, \DevenB \hdot \DoddB + ad\, \DevenB \hdot \DpoddB &+& bc\, \DpevenB \hdot \DoddB + bd\, \DpevenB \hdot \DpoddB & \Big) \cdot \A \\
+&\A \cdot \Big( & ab\, \DevenB \hdot \DoddB + a^{2}\, \DevenB \hdot \DpoddB &+& b^{2}\, \DpevenB \hdot \DoddB + ab\, \DpevenB \hdot \DpoddB & \Big) \cdot \B \\
+&\B \cdot \Big( & cd\, \DevenB \hdot \DoddB + d^{2}\, \DevenB \hdot \DpoddB &+& c^{2} \, \DpevenB \hdot \DoddB + cd\, \DpevenB \hdot \DpoddB & \Big) \cdot \A \\
+&\B \cdot \Big( & bd\, \DevenB \hdot \DoddB + ad\, \DevenB \hdot \DpoddB &+& bc\, \DpevenB \hdot \DoddB + ac\, \DpevenB \hdot \DpoddB & \Big) \cdot \B
\ea \right)
\ea
\eeq
\beq 
\label{eqn:Sduality1c}
\ba{ccccc}
\DpevenB \cdot \sh_{\Hbasis} \cdot \DoddB &\to& \Big(\A\,(c\DevenB+d\DpevenB) + \B\,(b\DevenB+a\DpevenB)\Big) \hdot \Big( ( a \DoddB + b\DpoddB) \A + (d\DoddB+c\DpoddB)\B \Big) \\
&=& \left(
\ba{ccccrc}
& \A \cdot \Big( & ac \, \DevenB \hdot \DoddB + bc\, \DevenB \hdot \DpoddB &+& ad\, \DpevenB \hdot \DoddB + bd\, \DpevenB \hdot \DpoddB & \Big) \cdot \A \\
+&\A \cdot \Big( & cd\, \DevenB \hdot \DoddB + c^{2}\, \DevenB \hdot \DpoddB &+& d^{2}\, \DpevenB \hdot \DoddB + cd\, \DpevenB \hdot \DpoddB & \Big) \cdot \B \\
+&\B \cdot \Big( & ab\, \DevenB \hdot \DoddB + b^{2}\, \DevenB \hdot \DpoddB &+& a^{2} \, \DpevenB \hdot \DoddB + ab\, \DpevenB \hdot \DpoddB & \Big) \cdot \A \\
+&\B \cdot \Big( & bd\, \DevenB \hdot \DoddB + bc\, \DevenB \hdot \DpoddB &+& ad\, \DpevenB \hdot \DoddB + ac\, \DpevenB \hdot \DpoddB & \Big) \cdot \B
\ea \right)
\ea
\eeq
\beq 
\label{eqn:Sduality1d}
\ba{ccccc}
\DpevenB \cdot \sh_{\Hbasis} \cdot \DpoddB &\to& \Big(\A\,(c\DevenB+d\DpevenB) + \B\,(b\DevenB+a\DpevenB)\Big) \hdot \Big( ( c \DoddB + d\DpoddB) \A + (b\DoddB+a\DpoddB)\B \Big) \\
&=& \left(
\ba{ccccrc}
& \A \cdot \Big( & c^{2}\, \DevenB \hdot \DoddB + cd\, \DevenB \hdot \DpoddB &+& cd\, \DpevenB \hdot \DoddB + d^{2}\, \DpevenB \hdot \DpoddB & \Big) \cdot \A \\
+&\A \cdot \Big( & bc\, \DevenB \hdot \DoddB + ac\, \DevenB \hdot \DpoddB &+& bd\, \DpevenB \hdot \DoddB + ad\, \DpevenB \hdot \DpoddB & \Big) \cdot \B \\
+&\B \cdot \Big( & bc\, \DevenB \hdot \DoddB + bd\, \DevenB \hdot \DpoddB &+& ac\, \DpevenB \hdot \DoddB + ad\, \DpevenB \hdot \DpoddB & \Big) \cdot \A \\
+&\B \cdot \Big( & b^{2}\, \DevenB \hdot \DoddB + ab\, \DevenB \hdot \DpoddB &+& ab\, \DpevenB \hdot \DoddB + a^{2}\, \DpevenB \hdot \DpoddB & \Big) \cdot \B
\ea \right)
\ea
\eeq

\subsection{Alternative flux relations}

The $\merge$ images of the $\sD_{i}\sD_{j} : \Hn{3}\to\Hn{3}$ expressions have the following expressions in terms of the $\sD_{i}\sD_{j}$ flux matrices.

\beq 
\label{eqn:W2fluxes1}
\ba{ccccc}
\DevenB_{2} \hdot \DoddB_{2}  &=& m_{2}\cdot \t{\A} \hdot \t{\A} \cdot n_{2} + m_{2}^{\prime}\cdot \t{\B} \hdot \t{\B} \cdot n_{2}^{\prime} \\
&=& \left(
\ba{crcl}
&\A \cdot \Big( \DevenB_{1} \cdot \t{\A} \hdot \DoddB_{1} &+& \DevenB_{1}^{\prime} \cdot \t{\B} \hdot \DoddB_{1}^{\prime}  \Big) \cdot \A \\
+&\B \cdot \Big( \DevenB_{1}^{\prime} \cdot \t{\A} \hdot \DoddB_{1} &+& \DevenB_{1} \cdot \t{\B} \hdot \DoddB_{1}^{\prime} \Big) \cdot \A \\
+&\A \cdot \Big( \DevenB_{1} \cdot \t{\A} \hdot \DoddB_{1}^{\prime} &+& \DevenB_{1}^{\prime} \cdot \t{\B} \hdot \DoddB_{1} \Big) \cdot \B \\
+&\B \cdot \Big( \DevenB_{1}^{\prime} \cdot \t{\A} \hdot \DoddB_{1}^{\prime} &+& \DevenB_{1}\cdot \t{\B} \hdot \DoddB_{1}  \Big) \cdot \B 
\ea \right)
\ea
\eeq
\beq 
\label{eqn:W2fluxes2}
\ba{ccccc}
\DevenB_{2}^{\prime} \hdot \DoddB_{2}^{\prime} &=& m_{2}\cdot \t{\B} \hdot \t{\B} \cdot n_{2} + m_{2}^{\prime}\cdot \t{\A} \hdot \t{\A} \cdot n_{2}^{\prime} \\
&=& \left(
\ba{crcl}
 &\A \cdot \Big( \DevenB_{1}^{\prime} \cdot \t{\A} \hdot \DoddB_{1}^{\prime} &+& \DevenB_{1} \cdot \t{\B} \hdot \DoddB_{1} \Big) \cdot \A \\
+&\B \cdot \Big( \DevenB_{1} \cdot \t{\A} \hdot \DoddB_{1}^{\prime} &+& \DevenB_{1}^{\prime} \cdot \t{\B} \hdot \DoddB_{1} \Big) \cdot \A \\
+&\A \cdot \Big( \DevenB_{1}^{\prime} \cdot \t{\A} \hdot \DoddB_{1} &+& \DevenB_{1} \cdot \t{\B} \hdot \DoddB_{1}^{\prime} \Big) \cdot \B \\
+&\B \cdot \Big( \DevenB_{1} \cdot \t{\A} \hdot \DoddB_{1} &+& \DevenB_{1}^{\prime} \cdot \t{\B} \hdot \DoddB_{1}^{\prime} \Big) \cdot \B 
\ea \right) 
\ea
\eeq
\beq 
\label{eqn:W2fluxes3}
\ba{ccccc}
\DevenB_{2} \hdot \DoddB_{2}^{\prime} &=&  m_{2}\cdot \t{\A} \hdot \t{\A} \cdot n_{2}^{\prime} + m_{2}^{\prime}\cdot \t{\B} \hdot \t{\B} \cdot n_{2}  \\
&=& \left(
\ba{crcl}
 &\A \cdot \Big( \DevenB_{1} \cdot \t{\A} \hdot \DoddB_{1}^{\prime} &+& \DevenB_{1}^{\prime} \cdot \t{\B} \hdot \DoddB_{1} \Big) \cdot \A \\
+&\B \cdot \Big( \DevenB_{1}^{\prime} \cdot \t{\A} \hdot \DoddB_{1}^{\prime} &+& \DevenB_{1} \cdot \t{\B} \hdot \DoddB_{1} \Big) \cdot \A \\
+&\A \cdot \Big( \DevenB_{1} \cdot \t{\A} \hdot \DoddB_{1} &+& \DevenB_{1}^{\prime} \cdot \t{\B} \hdot \DoddB_{1}^{\prime} \Big) \cdot \B \\
+&\B \cdot \Big( \DevenB_{1}^{\prime} \cdot \t{\A} \hdot \DoddB_{1} &+& \DevenB_{1} \cdot \t{\B} \hdot \DoddB_{1}^{\prime} \Big) \cdot \B 
\ea \right) 
\ea
\eeq
\beq 
\label{eqn:W2fluxes4}
\ba{ccccc}
\DevenB_{2}^{\prime} \hdot \DoddB_{2} &=&  m_{2}\cdot \t{\B} \hdot \t{\B} \cdot n_{2}^{\prime} + m_{2}^{\prime}\cdot \t{\A} \hdot \t{\A} \cdot n_{2} \\
&=& \left(
\ba{crcl}
 &\A \cdot \Big( \DevenB_{1}^{\prime} \cdot \t{\A} \hdot \DoddB_{1} &+& \DevenB_{1} \cdot \t{\B} \hdot \DoddB_{1}^{\prime} \Big) \cdot \A \\
+&\B \cdot \Big( \DevenB_{1} \cdot \t{\A} \hdot \DoddB_{1} &+& \DevenB_{1}^{\prime} \cdot \t{\B} \hdot \DoddB_{1}^{\prime} \Big) \cdot \A \\
+&\A \cdot \Big( \DevenB_{1}^{\prime} \cdot \t{\A} \hdot \DoddB_{1}^{\prime} &+& \DevenB_{1} \cdot \t{\B} \hdot \DoddB_{1} \Big) \cdot \B \\
+&\B \cdot \Big( \DevenB_{1} \cdot \t{\A} \hdot \DoddB_{1}^{\prime} &+& \DevenB_{1}^{\prime} \cdot \t{\B} \hdot \DoddB_{1} \Big) \cdot \B 
\ea \right) 
\ea
\eeq

\end{document}